\documentclass[structabstract]{aa}  

\usepackage{graphicx,txfonts,aalongtable,lscape,natbib}
\bibpunct{(}{)}{;}{a}{}{,} 

\begin{document}

\title{Trigonometric parallaxes of young field L dwarfs}

\author{M$.$ R$.$ Zapatero Osorio\inst{1}, V$.$ J$.$ S$.$ B\'ejar\inst{2,3},  P$.$ A$.$ Miles-P\'aez\inst{2,3}, 
        K$.$ Pe\~na Ram\'\i rez\inst{2,3}\thanks{ Currently at: Instituto de Astrof\'isica. Pontificia Universidad Cat\'olica de Chile (IA-PUC), E-7820436 Santiago, Chile.}, R$.$ Rebolo\inst{2,3,4}, 
   \and E$.$ Pall\'e\inst{2,3}}

\institute{Centro de Astrobiolog\'\i a (INTA-CSIC), Carretera de Ajalvir km 4, E-28850 Torrej\'on de Ardoz, Madrid, Spain.
   \and    Instituto de Astrof\'\i sica de Canarias, C/$.$ V\'\i a L\'actea s/n, E-38205 La Laguna, Tenerife, Spain.
   \and    Universidad de La Laguna, Tenerife, Spain.
   \and    Consejo Superior de Investigaciones Cient\'\i ficas (CSIC), Madrid, Spain.}

\date{Received 2013; accepted 2014}

\abstract{}{We aim to determine the trigonometric parallaxes and proper motions of a sample of ten field L0--L5 dwarfs with spectroscopic evidence for low-gravity atmospheres. The ten sources were located in color-absolute magnitude diagrams and in the Hertzsprung-Russell (HR) diagram for age and mass derivations, and were discussed in comparison with field and star cluster dwarfs of related spectral classification and with state-of-the-art solar-metallicity evolutionary models.}
{We obtained $J$ and $K_s$ imaging data using 2--4-m class telescopes with a typical cadence of one image per month between 2010 January and 2012 December, i.e., the data cover a time baseline of nearly three years. We also obtained low resolution optical spectra ($R$ $\sim$ 300, 500--1100 nm) using the 10-m Gran Telescopio de Canarias to assess the presence of lithium absorption in four targets and confirm their young age. The derived parallaxes and proper motions were combined with data from the literature to determine $T_{\rm eff}$, luminosity, and space velocities. All this information along with the lithium observations was used to assess the ages and masses of the sample. The astrometric curves were also examined for periodic perturbations indicative of unseen companions.}
{Trigonometric parallaxes and proper motions were derived to typical accuracies of a milliarcseconds (mas) and $\pm$10 mas\,yr$^{-1}$, respectively. All ten L dwarfs have large motions ($\mu \ge 70$ mas\,yr$^{-1}$), and are located at distances between 9 and 47 pc. They lie above and on the sequence of field dwarfs in the absolute $J$ and $K_s$ magnitude versus spectral type and luminosity versus effective temperature diagrams, implying ages similar to or smaller than those typical of the field. In the HR diagram, 2MASS J00332386$-$1521309 (L4), 2MASS\,J00452143$+$1634446 (L2), 2MASS\,J03552337$+$1133437 (L5),  2MASS\,J05012406$-$0010452 (L4), G\,196--3B (L3), 2MASS\,J17260007$+$1538190 (L3), and 2MASS\,J22081363$+$2921215 (L3) occupy locations compatible with most likely ages in the interval $\approx$10--500 Myr if they are single objects. All of these dwarfs (except for 2MASS\,J00332386$-$1521309) show strong lithium absorption at 670.8 nm, thus confirming the young ages and masses ranging from $\approx$11 through $\approx$45 M$_{\rm Jup}$ for this subsample. The detection of atomic lithium in the atmosphere of 2MASS\,J00452143$+$1634446 (L2) is reported for the first time.  The lack of lithium in 2MASS\,J00332386$-$1521309 (L4) is not compatible with its position in the HR diagram,  suggesting a spectral type earlier than L4. The remaining three dwarfs, 2MASS\,J02411151$-$0326587 (L0), 2MASS\,J10224821$+$5825453 (L1), and 2MASS\,J15525906$+$2948485 (L0) have locations in the HR diagram indicative of older ages and higher masses consistent with the observed lithium depletion previously published. The dynamical studies based on space velocities derived from our parallaxes and proper motions fully support the aforementioned results for  2MASS\,J00452143$+$1634446, 2MASS\,J03552337$+$1133437, G\,196--3B, 2MASS\,J10224821$+$5825453, and 2MASS\,J15525906$+$2948485. We did not find evidence for the presence of astrometric companions with minimum detectable masses typically $\ge$25 M$_{\rm Jup}$ and face-on, circular orbits with periods between 60--90 d and 3 yr around eight targets. }
{The astrometric and spectroscopic data indicate that about 60--70\%~of the field L-type dwarfs in our sample with evidence for low-gravity atmospheres are indeed young-to-intermediate-age brown dwarfs of the solar neighborhood with expected ages and masses in the intervals $\approx$10--500 Myr and $\approx$11--45 M$_{\rm Jup}$. The peaked-shape of the $H$-band spectra of L dwarfs, a signpost of youth, appears to be present up to ages of 120--500 Myr and intermediate-to-high gravities.}


\authorrunning{Zapatero Osorio et al$.$}
\titlerunning{Trigonometric parallaxes of young field L dwarfs}

\maketitle

\section{Introduction}
Trigonometric parallax is a vital parameter to understand the basic physical properties of new objects and to construct the population architecture of the solar neighborhood. To date there are more than 1200 spectroscopically confirmed  L- and T-type low-mass dwarfs in the nearby field\footnote{Based on the Dwarf Archive web site: http://dwarfArchives.org}, and the number of even cooler dwarfs is steadily increasing with the discoveries of late-T and Y-type objects \citep[and references therein]{cushing11,liu11,luhman12,lodieu12,kirk11,kirk12}. As indicated by the theory of stellar and substellar evolution (e.g., \citealt{burrows93,chabrier00a}), physical properties of substellar objects significantly change with time; therefore, at least two parameters among temperature, luminosity and age are required to determine the substellar mass. Many of the known Ls and Ts share an age similar to that of the Solar System and are therefore very low-mass stars and brown dwarfs. However, those with ages below a few ten Myr could have masses within the planetary regime\footnote{There is no wide consensus on the maximum mass of an exoplanet. Here, we adopt the deuterium burning mass limit as this maximum mass. Objects with masses between the deuterium and hydrogen burning mass limits are termed brown dwarfs irrespective of their formation process.}, i.e., below the deuterium burning mass  limit \citep{saumon96} defined at 12 times the mass of Jupiter (M$_{\rm Jup}$). Given the increasing number of ultracool (LTY) dwarf discoveries, the diversity of the L and T findings in age, chemical composition, and photometric and spectroscopic properties is becoming apparent. Distances provide crucial information for calculating luminosities, ages, and masses. The early parallax programs of \citet{dahn02}, \citet{tinney03}, and \citet{vrba04}, and the recent works based on optical and near-infrared data by \citet{burgasser08}, \citet{schilbach09}, \citet{marocco10}, \citet{andrei11}, \citet{dupuy12}, \citet{scholz12}, \citet{faherty12,faherty13}, \citet{marsh13}, \citet{beichman13}, and others have provided absolute parallaxes of over 150 L and T dwarfs  (less than 20\%~of all known L and T objects) and 8 Y dwarfs with metallicity approximately solar and subsolar. These studies have led to the precise location of the L--T spectrophotometric sequence in color-magnitudes diagrams as well as in the Hertzsprung-Russell (HR) diagram, and to the derivation of comprehensive relations between absolute magnitudes of ultracool dwarfs and spectral types. 

However, not all the details of the L--T dwarf sequence are fully understood, e.g., the large scatter that cannot be explained by multiplicity alone \citep{burrows06}, and the $J$- and $H$-band luminosity bumps near the L/T transition (e.g., \citealt{looper08}). \citet{marley10} suggested that, by decreasing cloudiness, objects become bluer in $J-K$ and brighten in $J$-band since the dust clearing makes the atmospheric 1 $\mu$m regions less opaque, which may account for the observed bump at the L/T transition. More recently, \citet[and references therein]{faherty12,faherty13} have claimed that some young ultracool dwarfs appear 0.2--1.0 mag underluminous at near-infrared wavelengths compared to equivalent spectral type objects, in clear contrast to what is predicted by the theory of stellar and substellar formation and evolution. These young dwarfs have red colors differing from those of their spectral counterparts, suggesting they have more dusty atmospheres.

Here we focus on a sample of ten field L dwarfs with a peculiar property: they show spectroscopic evidence for low-gravity atmospheres  (log\,$g$ $<$ 4.0--4.5 cm\,s$^{-2}$), i.e., young ages. According to \citet{cruz09}, many of them show spectroscopic features indicative of ages younger than the Pleiades. In addition, their very red near- and mid-infrared colors appear to differ from those of dwarfs with similar spectral classification, which has also been argued in favor of low gravities (a signpost of correlation between enhanced photospheric dust and youth). All ten L dwarfs have spectrophotometric distances estimated at less than 70 pc, making them suitable targets for trigonometric parallax determinations using ground-based facilities. Our main objectives are to characterize these objects properly in terms of age and mass, to provide strong relations between observed spectral features (like the presence of lithium in their atmospheres) and estimated ages/masses, and to contribute to the view of the solar vicinity population. In Section~\ref{targsel} we provide the description of the sample. Imaging and spectroscopic observations are described in Section~\ref{obs}. The astrometric analysis including trigonometric parallaxes and proper motions is given in Section~\ref{astroanalysis}.  Our measurements of the ten suspected ``young" L dwarfs are compared with other ultracool dwarfs in color-magnitude diagrams and with evolutionary stellar and substellar models in Section~\ref{discussion}, where we also provide age and mass estimates. Additionally, in Section~\ref{discussion} we explore the presence of astrometric companions around the targets and provide minimum masses of companions that could have been detected in our study given the quality of the data. Finally, conclusions are presented in Section~\ref{conclusions}.


\begin{figure}   
\center
\includegraphics[width=8.9cm]{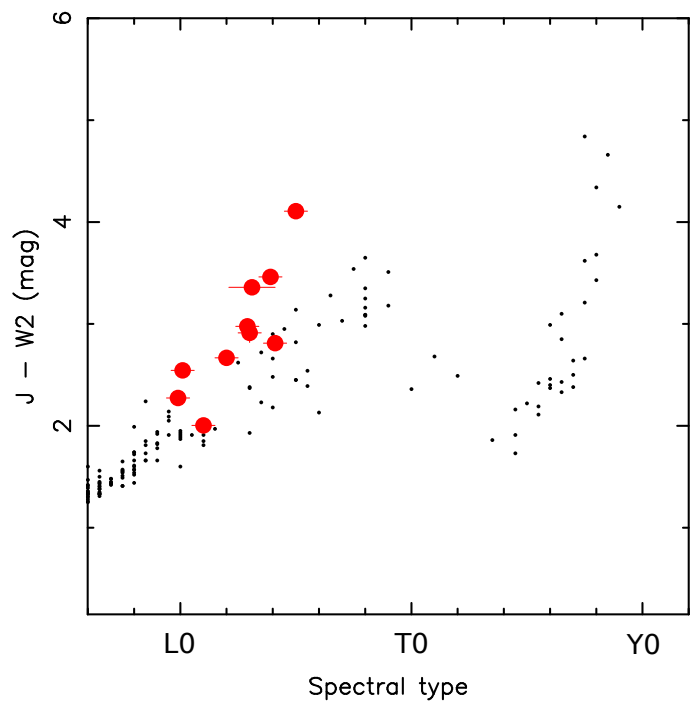}
\caption{$J-W2$ color as a function of spectral type for our ten targets (large-size dots) and field dwarfs (small dots). We used the 2MASS $J$-band data \citep{skrutskie06}. The field dwarf photometry was compiled by \citet{dupuy12}.
  \label{jw2}}
\end{figure}

\section{Target selection \label{targsel}}
Our list of targets comprises ten L0--L5 field dwarfs with spectroscopic evidence for low gravity atmospheres. Complete target names and spectral types based on optical spectra are shown in Table~\ref{targets}. In the following, we shall use abridged names to refer to program objects. All sources have near-infrared spectral classifications similar to the optical ones within 0.5--1 subtypes except for J0033$-$1521, which is classified as an L1 dwarf in the infrared \citep{allers13}. The sample is selected from the compilation of suspected young ultracool dwarfs in the field made by \citet{cruz09}; all of our targets are observable from the northern hemisphere and represent 44\%~of the total list built by these authors. According to their spectrophotometric distances, the ten targets are located within $\sim$70 pc from the Sun. They display notably weak CrH and FeH molecular absorption at around 861 and 870 nm,  and weak K\,{\sc i} and Na\,{\sc i} doublets at $\sim$770 and $\sim$820 nm. On the contrary, absorption due to metal oxides (e.g., Ti0 at 845 nm and VO at 745 nm) appear enhanced in our targets as compared to ``normal''  dwarfs of related spectral types. These properties, some already noted for young late-M dwarfs by \citet{martin96}, are attributable to reduced photospheric pressures typical of low gravity atmospheres. Consequently, all ten targets are believed to be undergoing self-gravitational contraction expected for ages typically below a few hundred Myr.  Eight targets (J0045$+$1634, J0241$-$0326, J0355$+$1133, J0501$-$0010, G\,196$-$3B, J1552$+$2948, J1726$+$1538, and J2208$+$2921) also have near-infrared spectral features supporting low gravity atmospheres \citep{mclean03,allers10,osorio10,faherty13,allers13}: the sharply peaked shape of their $H$-band spectra, a signpost of youth originally pointed out by \citet{lucas01}, and weaker near-infrared K\,{\sc i} lines than expected from the trend of high gravity dwarfs. Of the ten targets, half (G\,196$-$3B, J1726$+$1538, J2208$+$2921, J0501$-$0010, and J0355$+$1133) show Li\,{\sc i} absorption at 670.8 nm according to the literature \citep[and references therein]{rebolo98,cruz09}, thus confirming their brown dwarf mass (for a discussion on the lithium test and substellarity see \citealt{rebolo92,basri00,kirk08}).  In Table~\ref{summary} we provide the summary of the availability of optical and near-infrared spectra in the literature along with an indication of two youth spectroscopic features.

\addtocounter{table}{1}
\begin{table}
\small
\centering
\caption{Summary of optical and near-infrared spectra available in the literature, and presence of youth spectroscopic features. \label{summary}}
\begin{tabular}{llcccl}
\hline  \hline
\multicolumn{1}{l}{Object} & 
\multicolumn{1}{l}{SpT} & 
\multicolumn{1}{c}{Li\,{\sc i}\tablefootmark{a}} & 
\multicolumn{1}{c}{Peaked $H$\tablefootmark{b}} & 
\multicolumn{1}{c}{G\tablefootmark{c}} & 
\multicolumn{1}{l}{Ref.} \\ 
\hline
J0033$-$1521 & L4$\beta$  &  N    &  N   & FLD & 1,2 \\  
J0045$+$1634 & L2$\beta$  &  N,Y  &  Y   & VL  & 1,2,3\\  
J0241$-$0326 & L0$\gamma$ &  N    &  Y   & VL  & 1,2,3 \\    
J0355$+$1133 & L5$\gamma$ &  Y    &  Y   & VL  & 1,2,3,4 \\  
J0501$-$0010 & L4$\gamma$ &  Y    &  Y   & VL  & 1,2,5 \\  
G\,196--3B   & L3$\beta$  &  Y    &  Y   & VL  & 1,2,3,5,6,7 \\  
J1022$+$5825 & L1$\beta$  &  N    &  N   & FLD & 1,2 \\  
J1552$+$2948 & L0$\beta$  &  N    &  Y   & INT & 1,2 \\    
J1726$+$1538 & L3$\beta$  &  N,Y  &  Y   & INT & 8,1,2 \\  
J2208$+$2921 & L3$\gamma$ &  Y    &  Y   & VL  & 1,2 \\  
\hline
\end{tabular}
\tablefoot{
\tablefoottext{a}{Lithium at 670.8 nm. For a couple of objects, some authors claim detection (``Y'') and others do not (``N''), likely depending on the quality and spectral resolution of the data.}
\tablefoottext{b}{``N'' stands for no detection of this feature in the near-infrared spectrum.} 
\tablefoottext{c}{Gravity class indicator from \citet{allers13}; very low (VL), intermediate (INT), and high (FLD) gravity.} \\
References. (1) \citealt[and references therein]{cruz09}; (2) \citealt{allers13}; (3) this paper; (4) \citealt{faherty13}; (5) \citealt{allers10}; (6) \citealt{rebolo98}; (7) \citealt{osorio10}; (8) \citealt{schweitzer01}.
}
\end{table}

The near- and mid-infrared photometry, and \citet{cruz09} optical spectral classification of all targets are summarized in Table~\ref{targets}.  Regarding the Wide-field Infrared Survey Explorer ({\sl WISE}) data, we considered only magnitudes with associated signal-to-noise ratio of three and higher, i.e., photometric error bars $\le$0.4 mag. Most sources are detected in the {\sl WISE} $W1$ (3.3526 $\mu$m), $W2$ (4.6028 $\mu$m), and $W3$ (11.5608 $\mu$m) bands, and there is only one detection in the $W4$ (22.0883$\mu$m) filter as indicated in Table~\ref{targets} (for a description of {\sl WISE} see \citealt{wright10}). G\,196$-$3B (located at 16\arcsec~from its primary star) remains unresolved in the {\sl WISE} public archive. The {\sl Spitzer} photometry, available for six objects in the sample, is taken from \citet{luhman09} and \citet{osorio10}. In Figure~\ref{jw2} we depict the $J-W2$ color as a function of M, L, and T spectral types. For G\,196$-$3B we assumed that the $W2$  magnitude is similar to the {\sl Spitzer} $[4.5]$ data. This is a reasonable assumption for the L dwarfs as discussed in \citet{osorio11}: the widths of the {\sl WISE W2} and {\sl Spitzer} $[4.5]$ bands are alike, and although the {\sl WISE} filter is slightly shifted to redder wavelengths, there is no strong molecular absorption at these frequencies in the spectra of M and L dwarfs. Furthermore, \citet{wright11} also noted that the color term between {\sl Spitzer} $[4.5]$ and {\sl WISE W2} is rather small  ($[4.5] - W2$ = 0.054 mag) and seems to have no trend with spectral type even for the T dwarfs.

Our targets are shown in comparison with tens of field dwarfs compiled by \citet[Tables 12 and 13]{dupuy12} in Figure~\ref{jw2}. Three program objects (G\,196$-$3B, J0355$+$1133, and J0501$-$0010) appear to be redder by 0.5--1 mag than the field dwarfs of related types, while J1022$+$5825 sits on the average $J-W2$ index of ``normal'' dwarfs. The remaining targets have colors along the upper red envelope of the field. This reddish nature was already discussed in the literature \citep{osorio10,faherty13}, and it is believed to be caused by thick condensate clouds in low gravity atmospheres and/or the presence of circum(sub)stellar material. With $(J-W2)$\,=\,4.11 mag, which seems typical of the late-Ts as illustrated in Figure~\ref{jw2}, and $J-K_s$\,=\,2.52 mag, J0355$+$1133 and the recently discovered WISEP\,J004701.06$+$680352.1 \citep{gizis12},  PSO\,J318.5338$-$22.8603 \citep{liu13a}, and 2MASS\,J01225093$-$2439505\,B \citep{bowler13} turn out to be the reddest field L dwarfs detected so far. Understanding the nature of the population of unusually red field L dwarfs  in the solar vicinity is one of the objectives of this paper. 


\section{Observations \label{obs}}
\subsection{Near-infrared images}
Near-infrared images were acquired with cameras installed on 2-m and 4-m class telescopes. The great bulk of the data were taken with the prime focus instrument OMEGA2000 of the 3.5-m telescope on the Calar Alto Observatory (Almer\'\i a, Spain). This camera has one 2048\,$\times$\,2048 pixels HgCdTe HAWAII-2 array projecting a wide field of view (15\farcm4\,$\times$\,15\farcm4) on the sky. The OMEGA2000 pixel size is 0\farcs45. Some data were also collected with the NOTCam instrument based on the Rockwell Science Center HAWAII array with 1024\,$\times$\,1024 pixels in HgCdTe and covering a field of view of 4\arcmin\,$\times$\,4\arcmin. NOTCam has a pixel size of 0\farcs234 on the sky and is attached to the Cassegrain focus of the 2.5-m Nordic Optical Telescope on the Roque de los Muchachos Observatory (La Palma Island, Spain). To a lesser extent, the third instrument used for this project was the Long-slit Intermediate Resolution Infrared Spectrograph (LIRIS) available at the Cassegrain focus of the 4.2-m William Herschel Telescope (Roque de los Muchachos Observatory). LIRIS uses a 1024\,$\times$\,1024 HAWAII detector with a pixel scale of 0\farcs25, yielding a field of view of 4\farcm27\,$\times$\,4\farcm27. All three instruments cover the wavelength range from 0.8 through 2.5 $\mu$m. The OMEGA2000 and LIRIS images were collected with the $K_s$ filter (bandwidth of 1.99--2.31 $\mu$m), while the majority of the NOTCam data were obtained using the $J$ filter (bandwidth of 1.17--1.33 $\mu$m). 

For a proper sampling of the parallax amplitude and objects' motion, OMEGA2000 observations were typically carried out with a cadence of once per month from 2010 January through 2012 December (i.e., a time baseline of 3 yr). Because of a technical failure of the Calar Alto 3.5-m telescope, OMEGA2000 observations were interrupted for about eight months from 2010 August through 2011 March. NOTCam and LIRIS data were obtained during this time interval providing useful astrometric measurements that fill in the temporal gap left by the OMEGA2000 data. The log of the observations is shown in Table~\ref{obslog}, where we list the Universal Time (UT) observing dates, exposure times, instruments, filters, and the full-width-at-half-maximum (FWHM) of the reduced images. The seeing of the data ranges from 0\farcs5 through 2\farcs6, with a median value at 1\farcs0. Most images were collected with air masses in the interval 1.0--2.5 or zenith distances of $z$\,=\,0\degr -- 66\fdg4. There is a total of 262 observing epochs for the ten young field L dwarfs observable from northern observatories. 

For each epoch, the observing strategy consisted of acquiring frames following a muti-point dither pattern for a proper subtraction of the sky contribution. The dither pattern cycle of the NOTCam and LIRIS data was repeated a few times for each target to increase the signal-to-noise ratio of the final measurements. Individual frames were dark-current subtracted and divided by a flat-field to correct for the pixel-to-pixel variations. Sky contribution was removed from all individual images, which were finally registered and stacked together to produce deep data. This last step is necessary in order to retain a significant number of bright point-like sources in the NOTCam and LIRIS data since these two instruments provide relatively small fields of view. All image reduction steps were executed within the {\sc iraf}\footnote{{\sc iraf} is distributed by the National Optical Astronomy Observatories, which are operated by the Association of Universities for Research in Astronomy, Inc., under cooperative agreement with the National Science Foundation.} environment.

\addtocounter{table}{1}
\begin{table*}
\centering
\caption{Log of spectroscopic observations and lithium measurements. \label{spec}}
\begin{tabular}{llccccc}
\hline  \hline
\multicolumn{1}{l}{Object} & 
\multicolumn{1}{l}{SpT} & 
\multicolumn{1}{c}{UT Obs.~date} & 
\multicolumn{1}{c}{Exposure} & 
\multicolumn{1}{c}{Slit width} & 
\multicolumn{1}{c}{Air mass} & 
\multicolumn{1}{c}{$pEW$ (Li\,{\sc i})} \\
\multicolumn{1}{l}{} & 
\multicolumn{1}{r}{} & 
\multicolumn{1}{r}{} & 
\multicolumn{1}{c}{(s)} & 
\multicolumn{1}{c}{(\arcsec)} & 
\multicolumn{1}{r}{} & 
\multicolumn{1}{c}{(\AA)} \\
\hline
J0045$+$1634 & L2$\beta$  & 2012 Dec 14 & 2$\times$300  & 1.0 & 1.09 & 3.1$\pm$0.4 \\
J0241$-$0326 & L0$\gamma$ & 2012 Dec 15  & 2$\times$600 &  1.2   & 1.18     &  $\le$10       \\
J0355$+$1133 & L5$\gamma$ & 2012 Dec 13 & 2$\times$300  & 0.8 & 1.12 & 9.2$\pm$0.4 \\
G\,196--3B   & L3$\beta$  & 2012 Dec 16 & 3$\times$180      & 0.8 & 1.10 & 8.0$\pm$2.0 \\
\hline
\end{tabular}
\end{table*}

\subsection{Optical spectroscopy}
Aimed at addressing the presence and possible variability of the intensity of lithium in the atmospheres of J0045$+$1634, J0241$-$0326, J0355$+$1133, and G\,196$-$3B, we carried out spectroscopic observations using the Optical System for Imaging and low-Resolution Integrated Spectroscopy (OSIRIS) spectrograph \citep{cepa98} of the Gran Telescopio de Canarias (GTC) on Roque de los Muchachos Observatory (La Palma, Spain). Of the four spectroscopic targets, the two former ones have no lithium detection reported in the literature, while the two later ones display a strong absorption feature at 670.8 nm (see Table~\ref{summary}). OSIRIS consists of a mosaic of two Marconi CCD42-82 (2048\,$\times$\,4096 pix). All of our spectra were registered on the second detector, which is the default detector (in terms of cosmetics) for long-slit spectroscopy. We used a binning of 2\,$\times$\,2 providing a pixel size of 0\farcs254, and the grism R300R, yielding a spectral coverage of 500--1100 nm and a nominal dispersion of 7.68 \AA\,pix$^{-1}$. The slit widths were 0\farcs8, 1\farcs0, and 1\farcs2 depending on the seeing conditions. Typical seeing was 0\farcs7--1\farcs0 at optical wavelengths and weather conditions were hampered by thin cirrus (except for G\,196$-$3B). In Table~\ref{spec} we provide the log of the OSIRIS observations, which include UT observing dates, exposure times, slit widths, and air masses. The binning of the pixels along the spectral direction and the projection of the slits onto the detector yielded spectral resolutions of $R$\,=\,315 for J0355$+$1133 and G\,196$-$3B, and $R$\,=\,250 for J0045$+$1634 at 750 nm. An order blocking filter blueward of 450 nm was used; however, there may exist second-order contribution redward of 950 nm (particularly important for blue objects), which was accounted for by observing the spectrophotometric standard star using the broad $z$-band filter and the same spectroscopic configuration as that of the science targets. Spectra were typically acquired at parallactic angle and at two nodding positions along the slit separated by 10\arcsec~for a proper subtraction of the sky emission contribution.

Raw images were reduced with standard procedures including bias subtraction and flat-fielding within {\sc iraf}. A full wavelength solution from calibration lamps taken during the observing nights was applied to the spectra. The error associated with the fifth-order Legendre polynomial fit to the wavelength calibration is typically 10\%~the nominal dispersion. We corrected the extracted spectra for instrumental response using data of the spectrophotometric standard star G\,158$-$100 (white dwarf) observed on the same nights and with the same instrumental configuration as our targets. This standard star has fluxes available in \citet{filippenko84}. To complete the data reduction, target spectra were divided by the standard star to remove the contribution of telluric absorption; the intrinsic features of the white dwarf were previously interpolated and its spectrum was normalized to the continuum before using it for division into the science data. The spectrophotometric standard star was observed a few hours before the targets, therefore some telluric residuals may be present in the corrected spectra, particularly the strong O$_2$ band at 760.5 nm.

\begin{figure}   
\center
\includegraphics[width=9cm]{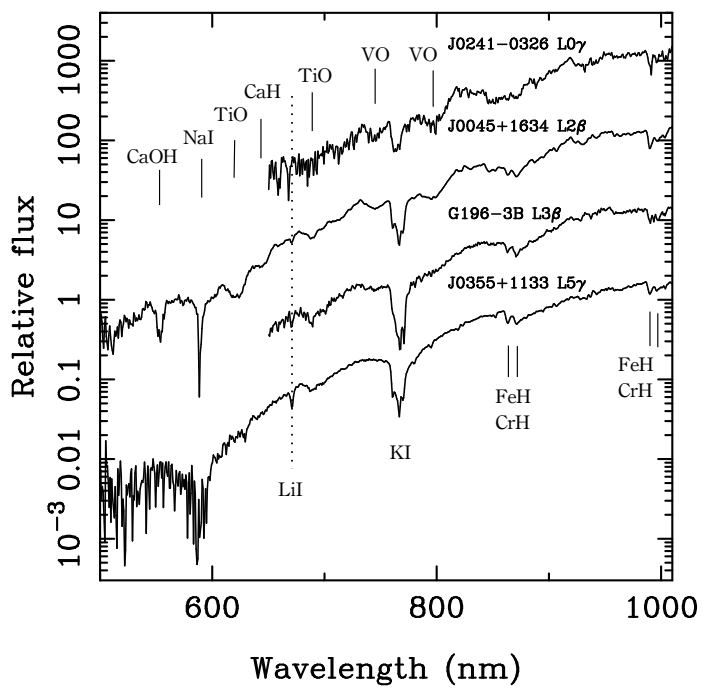} 
\caption{OSIRIS/GTC spectra of four of our targets (spectral resolution of about 300 at 750 nm). The most prominent atomic and molecular absorption features are labeled, the Li\,{\sc i} absorption line is indicated by a vertical dotted line. Telluric contribution is removed although some telluric residuals are present at 760.5 nm. The blue wavelengths of G\,196--3B and J0241$-$0326 spectra are not shown because they are very noisy (e.g., the feature of J0241$-$0326 immediately to the blue of the Li\,{\sc i} location is likely not real). Note that the relative flux axis is in logarithmic scale. All spectra are normalized to unity at 914--918 nm and are shifted by a constant for clarity.
\label{spectra}}
\end{figure}

The resulting reduced GTC spectra are depicted in Figure~\ref{spectra}. They are ordered by increasing spectral type and shifted by a constant for clarity. In order to clearly show the spectral features at blue wavelengths (e.g., the Na\,{\sc i} absorption at 589.3 nm), the relative fluxes of the spectra are plotted in logarithmic scale. For a detailed description of the atomic and molecular absorptions dependence on L spectral type see \citet{martin99}, \citet{kirk99}, and \citet{kirk05}.

\subsubsection{Lithium absorption}
The lithium absorption doublet at 670.8 nm is detected in J0045$+$1634, J0355$+$1133 and G\,196$-$3B. We measured the equivalent width of the spectral feature (note that the doublet is not resolved) with respect to the objects' relative continuum or pseudo-continuum modulated by the strong absorptions due to TiO, Na\,{\sc i} to the blue, and K\,{\sc i} to the red. We therefore refer to pseudo-equivalent widths ($pEW$). To compensate for the slightly different spectral resolution of the data, the integration of the line profile was always performed over the range 667.7--673.3 nm. Our measurements and their associated uncertainties are provided in Table~\ref{spec}. For J0241$-$0326, only an upper limit on the line $pEW$ was obtained. The Li\,{\sc i} $pEW$s of J0355$+$1133 and G\,196$-$3B are about 38\%~larger than those given in Table~1 by \citet{cruz09} because we selected a broad wavelength interval for the integration of the line profile that is appropriate for the GTC low-resolution spectra. We measured the Li\,{\sc i} $pEW$ over the Keck spectrum of J0355$+$1133 published by \citet{cruz09} and using the aforementioned method deriving $pEW$\,=\,9.5\,$\pm$\,1.0 \AA, in perfect agreement with the value obtained from the GTC data (see Table~\ref{spec}). 

The detection of lithium in absorption in the atmosphere of the L2$\beta$ dwarf J0045$+$1634 is reported for the first time. We measured $pEW$ (Li\,{\sc i})\,=\,3.1\,$\pm$\,0.4 \AA, which is compatible with the upper limit of $<$3 \AA~claimed by \citet{cruz09}. The Li\,{\sc i} strengths of J0045$+$1634, J0355$+$1133 and G\,196$-$3B compare with the $pEW$s of other L2--L5 field dwarfs and with the observed trend where Li\,{\sc i} $pEW$ increases with later subtypes \citep{kirk00}.

\section{Astrometric analysis \label{astroanalysis}}
To derive trigonometric parallaxes and proper motions, we selected the OMEGA2000 observations of 2010 January as the fundamental reference frames to which all other OMEGA2000 and LIRIS images are compared. The NOTCam data have their own reference frames corresponding to different epochs for each target; they were later referred to the OMEGA2000 reference frames by selecting observations taken on the same night or separated by less than three days. The reference frame of each program object is used to minimize any rotation, translation, and scaling changes between frames. Using the {\sc daofind} command within {\sc iraf} we identified all sources with photon peaks with detection above 8--10 $\sigma$, where $\sigma$ stands for the noise of the background, and FWHM resembling that of unresolved objects (i.e., extended sources were mostly avoided). In addition, we cared that the detected sources lied within the linear regime of the detectors response. For all targets and OMEGA2000 images, the number of sources identified per frame well exceeded 100 objects (in a few cases, the number of detected sources reached $\sim$700 objects). The centroids of detected objects were computed by estimating the $x$ and $y$ pixel positions of the best fitting one-dimensional Gaussian functions in each axis; typical associated errors are about 3--5\%~of a pixel or better. 

Pixel coordinates were transformed between different epochs using the {\sc geomap} routine within {\sc iraf}, which applied a polynomial of (typically) third order in $x$ and $y$ and computed linear terms and distortions terms separately. The linear term included an $x$ and $y$ shift, and $x$ and $y$ scale factor, a rotation, and a skew. The distortion surface term consisted of a polynomial fit to the residuals of the linear term. The $(x,y)$ astrometric transformation between observing epochs and the reference epoch was an iterative step, which included the rejection of objects deviating by more than 1.5--2 $\sigma$, where $\sigma$ corresponds to the dispersion of the transformation. Typical coordinates transformation dispersions ranged from 0.024 to 0.065 OMEGA2000 pixels (11--30 milliarcseconds, mas). The uncertainty is about 1.5--2 times larger for the NOTCam data likely because a moderate-to-high number of reference sources with poor centroid determinations were considered in the astrometric transformations (15--40 objects, the fields of view of OMEGA2000 and NOTCam are quite different in size), and because NOTCam may have some astrometric distortions that are not well ``erased'' by the third-order polynomials. During the total of three years of observations, we did not detect a significant change in the OMEGA2000 pixel size (maximum difference of 0.08\%) and the relative angle orientation of the frames (maximum difference of 0\fdg38 with a mean value at $-$0\fdg004\,$\pm$\,0\fdg004), thus proving the astrometric stability of OMEGA2000. Relative NOTCam pixel size and angle orientation were also stable within $\pm$0.4\%~and $\pm$0\fdg2. 

In absolute terms, using the 2MASS astrometry, which has an internal accuracy of about 80--100 mas \citep{skrutskie06}, and the OMEGA2000 reference frames we checked that the pixel size is 450.0\,$\pm$\,1.8 mas (average of ten measurements), where the uncertainty corresponds to the error of the mean. We also found that the reference frames were rotated with respect to the north-east orientation by 0\fdg28\,$\pm$\,0\fdg01. This has little impact in our astrometric analysis ($<$0.5\% in both axis), typically within the quoted error bars. We determined that the NOTCam pixel size is 234.3\,$\pm$\,1.0 mas, and that the NOTCam reference frames were rolled with respect to the corresponding OMEGA2000 references by angles ranging from $-$0\fdg14 to $-$0\fdg45. Those NOTCam frames with the largest rolled angles were conveniently derotated. 

Because our data were acquired at near-infrared wavelengths (mostly $K_s$), corrections by refraction due to the Earth's atmosphere and the large field of view of the detectors (particularly OMEGA2000) are expected to be small \citep{filippenko82}. Furthermore, since we are using relative astrometry, only the differential refraction is relevant, and this effect is in practice accounted for by using polynomial astrometric transformations of degree three and higher. According to \citet{fritz10}, we estimate the differential refraction errors to be significantly below 1 mas in our relative astrometry. Regarding the chromatic differential refraction, our targets are redder and show stronger water vapor absorption at near-infrared wavelengths than the vast majority of the reference sources used in the astrometric transformations. By convolving the response curve of the $K_s$ filter and the spectral energy distributions of a sample of GKML objects whose spectra were taken from the NASA Infrared Telescope Facility spectral library\footnote{http://irtfweb.ifa.hawaii.edu/$\sim$spex/IRTF\_Spectral\_Library/} \citep{cushing05,rayner09}, we found that GK(early)M stars have very similar effective wavelength in the $K_s$ band (2.130\,$\pm$\,0.003 $\mu$m), while for the L5-type dwarfs we derived an effective wavelength of 2.149 $\mu$m. Various works in the literature (e.g., \citealt[and references therein]{monet92,stone02,faherty12}) have demonstrated that the differential color refraction corrections are minimal at infrared wavelengths; we estimated them to be about a few mas for the largest zenith distances of our observations, i.e., smaller than the quoted astrometric uncertainties for individual images. For low zenith distances, the corrections are of the order of sub-mas. Therefore, we did not attempt to apply the differential chromatic refraction correction to the ($x,y$) positions in our astrometric analysis procedure.

\begin{figure*}
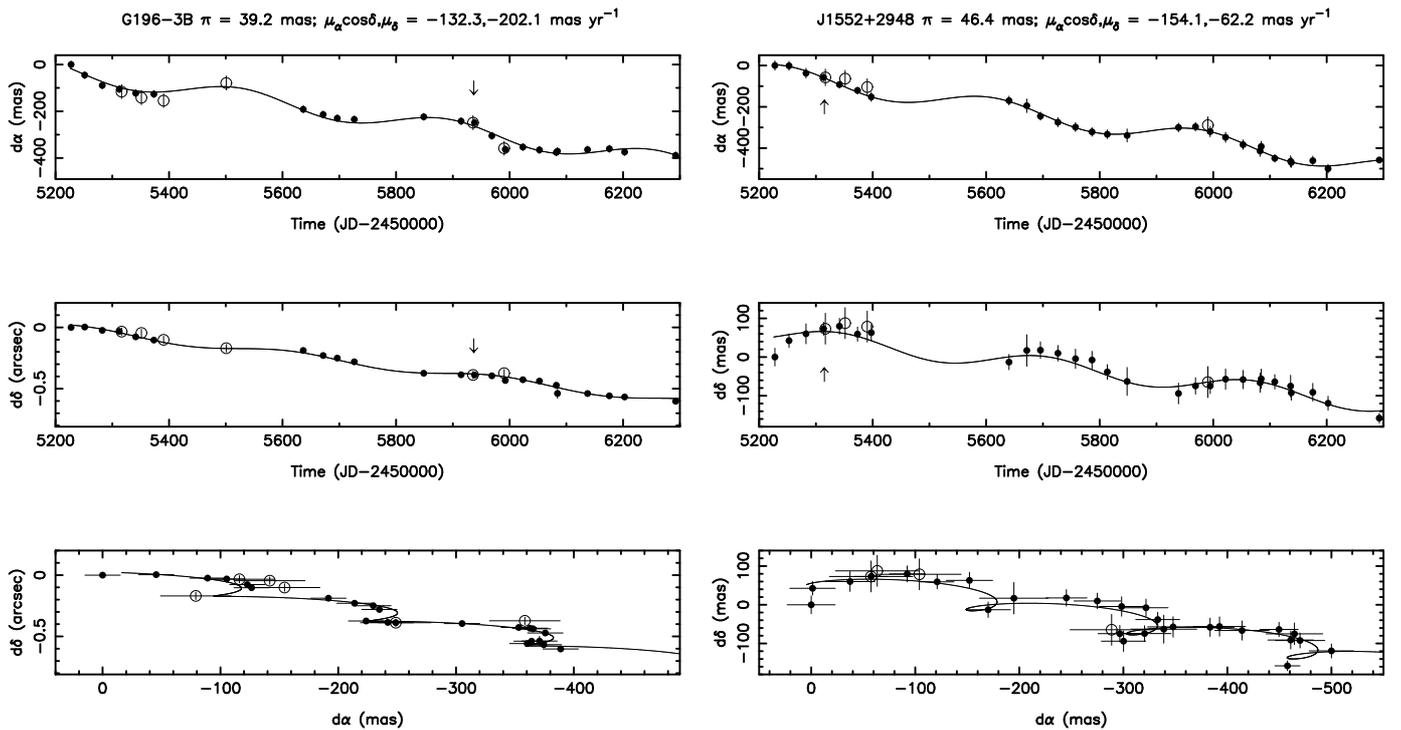
   
\center
\includegraphics[width=8.9cm]{g196.ps} ~~
\includegraphics[width=8.9cm]{j1552.ps}
\caption{Differential astrometric measurements obtained for G\,196$-$3B and J1552$+$2948 in the time interval from 2010 January through 2012 December. For each object, the top and middle panels depict the apparent trajectories of d$\alpha$ and d$\delta$, respectively, and the bottom panel illustrates d$\delta$ against d$\alpha$, where north is up and east is to the left. The reference epoch (2010 January) of the OMEGA2000 data corresponds to (d$\alpha$, d$\delta$)\,=\,(0, 0). The NOTCam data are normalized to the OMEGA2000 reference system at the observing epochs indicated by the arrows. OMEGA2000 measurements are plotted as solid circles, while NOTCam and LIRIS data (when available) are shown with open circles and triangles, respectively. The solid curves stand for the best fits to the proper motions and parallaxes. Astrometric error bars are plotted; in some panels, they have the size of the symbols.  The plots of the remaining targets are available in the online version of the paper. \label{astro1}}
\end{figure*}

\onlfig{
\begin{figure*}   
\centering
\includegraphics[width=8.9cm]{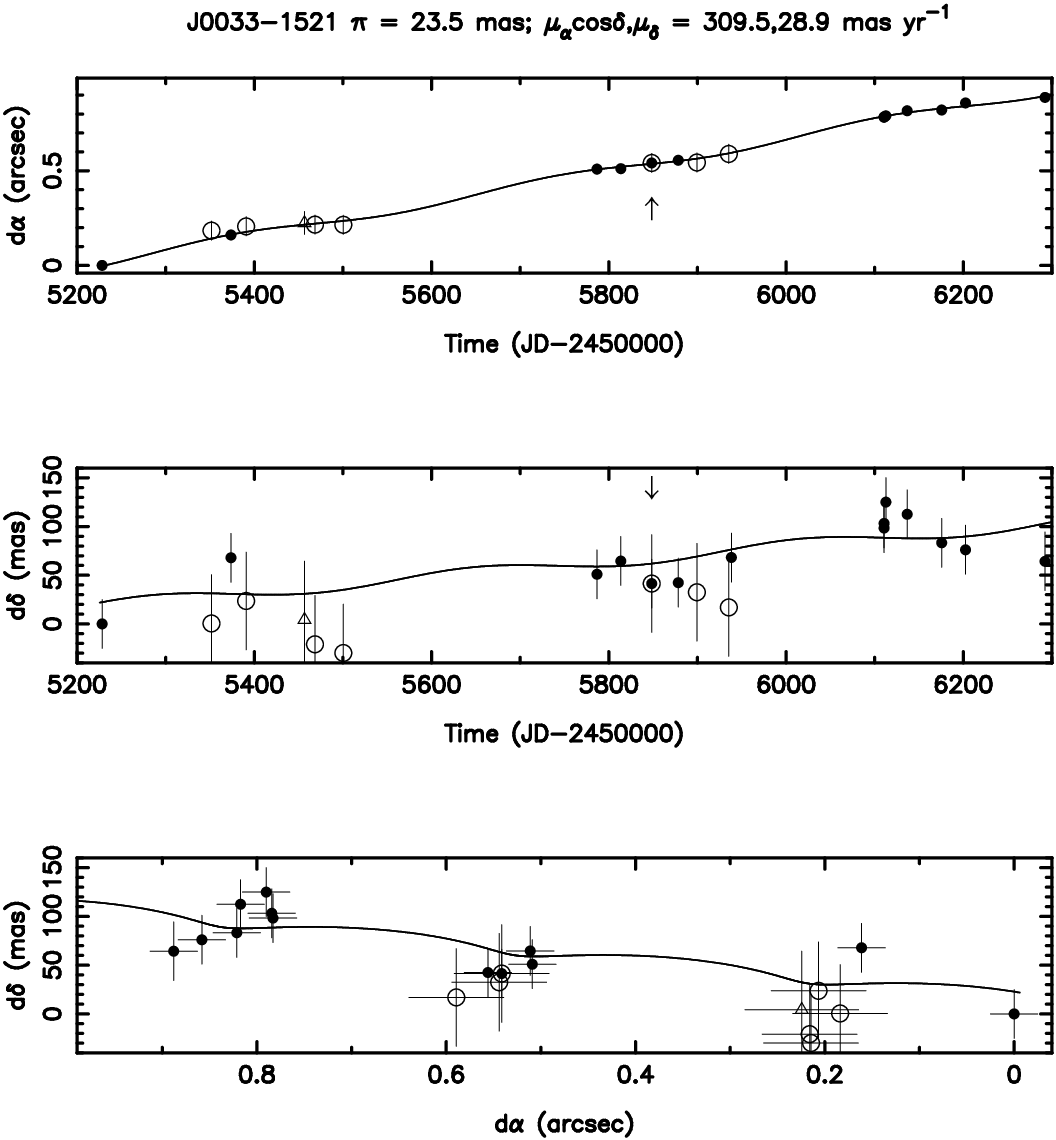} ~~
\includegraphics[width=8.9cm]{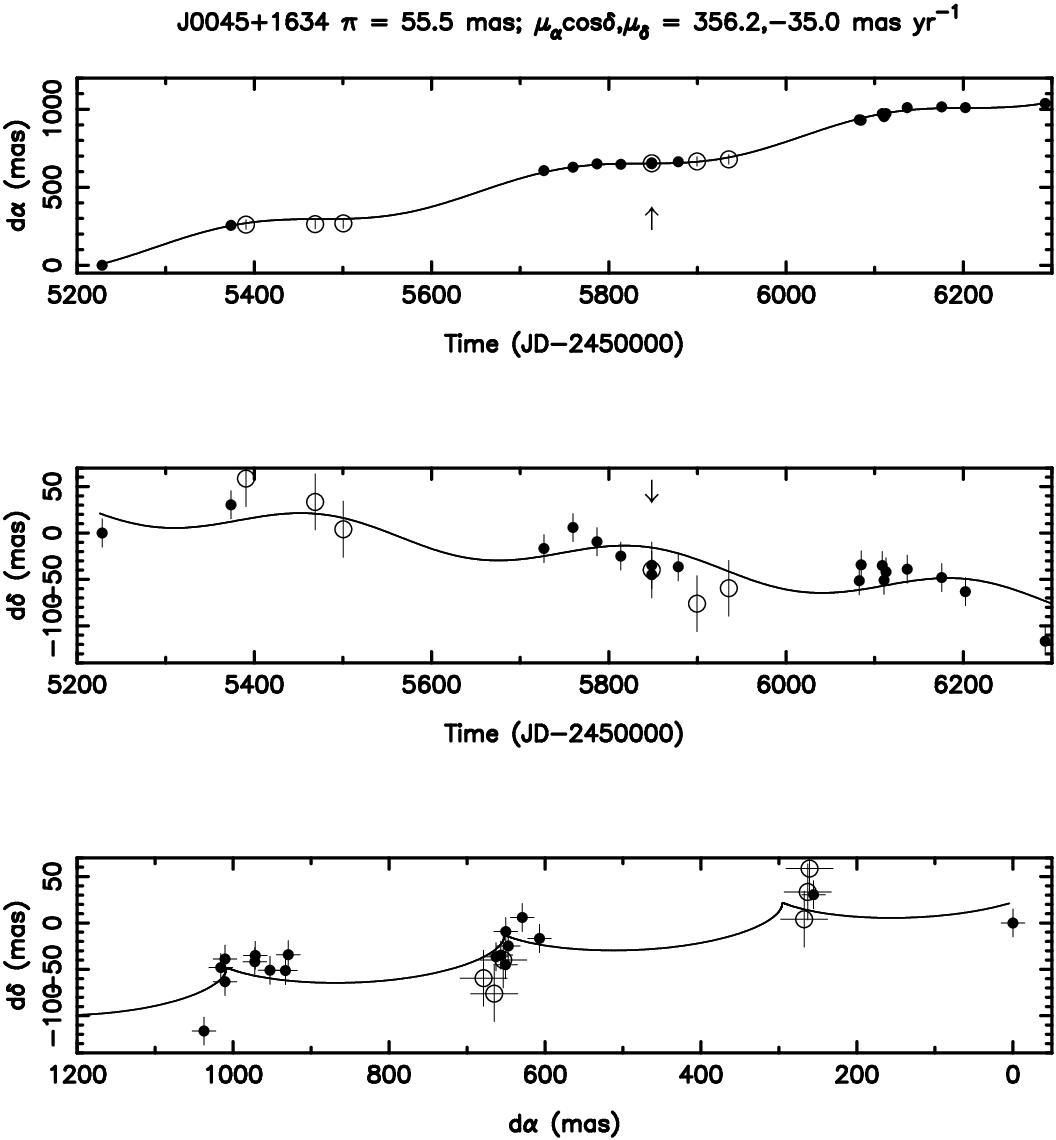} ~~
\includegraphics[width=8.9cm]{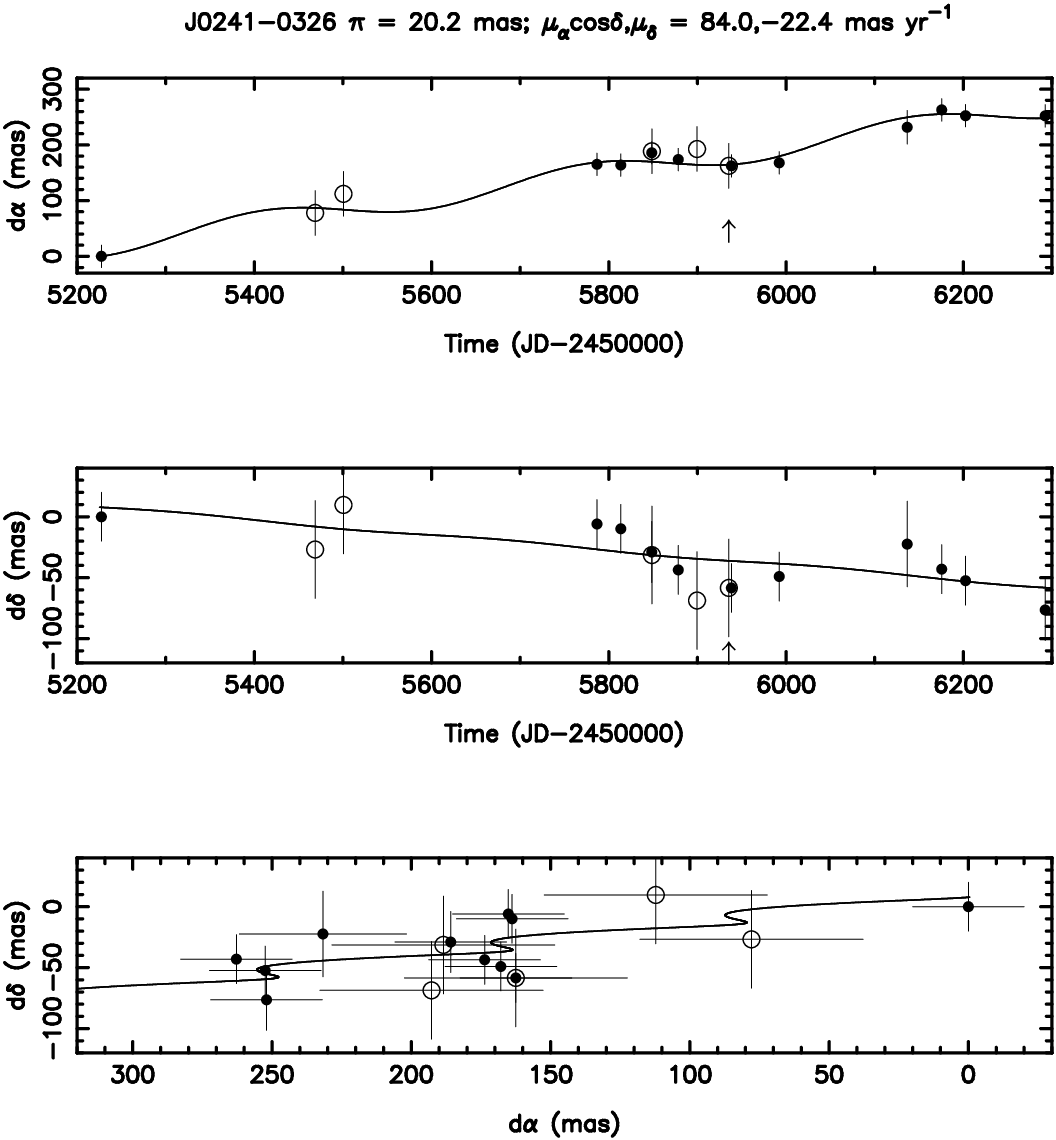}  ~~
\includegraphics[width=8.9cm]{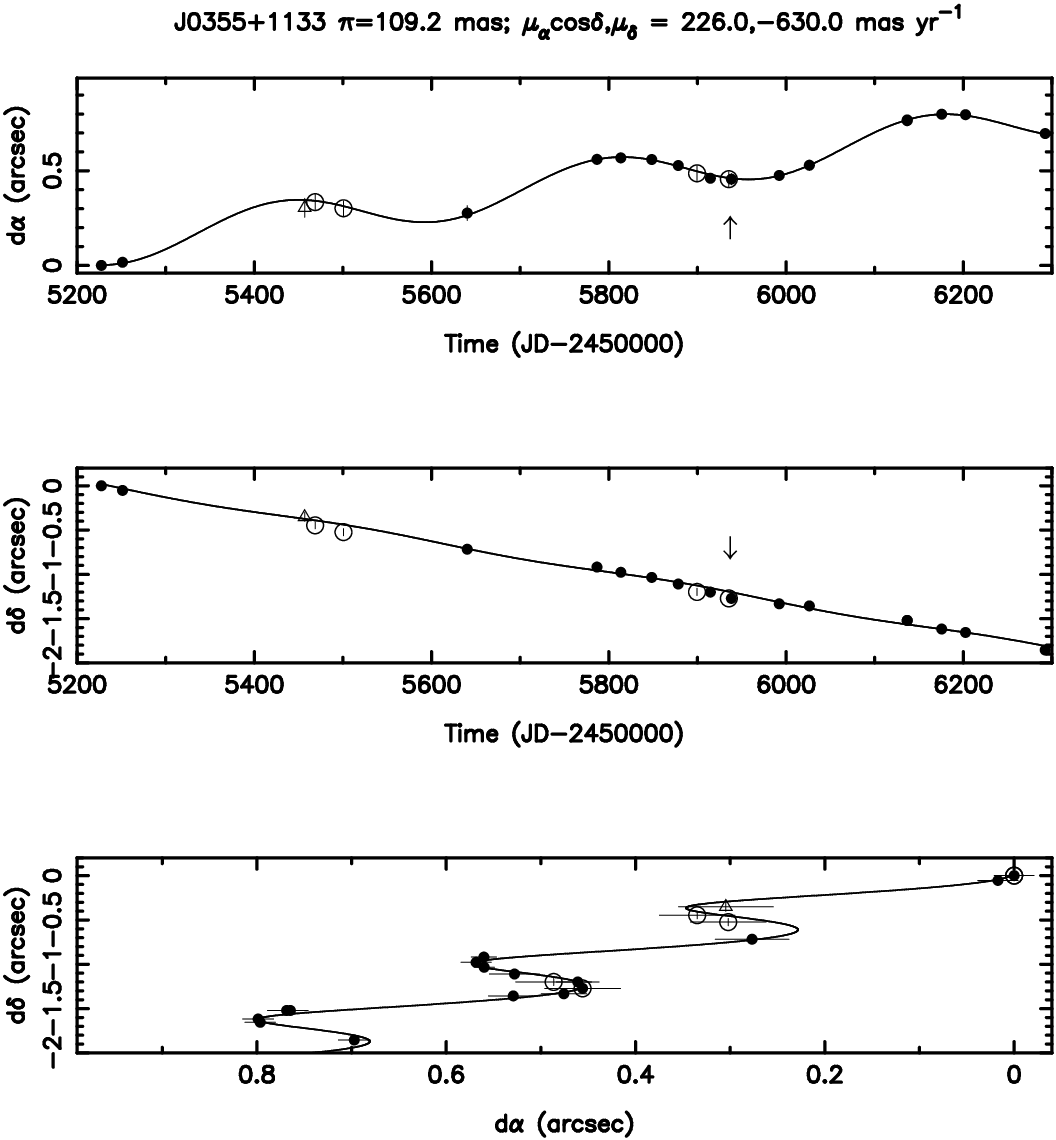} 
\caption{Differential astrometric measurements obtained for J0033$-$1521, J0045$+$1634, J0241$-$0326, and J0355$+$1133 in the time interval from 2010 January through 2012 December. For each object, the top and middle panels depict the apparent trajectories of d$\alpha$ and d$\delta$, respectively, and the bottom panel illustrates d$\delta$ against d$\alpha$, where north is up and east is to the left. The reference epoch (2010 January) of the OMEGA2000 data corresponds to (d$\alpha$, d$\delta$)\,=\,(0, 0). The NOTCam data are normalized to the OMEGA2000 reference system at the observing epoch indicated by the arrows. OMEGA2000 measurements are plotted as solid circles, while NOTCam and LIRIS data (when available) are shown with open circles and triangles, respectively. The solid curves stand for the best fits to the proper motions and parallaxes. Astrometric error bars are plotted; in some panels, they have the size of the symbols.  \label{astro2}}
\end{figure*}
}

\onlfig{
\begin{figure*}   
\center
\includegraphics[width=8.9cm]{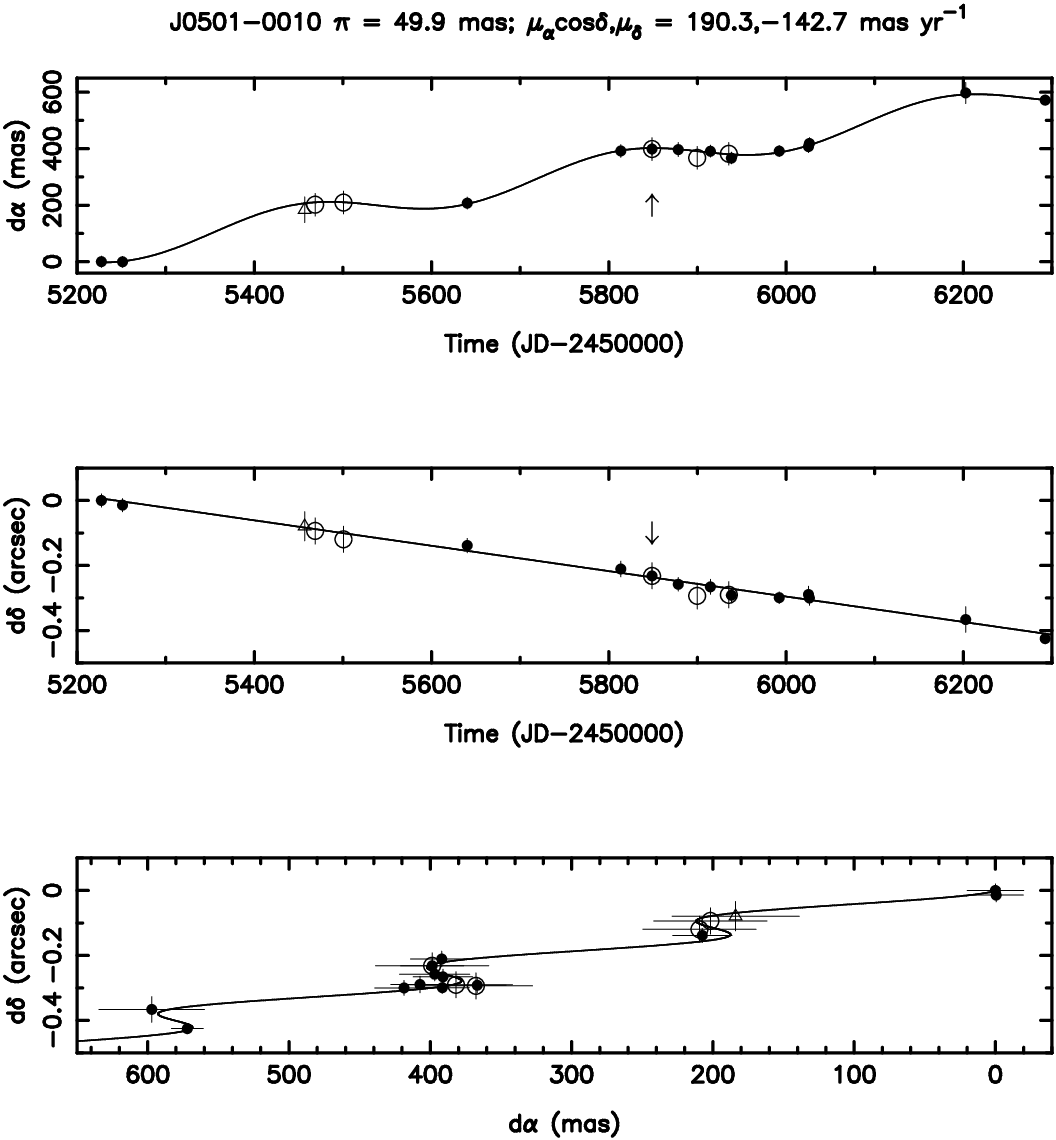} ~~
\includegraphics[width=8.9cm]{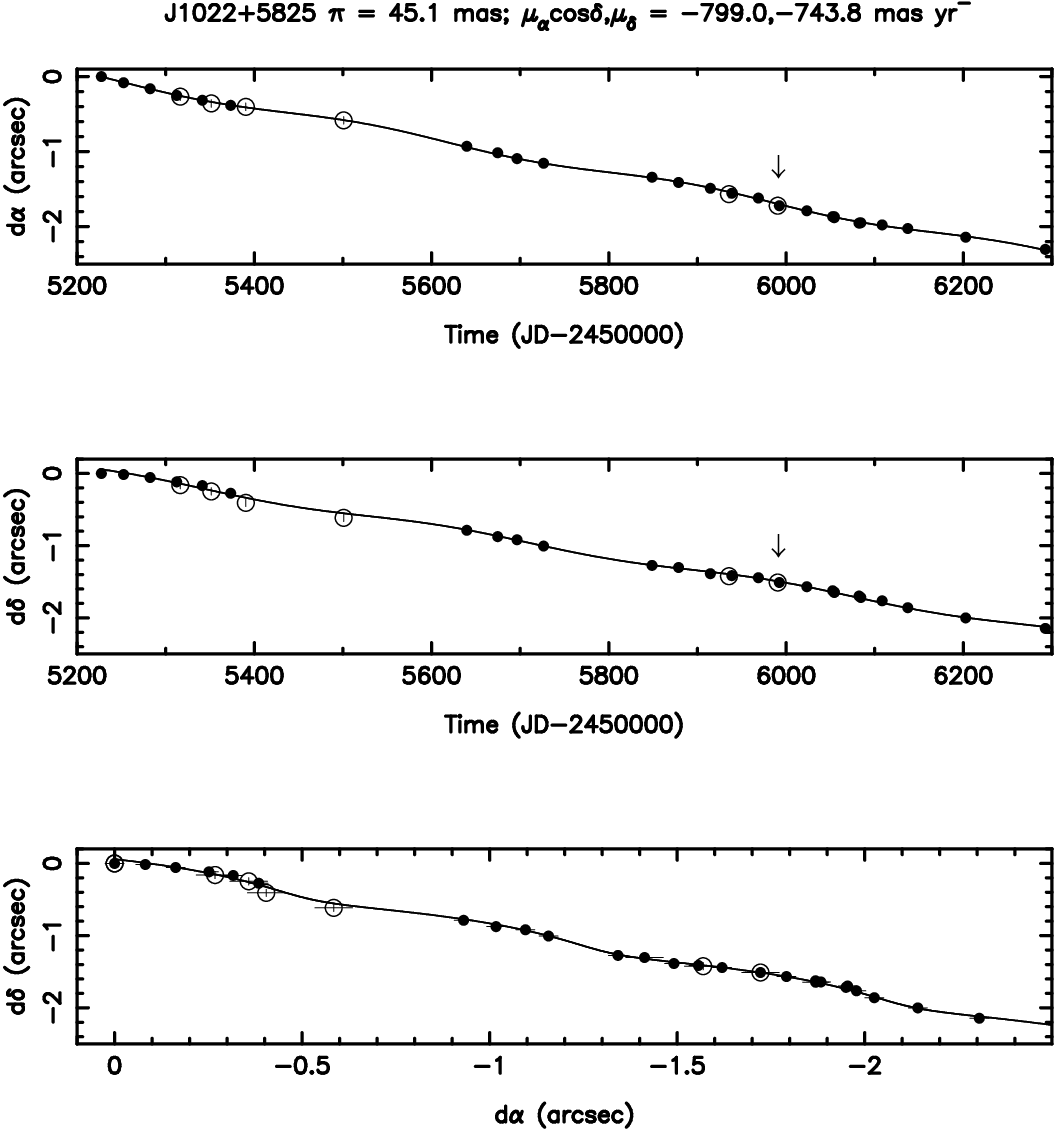} ~~
\includegraphics[width=8.9cm]{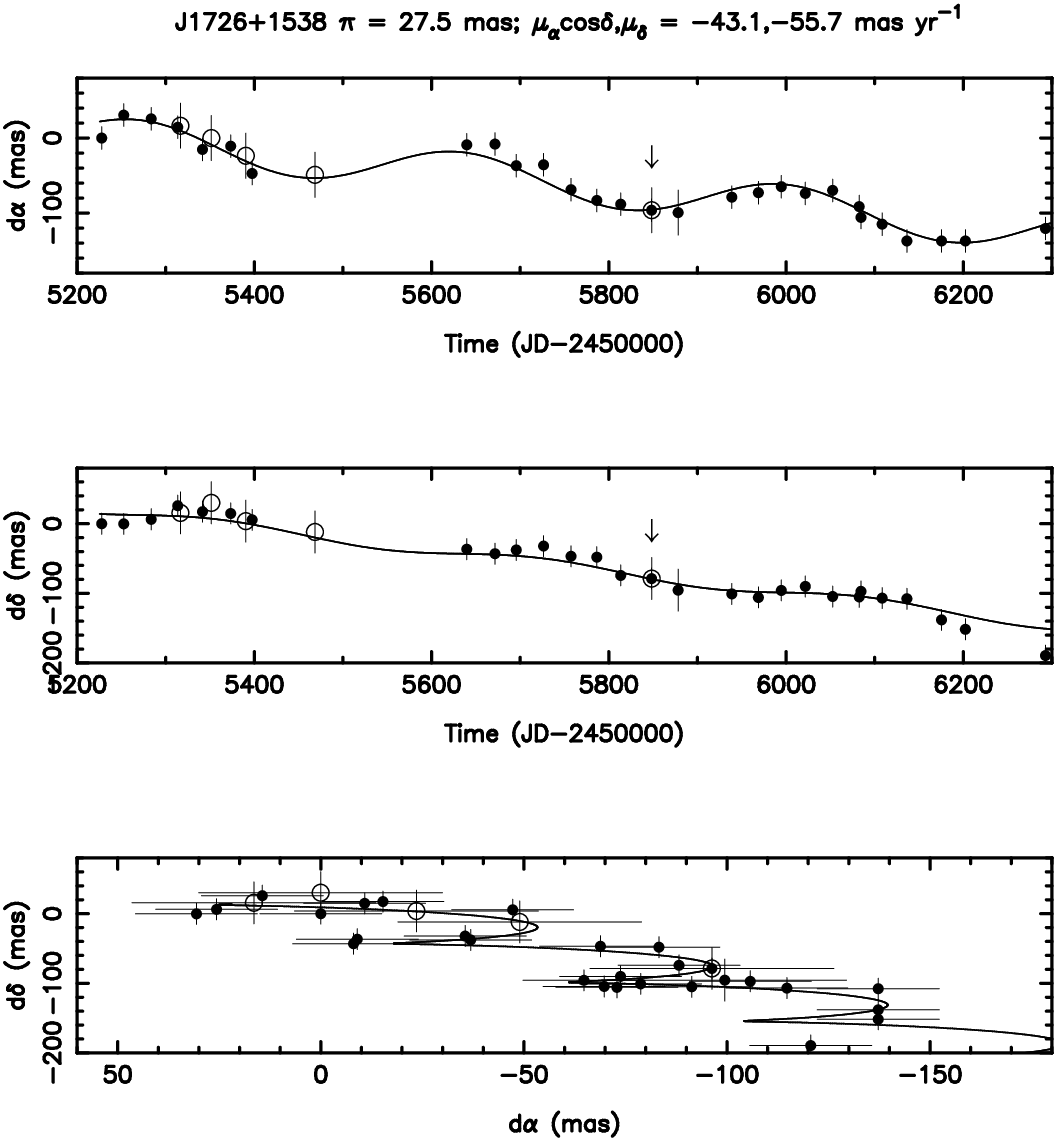} ~~
\includegraphics[width=8.9cm]{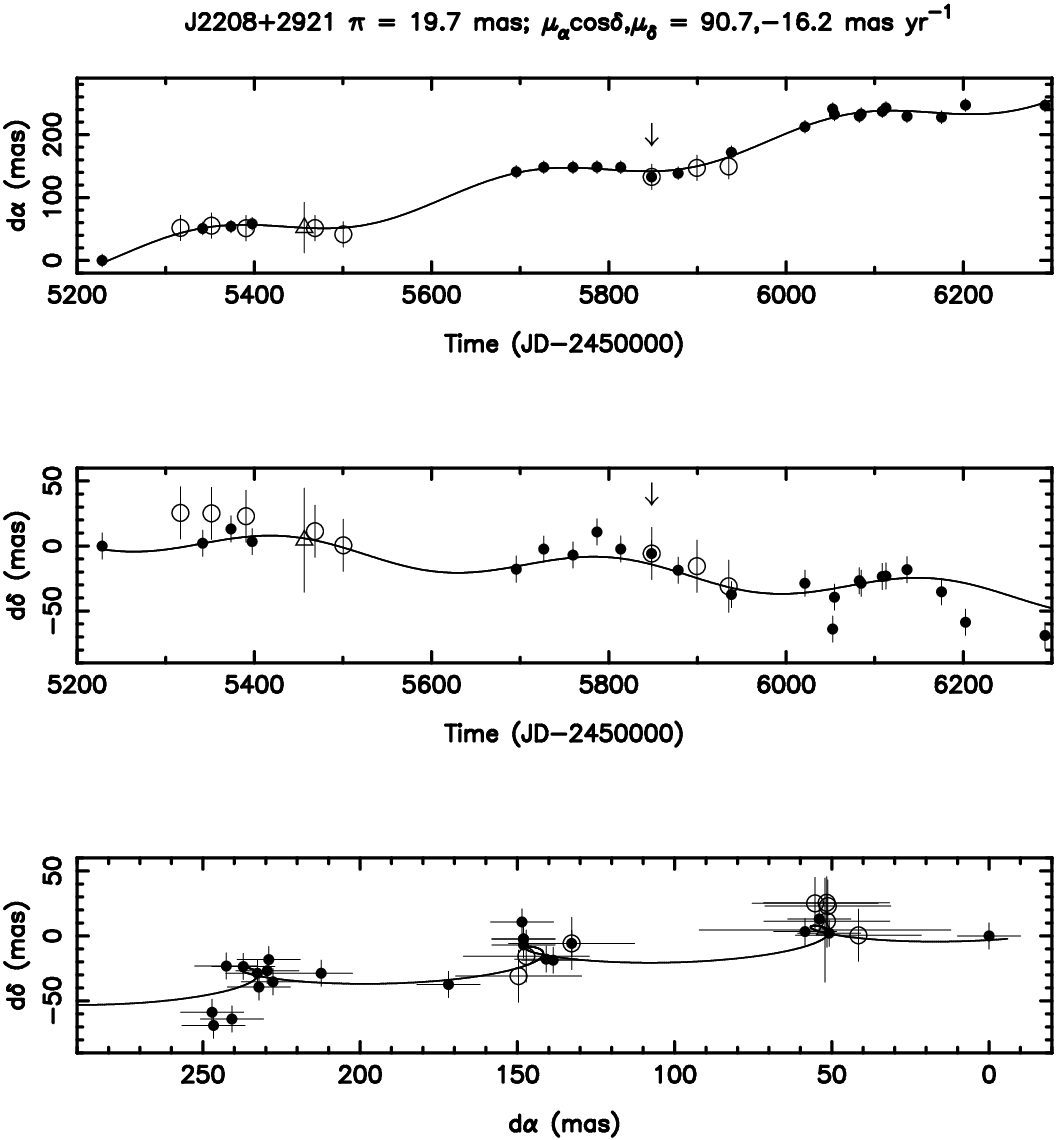} 
\caption{Differential astrometric measurements obtained for J0501$-$0010, J1022$+$5825, J1726$+$1538, and J2208$+$2921  in the time interval from 2010 January through 2012 December. For each object, the top and middle panels depict the apparent trajectories of d$\alpha$ and d$\delta$, respectively, and the bottom panel illustrates d$\delta$ as a function of d$\alpha$, where north is up and east is to the left. The reference epoch (2010 January) of the OMEGA2000 data corresponds to (d$\alpha$, d$\delta$)\,=\,(0, 0). Symbols as in Fig$.$ \ref{astro1}. \label{astro3}}
\end{figure*}
}

\subsection{Proper motion and parallax \label{method}}
In Table~\ref{astrodata} we provide all relative astrometric measurements of each target as a function of Julian Date. We also provide the corresponding air masses. Data (d$x$ and d$y$ were converted into d$\alpha$ and d$\delta$ using the corresponding plate scales) and their associated error bars are plotted against observing epoch in Figures~\ref{astro1}, \ref{astro2}, and~\ref{astro3} (the later two Figures are available in the online version of the manuscript). The epochs at which the NOTCam and OMEGA2000 astrometry was ``normalized'' are clearly indicated. The parallax can be seen as a wobble in the sources's position superimposed on its regular proper motion. The bottom panels corresponding to each target display the (d$\alpha$, d$\delta$) apparent relative trajectories, which on first order approximation depend on proper motion ($\mu_\alpha$, $\mu_\delta$), parallax ($\pi$), and observing time ($t$) according to the following mathematical expressions:
\begin{equation}
\label{eq1}
{\rm d}\alpha = \mu_\alpha ~ (t-t_o) + \pi ~ (f^\alpha_t - f^\alpha_o)
\end{equation}
\begin{equation}
\label{eq2}
{\rm d}\delta = \mu_\delta ~ (t-t_o) + \pi ~ (f^\delta_t - f^\delta_o)
\end{equation}
where the subscript $o$ indicates the reference epoch, and $f^\alpha$ and $f^\delta$ stands for the parallax factors in right ascension ($\alpha$) and declination ($\delta$), respectively. In our study, all the astrometric quantities are given in mas and the times $t$ and $t_o$ are measured in Julian Days. The parallax factors were computed following the equations given in \citet{green85} and obtaining the Earth barycenter from the DE405 Ephemeris\footnote{http://ssd.jpl.nasa.gov}. We applied the least squares fitting method to the set of Equations \ref{eq1} and~\ref{eq2} to derive the parallax and the proper motion for each target. At this stage only the OMEGA2000 data were fit since the NOTCam and LIRIS astrometry is affected by a larger uncertainty. The best-fit solutions are illustrated in Figures~\ref{astro1}, \ref{astro2}, and \ref{astro3} (the later two Figures are available in the online version of the paper). The NOTCam and LIRIS data nicely overlap with the best-fit curves, providing support to the derived parameters. Our results are shown in Table~\ref{par}. At this point we list relative parallaxes. 

\addtocounter{table}{1}
\begin{table*}
\centering
\caption{Proper motions, trigonometric parallaxes, effective temperatures, and luminosities. \label{par}}
\begin{tabular}{llrrrcrcr}
\hline  \hline
\multicolumn{1}{l}{Object} & 
\multicolumn{1}{l}{SpT} & 
\multicolumn{1}{c}{$\mu_\alpha$\,cos\,$\delta$} & 
\multicolumn{1}{c}{$\mu_\delta$} &
\multicolumn{1}{c}{$\pi_{\rm rel}$} & 
\multicolumn{1}{c}{Corr.} & 
\multicolumn{1}{c}{$\pi_{\rm abs}$} &
\multicolumn{1}{c}{$T_{\rm eff}$\tablefootmark{a}  } & 
\multicolumn{1}{c}{log\,$L/L_\odot$} \\
\multicolumn{1}{l}{} & 
\multicolumn{1}{r}{} & 
\multicolumn{1}{c}{(mas\,yr$^{-1}$)} & 
\multicolumn{1}{c}{(mas\,yr$^{-1}$)} &
\multicolumn{1}{c}{(mas)} & 
\multicolumn{1}{c}{(mas)} & 
\multicolumn{1}{c}{(mas)} &
\multicolumn{1}{c}{(K)} & 
\multicolumn{1}{c}{} \\
\hline
J0033$-$1521 & L4$\beta$  &  $+$309.5$\pm$10.4 &  $+$28.9$\pm$18.3 &  23.5$\pm$2.5 &  1.3  &    24.8$\pm$2.5  &  1720$\pm$55  &  $-$3.55$\pm$0.15    \\
J0045$+$1634 & L2$\beta$  &  $+$356.2$\pm$13.7 &  $-$35.0$\pm$10.9 &  55.5$\pm$2.0 &  1.8  &    57.3$\pm$2.0  &  1970$\pm$70  &  $-$3.53$\pm$0.08   \\
J0241$-$0326 & L0$\gamma$ &   $+$84.0$\pm$11.7 &  $-$22.4$\pm$8.6  &  20.2$\pm$2.6 &  1.3  &    21.4$\pm$2.6  &  2260$\pm$75 &  $-$3.75$\pm$0.17   \\
J0355$+$1133 & L5$\gamma$ &  $+$225.0$\pm$13.2 & $-$630.0$\pm$15.0 & 109.2$\pm$4.3 &  1.5  &   110.8$\pm$4.3  &  1615$\pm$50  &  $-$4.10$\pm$0.08  \\
J0501$-$0010 & L4$\gamma$ &  $+$190.3$\pm$9.5  & $-$142.8$\pm$12.5 &  49.9$\pm$3.7 &  1.1  &    51.0$\pm$3.7  &  1720$\pm$55  &  $-$4.00$\pm$0.12   \\
G\,196--3B   & L3$\beta$  &  $-$132.3$\pm$10.7 & $-$202.1$\pm$13.7 &  39.2$\pm$4.1 &  1.7  &    41.0$\pm$4.1  &  1840$\pm$65  &  $-$3.73$\pm$0.14   \\
J1022$+$5825 & L1$\beta$  &  $-$799.0$\pm$6.4  & $-$743.8$\pm$13.2 &  45.1$\pm$1.3 &  1.2  &    46.3$\pm$1.3  &  2110$\pm$75 &  $-$3.66$\pm$0.08    \\
J1552$+$2948 & L0$\beta$  &  $-$154.1$\pm$5.3  &  $-$62.2$\pm$10.6 &  46.5$\pm$0.9 &  1.3  &    47.7$\pm$0.9  &  2260$\pm$75  &  $-$3.63$\pm$0.07    \\
J1726$+$1538 & L3$\beta$  &   $-$43.1$\pm$7.1  &  $-$55.7$\pm$5.2  &  27.5$\pm$2.9 &  1.2  &    28.6$\pm$2.9  &  1840$\pm$65  &  $-$3.77$\pm$0.15    \\
J2208$+$2921 & L3$\gamma$ &   $+$90.7$\pm$3.0  &  $-$16.2$\pm$3.7  &  20.2$\pm$0.7 &  1.0  &    21.2$\pm$0.7  &  1840$\pm$65  &  $-$3.71$\pm$0.10    \\
\hline
\end{tabular}
\tablefoot{
\tablefoottext{a}{Temperature errors correspond to an uncertainty of half a subtype in the spectral classification. The scatter of the \citet{stephens09} temperature scale of M6--T8 field dwarfs is $\pm$100 K, not included here.}
}
\end{table*}

Another method to measure proper motions independently of parallax is to compare the astrometry of the targets taken 1, 2, and 3 years apart and at approximately the same time of the year (within $\pm$20 days). Displacements due to parallax will be minimal, in any case they will be small compared with the large motion of the targets and can be ignored. We checked that the proper motions obtained from both methods agree with each other within the error bars. The uncertainties associated to the proper motions shown in Table~\ref{par} correspond to the standard deviations of the motions determined by this second method. The parallax errors account for the proper motion uncertainties in Equations \ref{eq1} and~\ref{eq2}. 

\begin{figure*}
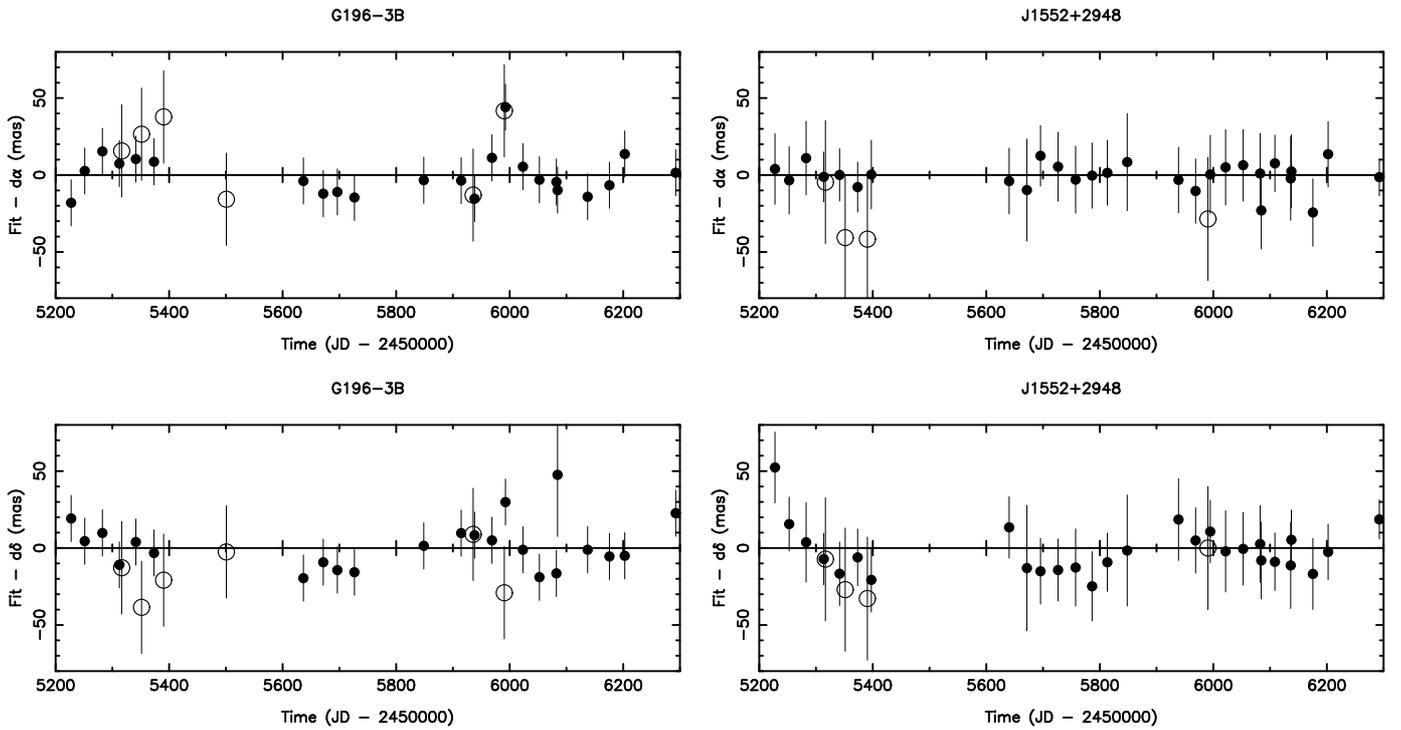
   
\center
\includegraphics[width=8.9cm]{g196_res.ps}  ~~
\includegraphics[width=8.9cm]{j1552_res.ps} 
\caption{Astrometric residuals of G\,196$-$3B and J1552$+$2948 as a function of the observing epoch. Symbols as in Fig$.$ \ref{astro1}. The plots of the remaining targets are available in the online version of the paper. \label{res1}}
\end{figure*}

\onlfig{
\begin{figure*}   
\center
\includegraphics[width=8.9cm]{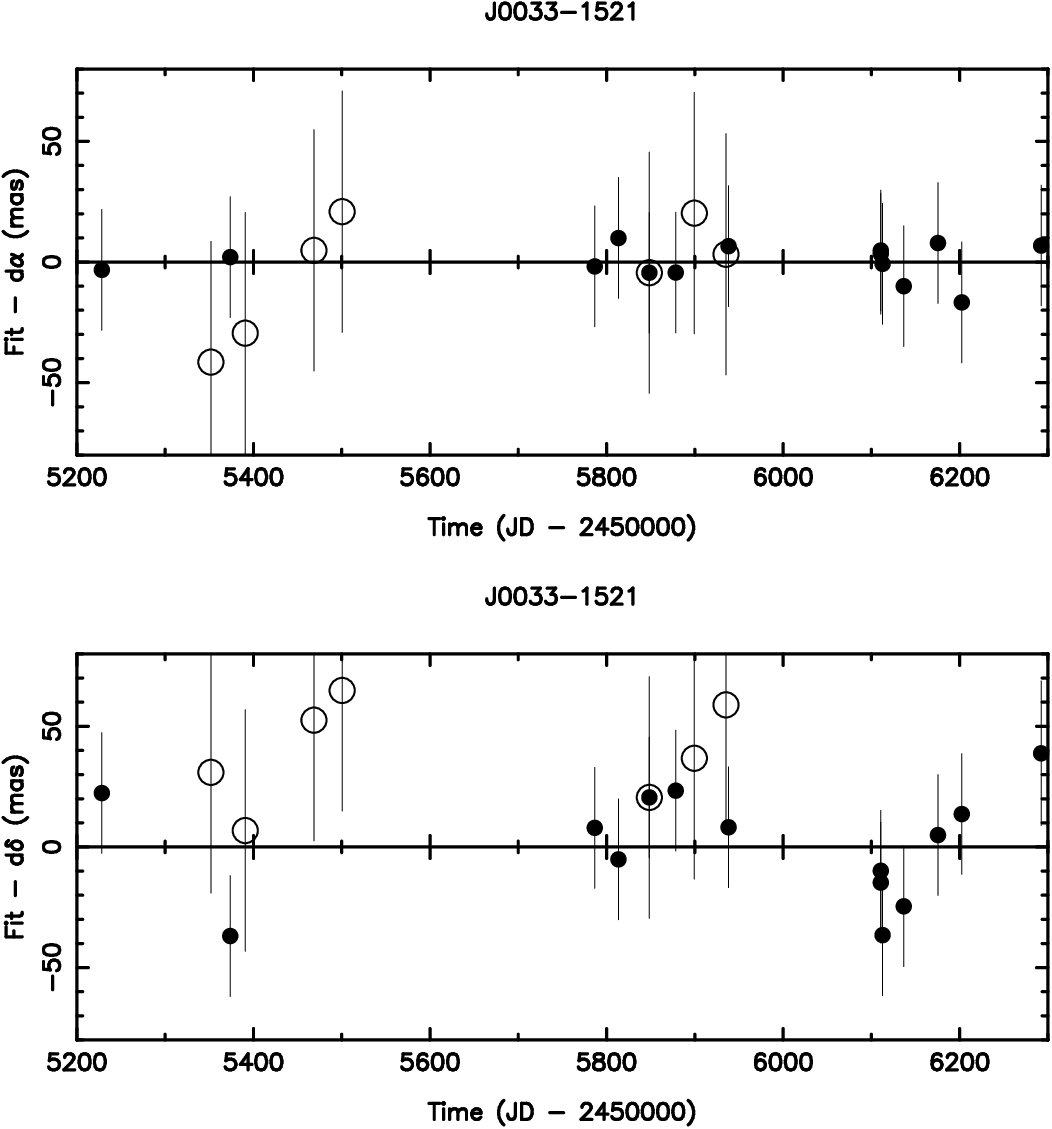} ~~
\includegraphics[width=8.9cm]{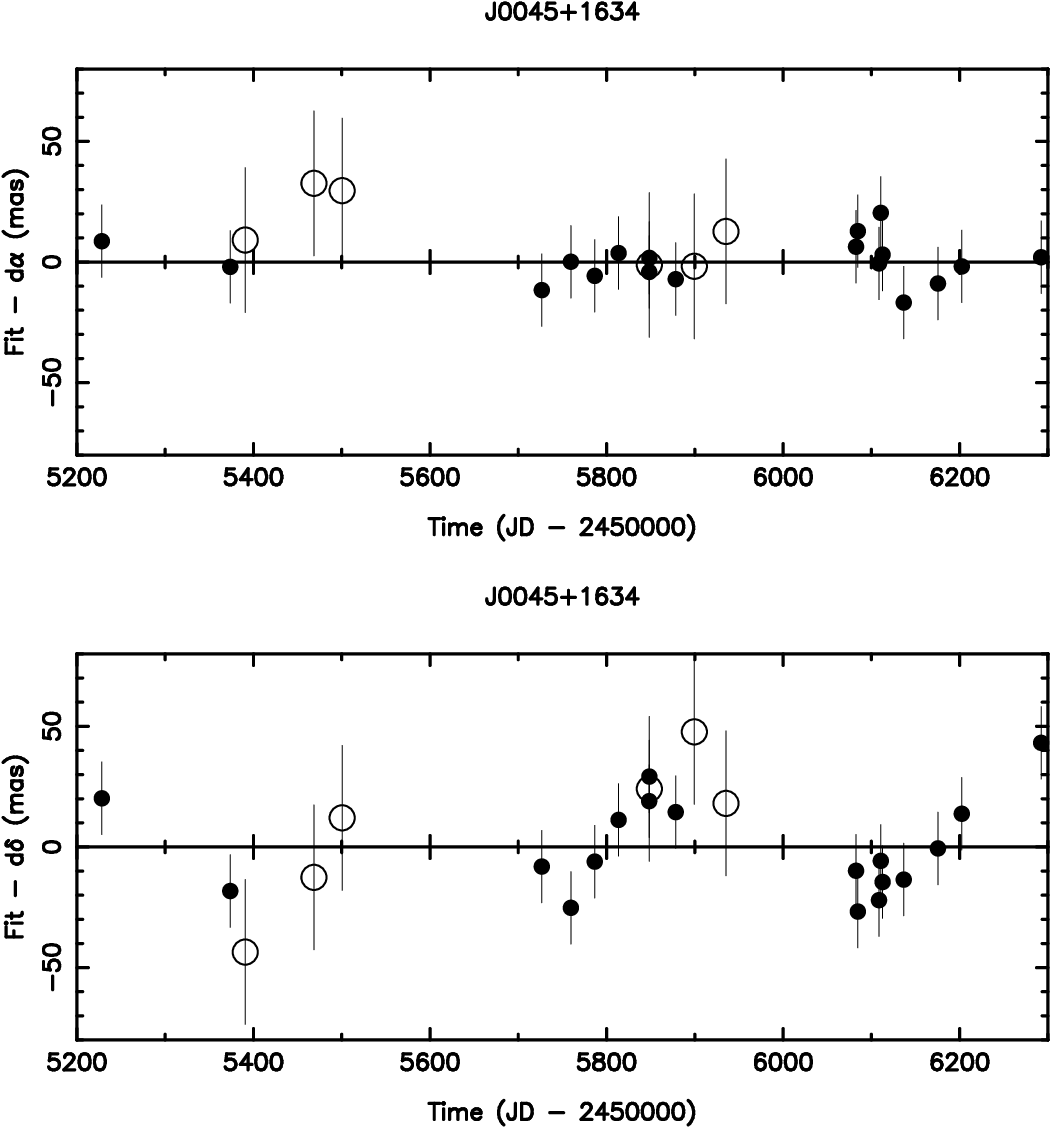} ~~
\includegraphics[width=8.9cm]{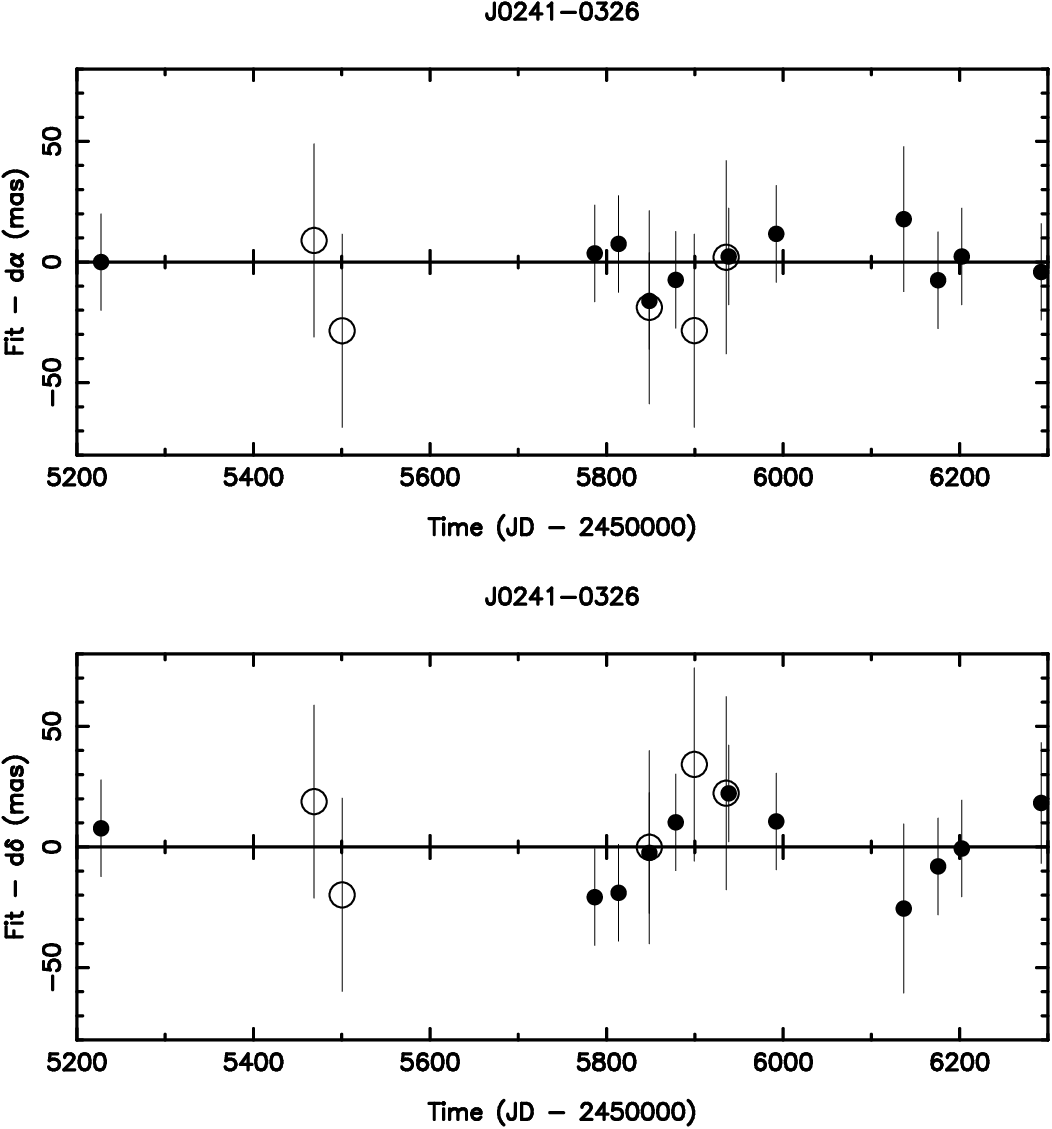}  ~~
\includegraphics[width=8.9cm]{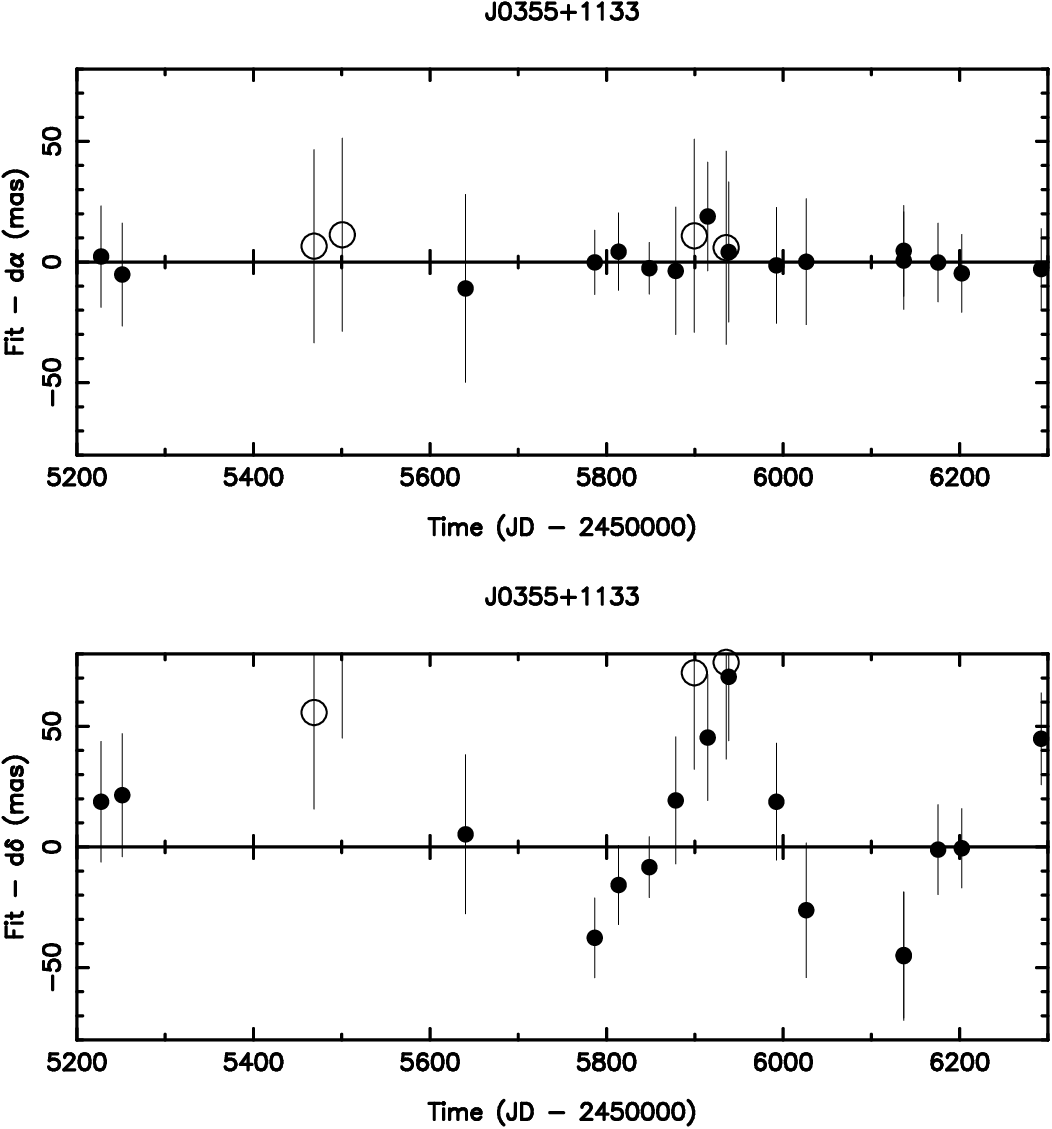} 
\caption{Astrometric residuals of J0033$-$1521, J0045$+$1634, J0241$-$0326, and J0355$+$1133 as a function of the observing epoch. Symbols as in Fig$.$ \ref{astro1}. \label{res2}}
\end{figure*}
}

\onlfig{
\begin{figure*}   
\center
\includegraphics[width=8.9cm]{j0501_res.ps} ~~
\includegraphics[width=8.9cm]{j1022_res.ps} ~~
\includegraphics[width=9cm]{j1726_res.ps} ~~
\includegraphics[width=9cm]{j2208_res.ps} 
\caption{Astrometric residuals of J0501$-$0010, J1022$+$5825, J1726$+$1538, and J2208$+$2921 as a function of the observing epoch. Symbols as in Fig$.$ \ref{astro1}. \label{res3}}
\end{figure*}
}

\begin{figure*}
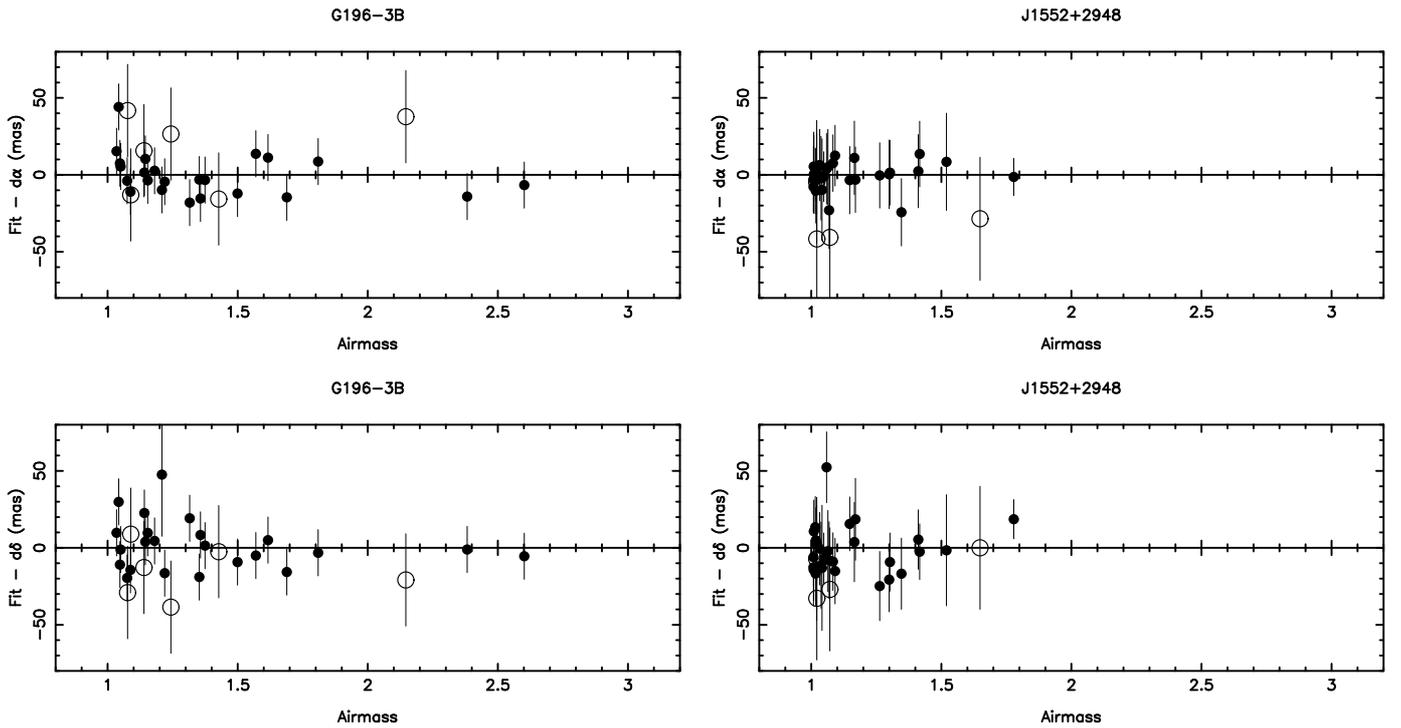
   
\center
\includegraphics[width=8.9cm]{g196_air.ps} ~~
\includegraphics[width=8.9cm]{j1552_air.ps} 
\caption{Astrometric residuals of G\,196$-$3B and J1552$+$2948 as a function of the observing air mass. Symbols as in Fig$.$ \ref{astro1}. Other targets are shown in the online version of the paper. \label{airmass1}}
\end{figure*}

\onlfig{
\begin{figure*}   
\center
\includegraphics[width=8.9cm]{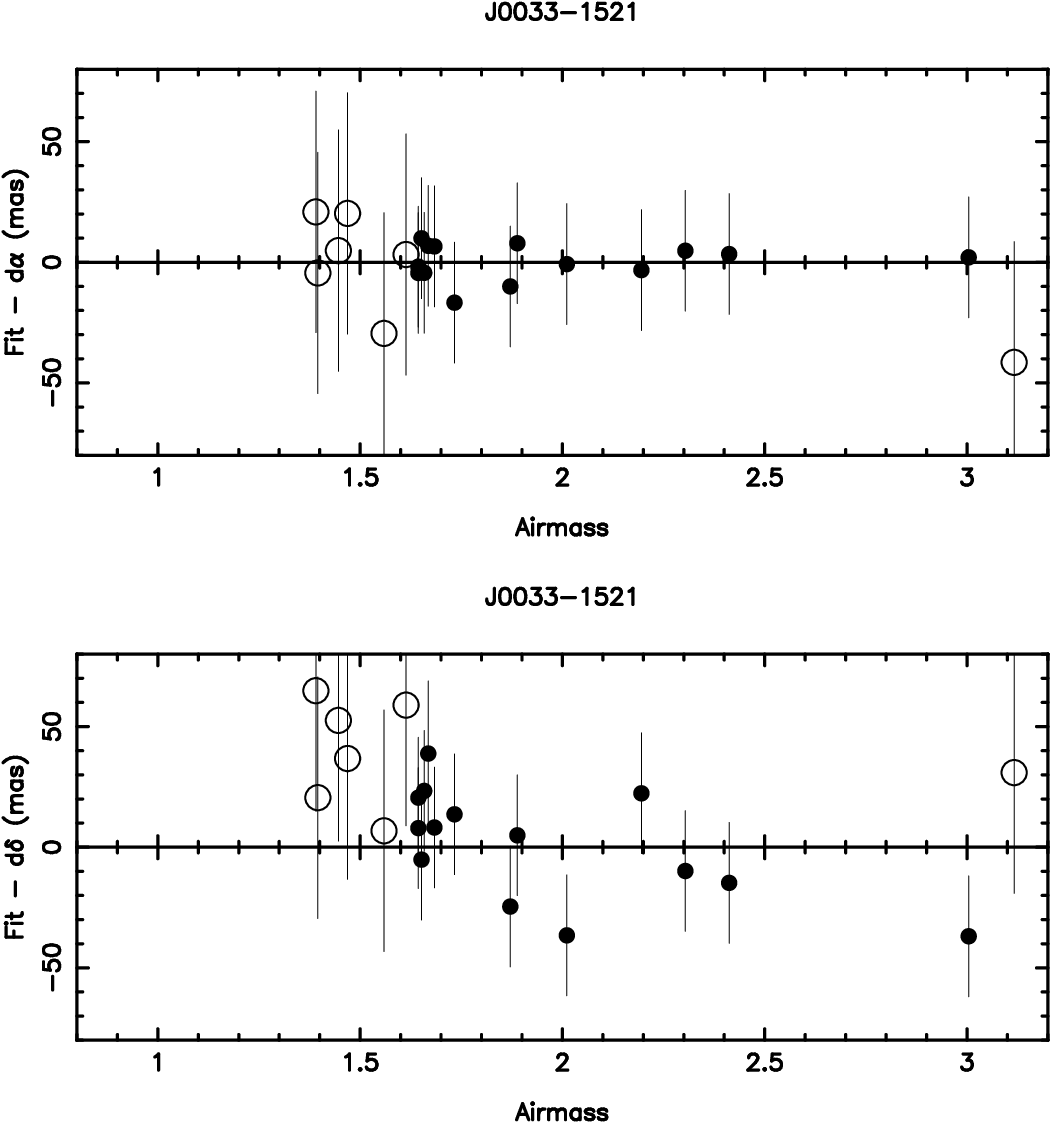} ~~
\includegraphics[width=8.9cm]{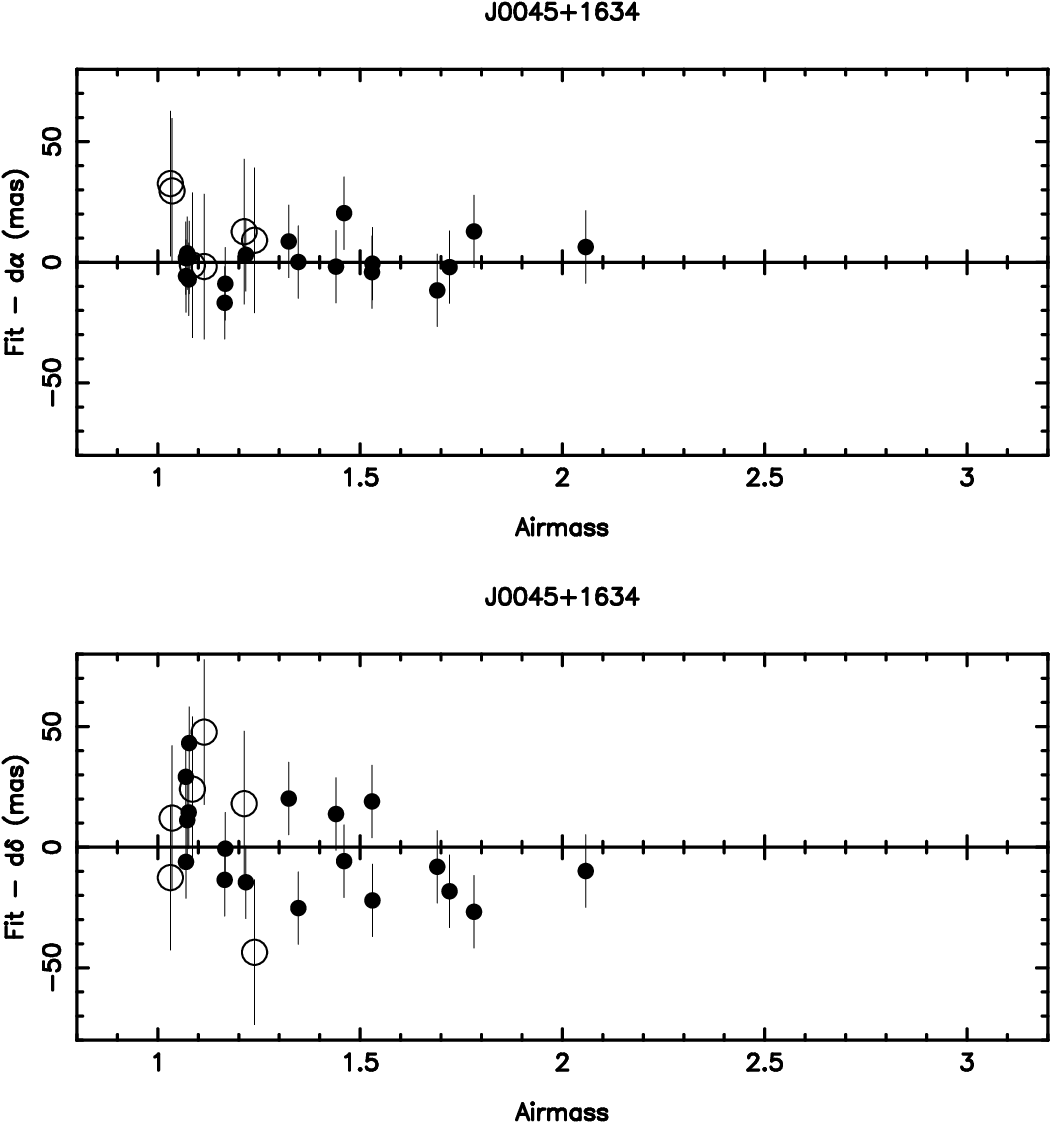} ~~
\includegraphics[width=8.9cm]{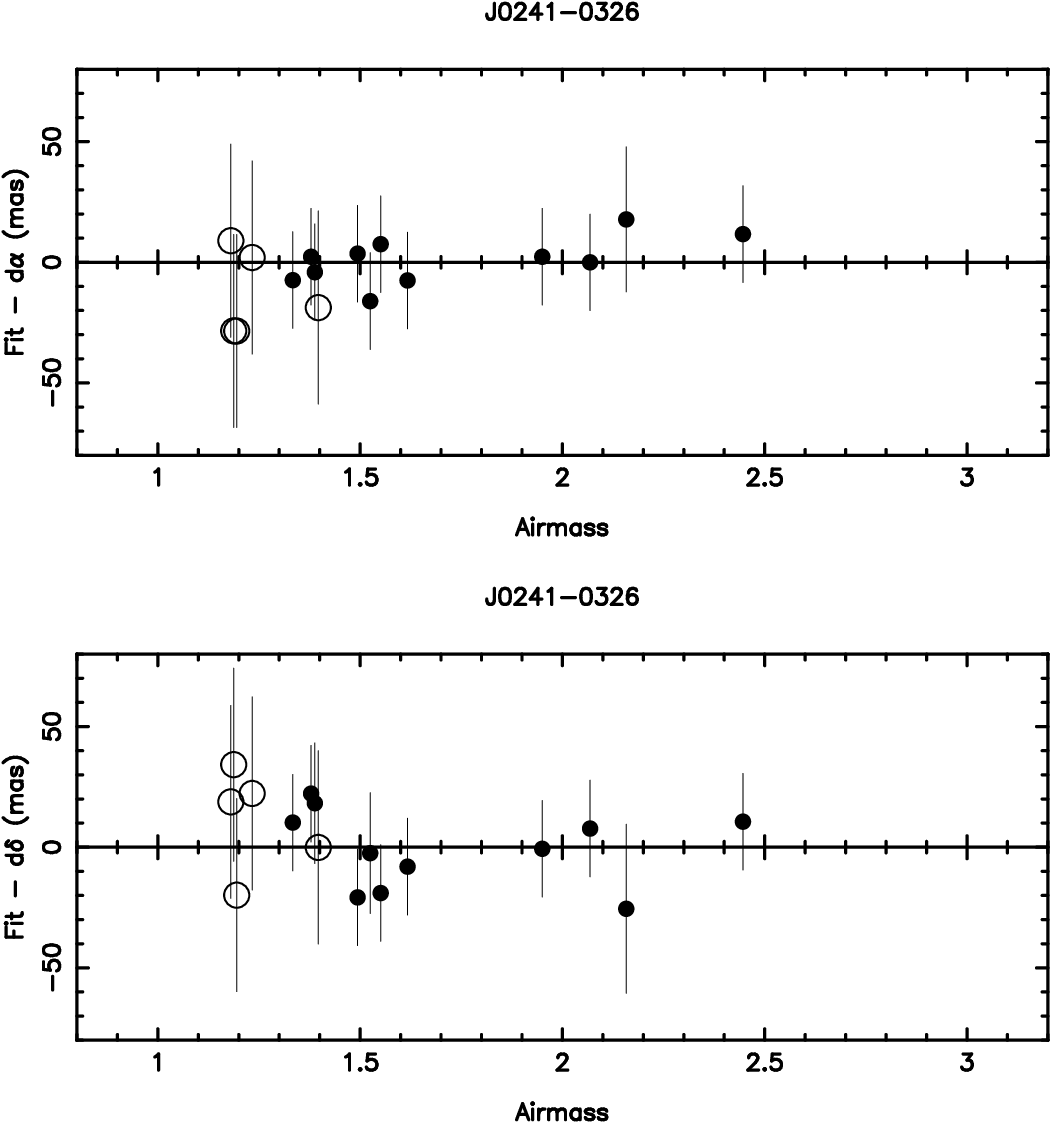} ~~
\includegraphics[width=8.9cm]{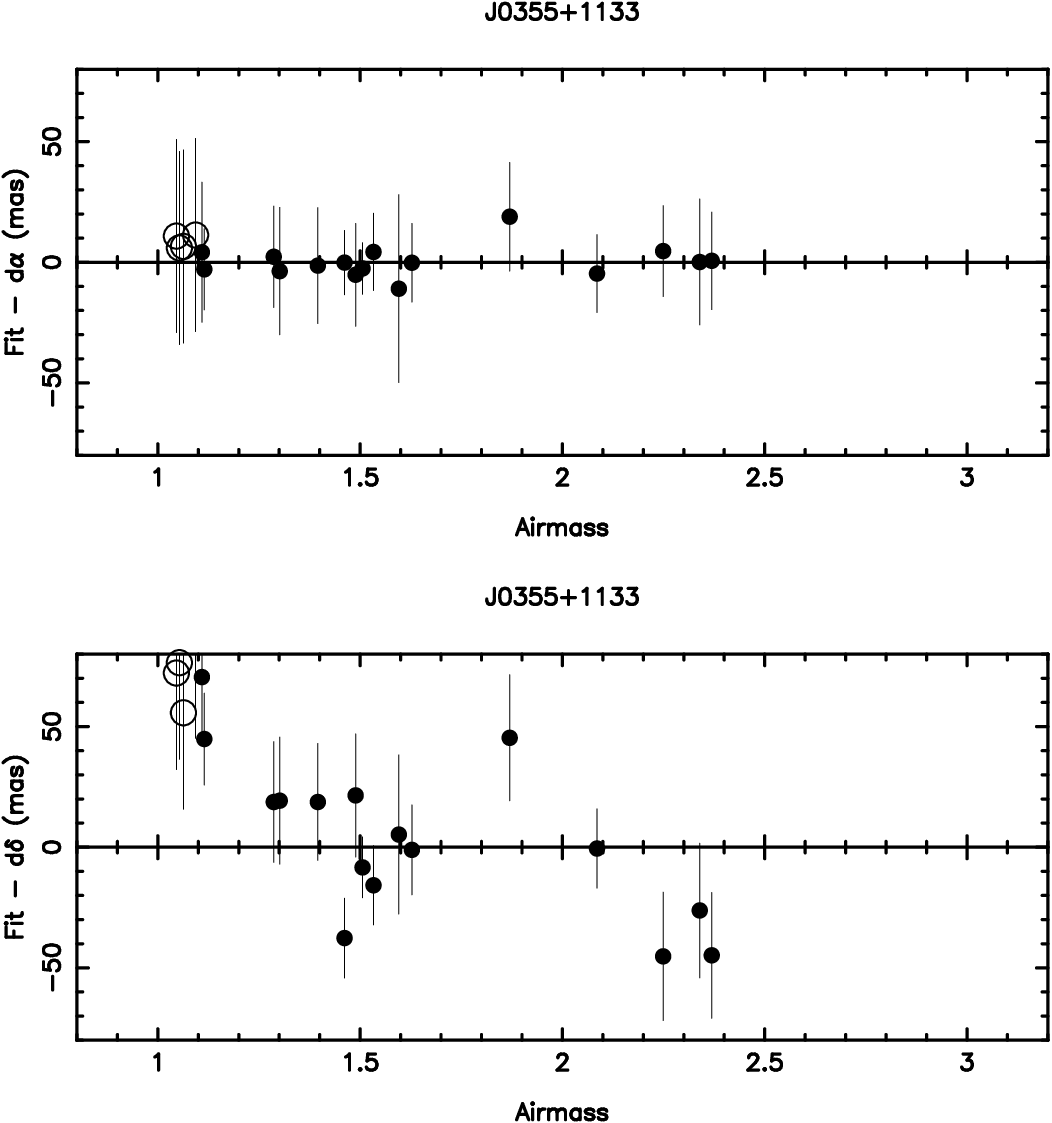}
\caption{Astrometric residuals of J0033$-$1521, J0045$+$1634, J0241$-$0326, and J0355$+$1133  as a function of the observing air mass. Symbols as in Fig$.$ \ref{astro1}. \label{airmass2}}
\end{figure*}
}

\onlfig{
\begin{figure*}   
\center
\includegraphics[width=8.9cm]{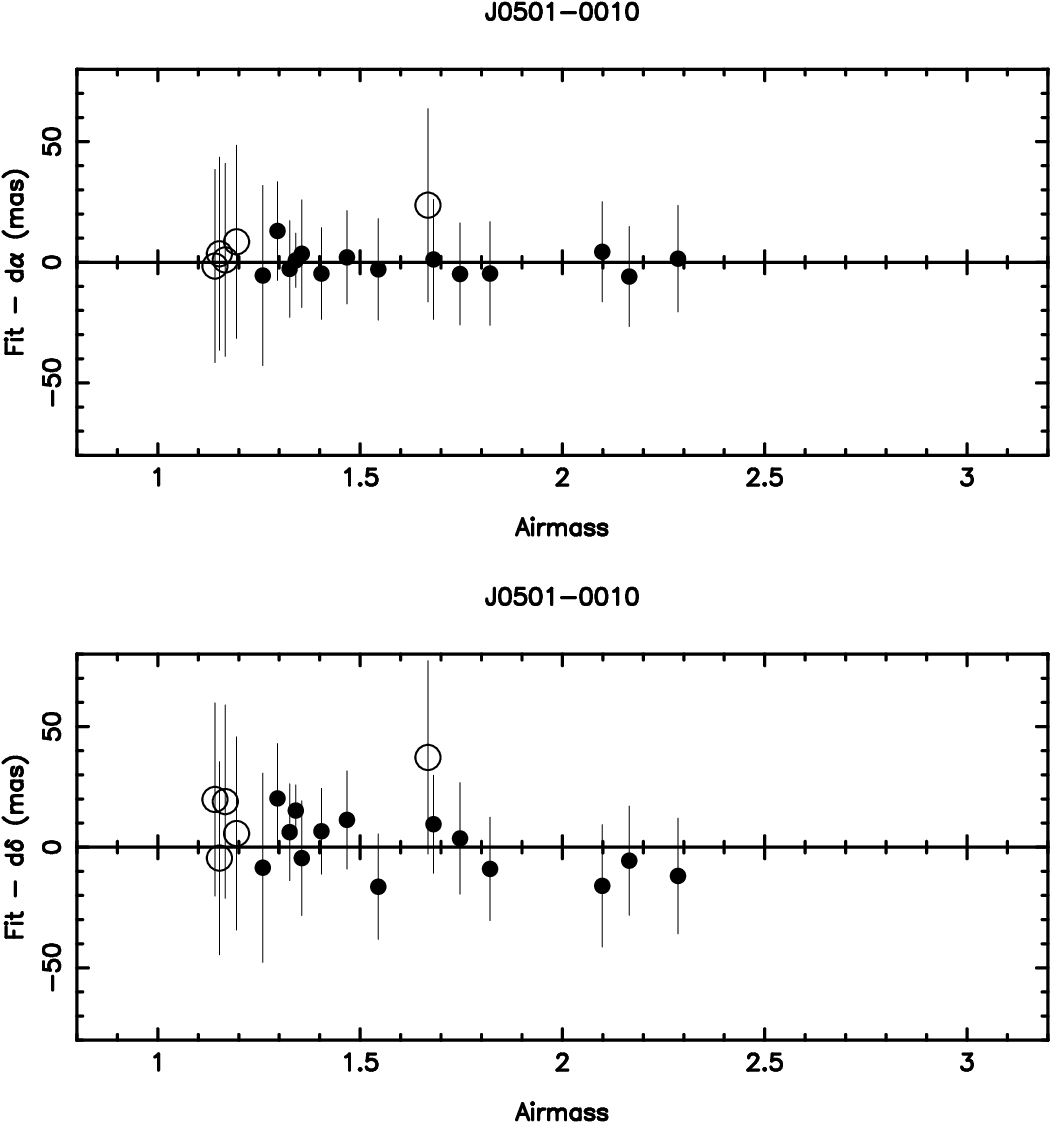} ~~
\includegraphics[width=8.9cm]{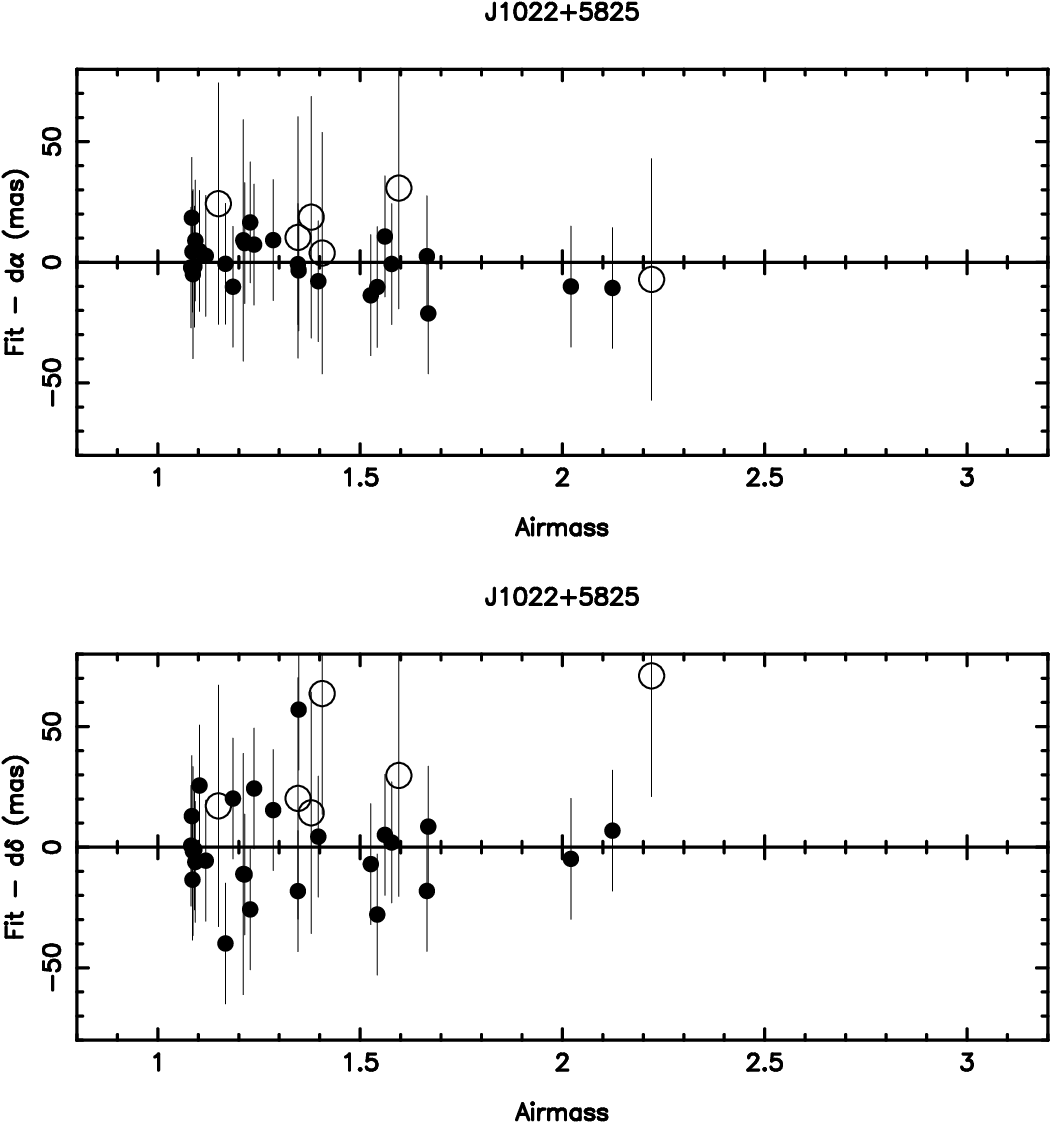} ~~
\includegraphics[width=8.9cm]{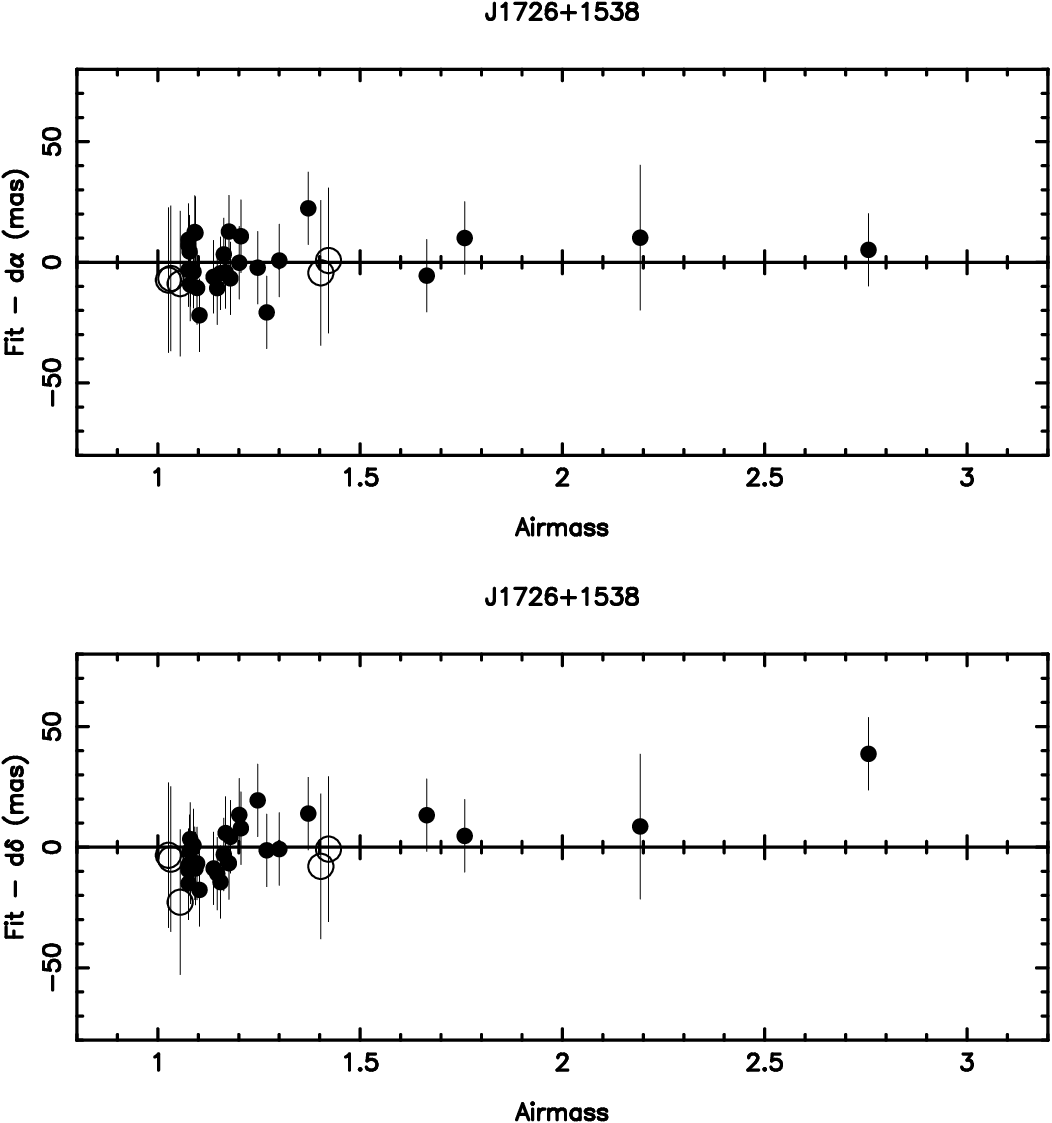} ~~
\includegraphics[width=8.9cm]{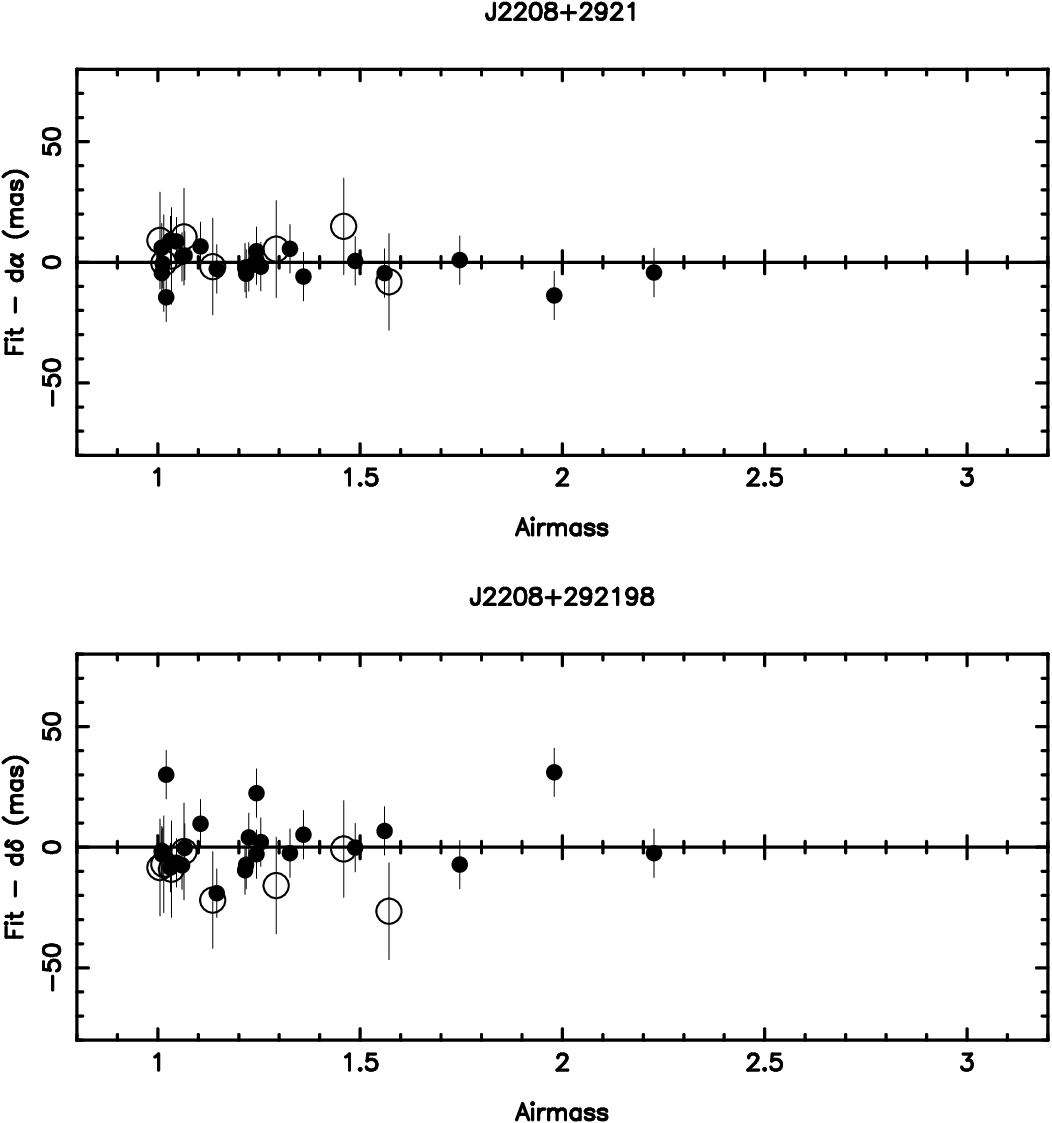}
\caption{Astrometric residuals of J0501$-$0010, J1022$+$5825, J1726$+$1538, and J2208$+$2921  as a function of the observing air mass. Symbols as in Fig$.$ \ref{astro1}. \label{airmass3}}
\end{figure*}
}

The astrometric residuals after subtracting the derived parallaxes and proper motions from the observations are shown in Figures~\ref{res1}, \ref{res2}, and~\ref{res3} (the later two Figures are available in the online version). The dispersions of the monthly OMEGA2000 d$\alpha$ residuals per object are given in Table~\ref{residuals}. The mean scatter of the residuals is $\pm$8.6 mas for $\alpha$ and $\pm$18.1 mas for $\delta$ (OMEGA2000), below the typical error bars assigned to the relative astrometry shown in Table~\ref{astrodata}, suggesting that individual uncertainties may be slightly overestimated. We note that the mean dispersion is constant over the three year time baseline. As for the NOTCam data, although they were not included in the least squares fitting step, residuals have a typical scatter of $\pm$19.5 mas ($\alpha$) and $\pm$31.1 mas ($\delta$), about a factor of two larger than those of OMEGA2000. 

To evaluate the impact of the color atmospheric differential refraction term in our analysis, we plotted the astrometric residuals as a function of air mass in Figures~\ref{airmass1}, \ref{airmass2}, and~\ref{airmass3} (the later two Figures are available in the online version of the paper). Despite having similar spectral classification, our targets are depicted separately because of their different coordinates and reference frame air masses. The distribution of the data points (both in $\alpha$ and $\delta$) in Figures~\ref{airmass1}, \ref{airmass2}, and~\ref{airmass3} is flat around a null value from air mass 1 through 2.5. Even the few observations taken with air mass between 2.5 and $\sim$3 appear to be well reproduced by the astrometric fits, thereby suggesting that chromatic differential refraction is not a major source of uncertainty in our study.

\begin{table*}
\centering
\caption{Dispersion of astrometric residuals, ages, masses, radii, surface gravities, explored periods, minimum masses of astrometric companions.
\label{residuals}}
\begin{tabular}{lcrcrcrcr}
\hline  \hline
\multicolumn{1}{l}{Object} & 
\multicolumn{1}{c}{SpT} & 
\multicolumn{1}{c}{$\Delta ({\rm d}\alpha$)} & 
\multicolumn{1}{c}{Age range\tablefootmark{a}} &
\multicolumn{1}{c}{Mass range\tablefootmark{b}} & 
\multicolumn{1}{c}{$R/R_\odot$} &
\multicolumn{1}{c}{log $g$\tablefootmark{c}} &
\multicolumn{1}{c}{Periods\tablefootmark{d}} & 
\multicolumn{1}{c}{M$_c$\tablefootmark{e}} \\
\multicolumn{1}{l}{} & 
\multicolumn{1}{c}{} & 
\multicolumn{1}{c}{(mas)} & 
\multicolumn{1}{c}{(Myr)} &
\multicolumn{1}{c}{(M$_{\rm Jup}$)} & 
\multicolumn{1}{c}{} & 
\multicolumn{1}{c}{(cm\,s$^{-2}$)} & 
\multicolumn{1}{c}{(d)} & 
\multicolumn{1}{c}{(M$_{\rm Jup}$)} \\
\hline
J0033$-$1521\tablefootmark{f}   & L4$\beta$  &  7.4  &  $\le$10  &$<$4--12 (4) & 0.19$\pm$0.05 & ---                &   76--1095 & ---  \\  
J0045$+$1634                    & L2$\beta$  &  8.8  &  10--100  & 12--25 (15) & 0.15$\pm$0.02  & 4.05--4.66  (4.28) &   59--1095 &  5.6  \\  
J0241$-$0326                    & L0$\gamma$ &  9.5  & $\ga$500  &$\ge$60 (80) & 0.09$\pm$0.02  & $\ge$5.12   (5.45) &   97--1095 & 56.9  \\    
J0355$+$1133                    & L5$\gamma$ &  6.4  &  50--500  & 15--40 (23) & 0.11$\pm$0.02  & 4.37--5.07  (4.68) &   66--1095 &  2.4  \\  
J0501$-$0010                    & L4$\gamma$ &  5.2  &  50--500  & 13--45 (25) & 0.11$\pm$0.02  & 4.27--5.19  (4.73) &   76--1095 &  4.7  \\  
G\,196--3B                      & L3$\beta$  & 13.9  &  10--300  & 11--45 (15) & 0.13$\pm$0.03  & 4.04--5.07  (4.36) &   46--1095 & 15.0  \\  
J1022$+$5825                    & L1$\beta$  &  9.6  & 100--1000 & 35--72 (50) & 0.11$\pm$0.02  & 4.76--5.36  (5.05) &   41--1095 & 16.4  \\  
J1552$+$2948                    & L0$\beta$  &  8.9  & $\ga$500  &$\ge$65 (75) & 0.10$\pm$0.01  & $\ge$5.13   (5.32) &   38--1095 & 18.5  \\    
J1726$+$1538                    & L3$\beta$  & 10.4  &  10--300  & 11--45 (20) & 0.13$\pm$0.03  & 4.07--5.12  (4.52) &   38--1095 & 10.0  \\  
J2208$+$2921                    & L3$\gamma$ &  6.1  &  10--300  & 11--45 (15) & 0.14$\pm$0.03  & 4.05--4.99  (4.34) &   46--1095 & 12.7  \\  
\hline
\end{tabular}
\tablefoot{
\tablefoottext{a}{Derived from the luminosity--$T_{\rm eff}$ plane. All objects are assumed to be single.} 
\tablefoottext{b}{Most likely mass value from the luminosity--$T_{\rm eff}$ plane is given in parenthesis. All objects are assumed to be single.}
\tablefoottext{c}{Value computed for the most likely mass is given in parenthesis.}
\tablefoottext{d}{Range of periods explored in the time-series analysis of the monthly astrometric residuals.}
\tablefoottext{e}{Minimum detectable companion masses (see Section~\ref{companions}).}
\tablefoottext{f}{ The lack of lithium in its atmosphere is not consistent with such young age and low mass (see Section~\ref{litio}). Surface gravity and minimum detectable companion masses are not computed for this object.}
}
\end{table*}
 
Our reference objects are almost all stars in the Galaxy, each one with its own distance and proper motion. This introduces a systematic error in the parallax and proper motion determination that must be considered. On the one hand, we expect that the motions of the reference objects are randomly orientated; therefore, their effect will be reduced and we did not correct for it. On the other hand, finite distances diminish part of the true parallax of our targets. To convert relative parallaxes into absolute parallaxes we followed a procedure similar to the one described in \citet{faherty12}. Using the 2MASS colors \citep{skrutskie06} of the reference objects we obtained their photometric distances by assuming that all of them are main sequence stars. We adopted the color--bolometric correction--spectral type relations given in \citet{johnson66} for BAFGK stars and in \citet{kirk93} for late-K and M stars. The defined relations are valid for colors in the interval $J-K_s$ = $-$0.2 to 1.53 mag. We adopted the mode of the distribution of reference objects distances as the correction to be added to the relative parallax that comes directly from our fits to obtain the absolute parallax. In Table~\ref{par} we list the values of these corrections derived for our target fields. The average relative-to-absolute parallax correction for the full list of targets is 1.34\,$\pm$\,0.28 mas ranging from 1.00 to 1.84 mas. We note that this correction is typically of the same order as the astrometric uncertainty associated with our derivation of the relative parallax. Absolute parallaxes for all ten targets are also provided in Table~\ref{par}. Their error bars result from the quadrature sum of the uncertainties of the relative parallaxes and the correction factors; they are dominated by the errors of the relative parallaxes.

\begin{figure*}
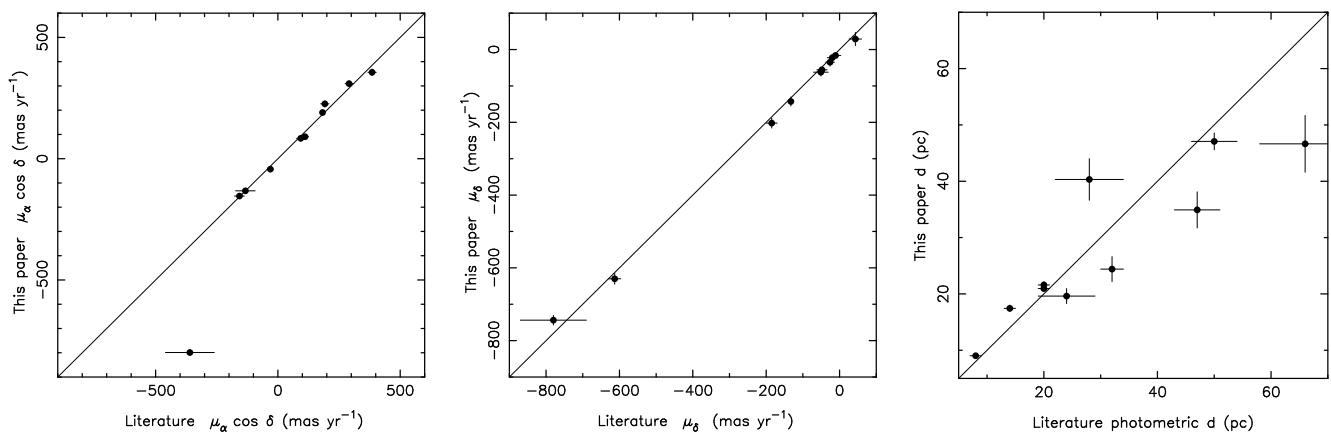
   
\center
\includegraphics[width=5.5cm]{pm1.ps}~~~~
\includegraphics[width=5.5cm]{pm2.ps}~~~~
\includegraphics[width=5.5cm]{d.ps}
\caption{Comparison of our measurements with data from the literature \citep{faherty09,cruz09}. The solid diagonal line in the three panels corresponds to exact equality. \label{comp}}
\end{figure*}

\subsection{Comparison with the literature\label{literature}}
All of our targets have published proper motions \citep{jameson08,casewell08,faherty09,faherty12,osorio10}. The two left panels of Figure~\ref{comp} display the comparison between the literature data and the measurements provided here. With one exception, we find that our values agree with the published ones within 1--1.5\,$\sigma$ the uncertainties, thus validating our method. The deviating object is J1022$+$5825, whose $\mu_\alpha\,{\rm cos}\,\delta$ shows discrepant measurements. Our value is a factor of 2.2 times larger in amplitude than the literature one \citep{faherty09}. While the agreement in $\mu_\delta$ is at the level of 1\,$\sigma$, we cannot reconcile the published $\mu_\alpha\,{\rm cos}\,\delta$ with our data. 

Trigonometric parallaxes are given for the first time for eight objects in our sample. Only J0355$+$1133 and J0501$-$0010 have astrometric distances available in the literature \citep{faherty12,faherty13}. A third object, J0033$-$1521, is in the target list of \citet[PARSEC program]{andrei11}, but to the best of our knowledge no parallax measurement is available so far. For J0355$+$1133, our parallax derivation ($\pi$ = 110.8\,$\pm$\,4.3 mas) and the value of \citet[$\pi$ = 122\,$\pm$\,13 mas]{faherty13} are consistent to within 1\,$\sigma$, although our measurement locates this L dwarf at a farther distance ($d$ = 9.03\,$\pm$\,0.34 pc versus 8.20\,$\pm$\,0.88 pc), in better agreement with the very recent result (9.10\,$\pm$\,0.10 pc) of \citet{liu13b}. Regarding J0501$-$0010, the parallax measurements are discrepant at $>$5\,$\sigma$, which translates into significantly different distances: $d$ = 19.60\,$\pm$\,1.32 pc (ours) and $d$ = 13.09\,$\pm$\,0.82 pc \citep{faherty12}. This difference contrasts with the similarity of the proper motions measured by different groups \citep[and this paper]{casewell08,faherty09}. Our value for the relative-to-absolute parallax correction (1.1\,$\pm$0.28 mas) agrees well with \citet{faherty12} determination (1.2 mas), we thus conclude that the correction is not the source of the discrepancy. The fact that our trigonometric parallaxes locate these objects at larger distances has implications in the discussion of the following sections. 

Spectrophotometric distances were computed for all targets by \citet{cruz09}. These computations and our derivations are compared in the right panel of Figure~\ref{comp}.  \citet{cruz09} spectrophotometric distance of J0501$-$0010 agrees with our astrometric measurement at the 1-$\sigma$ level. 

\section{Discussion \label{discussion}}

\subsection{Magnitude--spectral type diagrams}
\begin{figure}   
\center
\includegraphics[width=8.5cm]{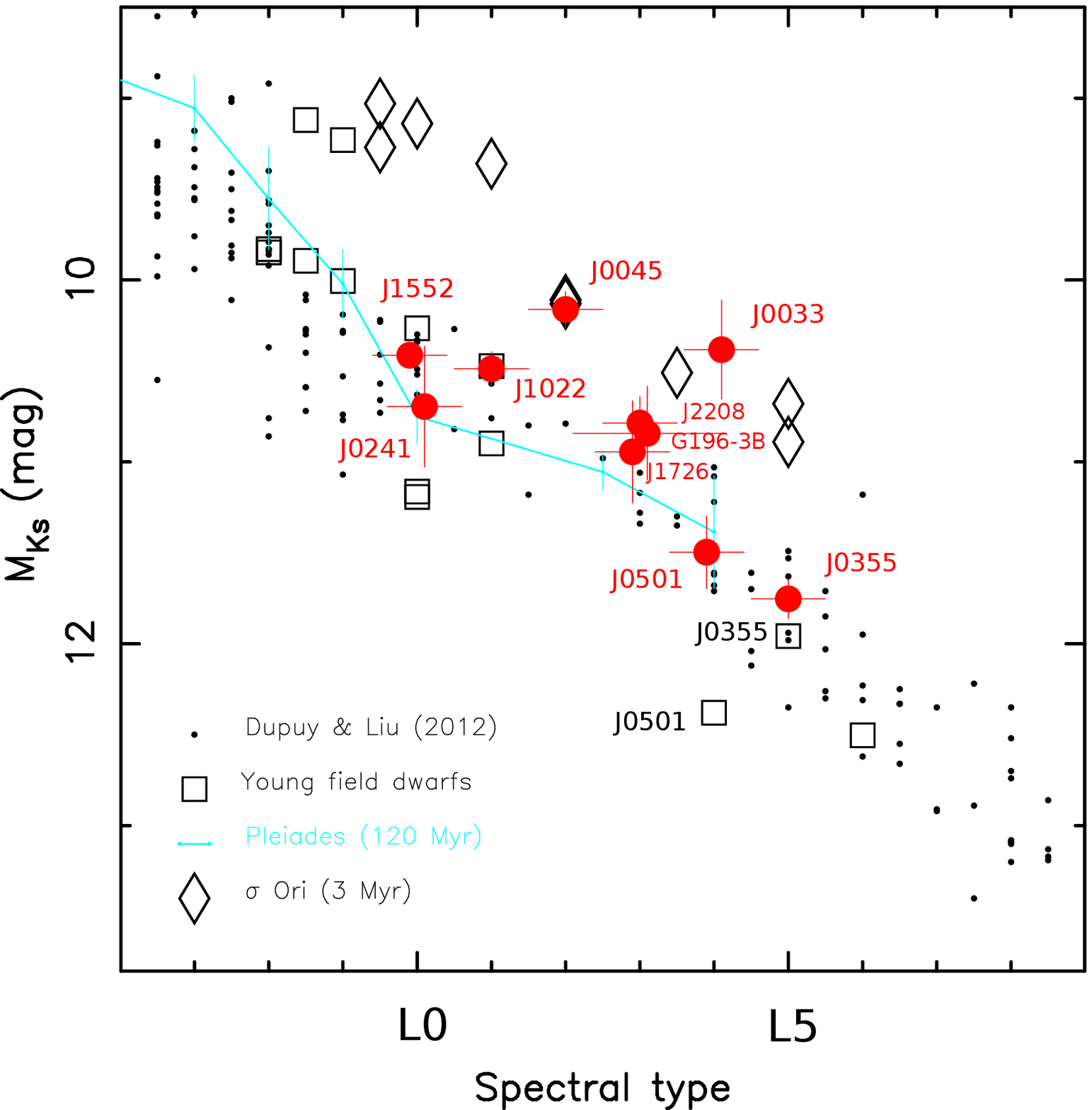}
\includegraphics[width=8.5cm]{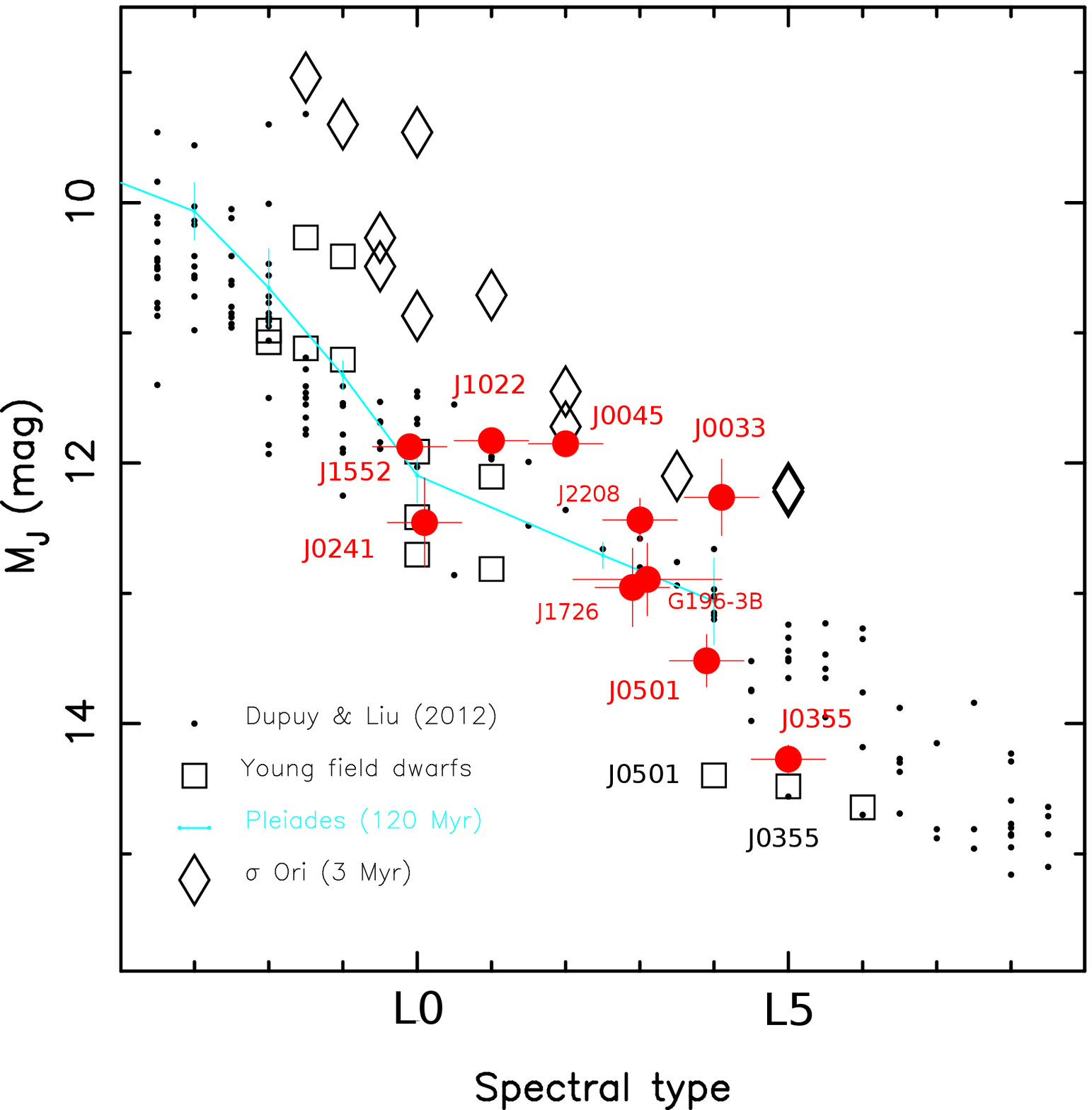}
\caption{Absolute $K_s$ (top) and $J$ (bottom) magnitudes as a function of optical spectral type for objects of various ages. Our sample is plotted as filled circles with error bars. All objects are labeled. The field sequence of dwarfs is defined by the sample of \citet[small dots]{dupuy12}. Young field dwarfs with parallax measurements given by \citet{faherty12,faherty13} are shown with open squares. The mean sequence of Pleiades dwarfs (120 Myr) is depicted by a solid line (it goes from the mid-Ms to L4). The youngest objects ($\sigma$~Ori cluster, $\sim$3 Myr) are illustrated with open diamonds. We do not show the error bars of all objects from the literature for the clarity of the figure. Also for clarity, our targets are slightly shifted in spectral type. Photometry is in the 2MASS system. \label{k_spt}}
\end{figure}

The derived trigonometric parallaxes were used to determine the distance modulus of each object and to obtain absolute magnitudes from apparent magnitudes. In Figure~\ref{k_spt} we show the location of all targets in the M$_J$ and M$_{Ks}$ versus spectral type diagrams  (2MASS photometry). To put our sample in context, the  two panels of Figure~\ref{k_spt} are completed with sources from the literature. Here, we assume that our targets have solar metallicity. The field sequence (dwarfs with known trigonometric parallaxes) is taken from the recent work by \citet{dupuy12}. Because the age and metallicity of field dwarfs are not well constrained, the sequence displays a significant scatter  in both $J$ and $K_s$ diagrams; equal-mass binarity might account for part of the scatter, but it is not enough (e.g., see \citealt[and references therein]{burrows06}). Nevertheless, our sample nicely follows the trends delineated by the field objects. 

Young cool dwarfs of known age, metallicity and distance are also shown in Figure~\ref{k_spt}. Pleiades substellar members confirmed spectroscopically and astrometrically (i.e., their proper motions are compatible with that of the stellar cluster) are discussed in \citet{bihain10}. The widely accepted age and distance of the Pleiades cluster is 120 Myr and 120 pc \citep{stauffer98,martin98,vanLeeuwen09a,vanLeeuwen09b} as derived from the lithium depletion boundary \citep{rebolo92} and the Hipparcos catalog \citep{perryman97}, respectively. Per spectral type interval, we averaged the absolute magnitudes of Pleiades members listed in Table~A.1 of \citet{bihain10} and plotted them along with their associated standard deviations in Figure~\ref{k_spt}: the Pleiades sequence extends from mid-M through L4. For the M types, the cluster sequence appears slightly overluminous compared to the field (as expected for young ages); however, it readily converges toward cooler spectral types eventually approaching the field sequence. This ``overlapping'' property, particularly for the L types, was already pointed out by \citet{bihain06}, and it is likely related to the fact that L-type objects (including brown dwarfs and low-mass stars) with ages above $\sim$100 Myr tend to have very similar size (0.08--0.12 $R_\odot$) independently of mass and age (see discussion in \citealt{luhman12rev}). 

For comparison purposes, we also plotted in Figure~\ref{k_spt} $\sigma$~Orionis member candidates with spectroscopy available in the literature (see compilation made by \citealt{pena12}); this cluster sequence extends down to L5. With an age estimated at $\sim$3 Myr \citep{osorio02,sherry08} and a distance of 352 pc \citep{perryman97}, $\sigma$~Orionis L-type substellar members appear overluminous by about 1 mag in $J$ and $K_s$ with respect to the Pleiades and the field sequences. This is consistent with the predicted changes in luminosity by evolutionary models \citep{baraffe03}; furthermore, at these young ages, L-type objects (low-mass brown dwarfs and planetary mass objects) have radii a factor of 2--3 times larger than their spectroscopic counterparts at the age of the Pleiades (mainly brown dwarfs). The ``overlapping'' feature between the 3-Myr isochrone and the Pleiades isochrone in color-magnitude or magnitude-spectral type diagrams might happen at cooler temperatures (or spectral types $>$L5) because free-floating planetary mass objects of a few Myr would have similar size as the 100-Myr low-mass brown dwarfs of related temperatures. As seen in Figure~\ref{k_spt}, most of our sources lie between the $\sigma$~Orionis and the field sequences, making it difficult to derive precise ages for our targets using magnitude--spectral type diagrams. For age estimates, we address the reader to Section~\ref{hrdiagram}.

Recently, \citet{faherty12} have claimed that young sources later than M9--L0 appear underluminous in $M_J$, $M_H$, and/or $M_K$ compared to equivalent spectral type objects in the field. Other groups have also claimed that some young, cool substellar companions to stars and brown dwarfs fall below the theoretical isochrones corresponding to their ages (e.g., see \citealt[and references therein]{metchev06,mohanty07,faherty12}). This is an unexpected feature since according to evolutionary models young brown dwarfs are always more luminous than older ones at a given temperature: if there were no major changes in their spectral energy distribution due to, e.g., low gravity, young ultracool dwarfs should appear brighter at all bands or wavelengths. However, all of the young L-type sources with the reported ``underluminous'' issue have very red colors likely indicating atmospheres with a high content of condensates. Dust absorption and scattering (and possibly dust emission) processes make them appear fainter and brighter at short and long wavelengths, respectively, but their net integrated luminosity must be higher than that of older dwarfs of related temperatures (see Section \ref{hrdiagram}). The color/dustiness differences are illustrated in Figure~\ref{k_spt}: the  reddest sources in our sample (G196$-$3B, J0355$+$1133, and J0501$-$0010) appear comparatively brighter in $K_s$ than in $J$ with respect to the field of ``normal'' dwarfs. 

In Figure~\ref{k_spt} we also plotted the young objects of \citet{faherty12,faherty13}. For the two sources in common, J0501$-$0010 and J0355$-$1133, our parallax measurements locate them within the scatter of the field sequence and definitively at brighter positions than \citet{faherty09,faherty12} data. The two locations of J0501$-$0010 are notoriously discrepant, suggesting that one of the distance determinations is wrong or that the error bars are largely underestimated. In general, none of our ten supposedly young L dwarfs appear significantly underluminous with respect to the field sequence in Figure~\ref{k_spt}; they tend to lie on the sequence or slightly above it.

\subsection{HR diagram: ages and masses\label{hrdiagram}}
To locate our targets in the HR diagram, observables must be converted into luminosities and effective temperatures ($T_{\rm eff}$). With the trigonometric parallaxes, absolute magnitudes are easily derived. We then applied bolometric corrections (BC) at different wavelengths to derive absolute luminosities. There are BCs determined for field cool dwarfs (e.g., \citealt{golimowski04}). However, since some of our targets show infrared colors differing from those of ``normal'' field dwarfs, we decided to use BCs specially derived for young L-type sources with reddish color properties. This is in agreement with the statements made by \citet{luhman12rev} and \citet{faherty12} that low surface gravity L dwarfs require a new set of BC/absolute magnitude calibrations. In \citet{todorov10}, these authors determined the appropriate values of BC$_{Ks}$ for three M9.5--L0 young substellar objects for which complete spectral energy distributions have been measured and are available in the literature: one object in Taurus and two young field dwarfs, one of which is J0241$-$0326 (L0). The average BC$_{Ks}$ is $+3.40$ mag, being BC$_{Ks}$\,=\,3.41 mag for J0241$-$0326. As stated in \citet{todorov10}, this BC is larger than that of ``normal'' dwarfs by $\sim$0.2 mag. Similarly, \citet{osorio10} found differing BCs when studying G\,196$-$3B, also in our target list. These authors derived BC$_J$\,=\,1.16 and BC$_{Ks}$\,=\,3.22 mag (with an uncertainty of $\pm$0.10 mag) for the L3 dwarf. We finally used \citet{todorov10} BCs for the L0--L2 objects in our sample, and \citet{osorio10} BCs for the L3--L5 sources. These BCs combined with absolute magnitudes and $M_{\rm bol}$\,=\,4.73 mag for the Sun yield the luminosities shown in Table~\ref{par}. The quoted errors account for the uncertainties in observed magnitudes, parallaxes and BCs. 

For comparison purposes, we also derived luminosities using the near-infrared BCs available in the literature for field dwarfs \citep{dahn02,golimowski04}. For half of the sample we found that the two luminosity derivations agree within $\pm$0.05 dex. The other half (J0033$-$1521, J0355$+$1133, J0501$-$0010, G\,196$-$3B, and J1726$+$1538) shows luminosities about 0.15 dex systematically fainter than the values derived from the BCs of young sources. These are not large differences and do not have significant impact in the following discussion, except for the fact that lower luminosities would make objects slightly older and more massive.

\begin{figure}
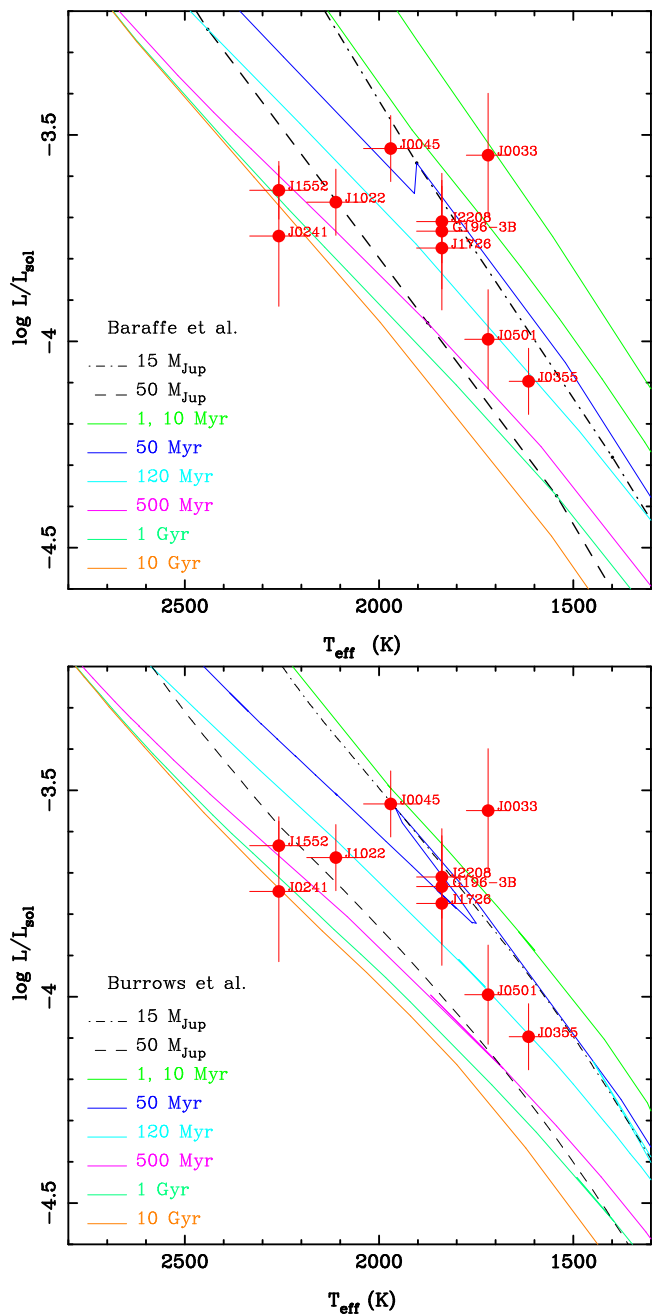
   
\center
\includegraphics[width=8.5cm]{lum_baraffe_r3.ps}
\includegraphics[width=8.5cm]{lum_burrows_r3.ps}
\caption{HR diagram of our targets (solid dots with error bars). The isochrones by \citet[top]{baraffe03} and \citet[bottom]{burrows93} covering the age interval 0.01--10 Gyr  are shown with solid lines. Two tracks of constant mass (15 and 50 M$_{\rm Jup}$) are depicted with dashed lines.
\label{lumteff}}
\end{figure}

The $T_{\rm eff}$ of the sample objects was determined using the optical spectral types summarized in Table~\ref{targets} and the temperature--spectral type relationship by \citet{stephens09}, which is a revised version of \citet{golimowski04} temperature calibration based on measured bolometric luminosities. The derived temperatures and their errors are listed in Table~\ref{par}. Errors take into account an uncertainty of half a subtype in the spectral classification. According to \citet{stephens09}, the scatter of the $T_{\rm eff}$--spectral type relationship is about $\pm$100 K (not included in the Table). This temperature scale yields $T_{\rm eff}$ values within the dispersions quoted by other $T_{\rm eff}$ scales built for high-gravity L dwarfs using the spectral fitting method (e.g., \citealt[and references therein]{cushing08,stephens09,testi09}). One object, J1726$+$1538, is in common with the work by \citet{schweitzer01}, where the authors analyzed the Keck high- and low-resolution optical spectra of the L3 dwarf and compared them to theoretical model atmospheres, which included dust condensation and dust opacities, to measure surface temperature and gravity (log\,$g$). From the study of the pseudo-continuum slope, \citet{schweitzer01} obtained $T_{\rm eff}$\,=\,1900 K and log\,$g$ = 6.0 (cm\,s$^{-2}$) for J1726$+$1538; this agrees with \citet{stephens09} relation. However, the comparison of the theory of model atmospheres with the observations of the Cs\,{\sc i} lines yielded a warmer temperature ($T_{\rm eff}$\,=\,2100 K) and a lower surface gravity (log\,$g$ = 5.5). The authors argued that the discrepancy between temperatures derived from low- and high-resolution spectra is likely due to the treatment of the dust opacity and dust settling issues. Alkali lines are well known to strongly depend on both atmospheric temperature and pressure (or gravity). The analysis of a few atomic lines may not be sufficient to disentangle the degeneracy caused by all three free parameters (temperature, gravity, dust), provided that metallicity is fixed to solar values. 

{Objects with later spectral types have cooler temperatures ranging from $\sim$2300 K (L0) to $\sim$1600 K (L5).  In Figure~\ref{lumteff} we plot the derived $T_{\rm eff}$'s and luminosities (as obtained from the BCs of young sources). Overplotted are the solar metallicity evolutionary models by \citet{burrows93} and \citet{baraffe03}, including isochrones from 10 Myr through 10 Gyr and two tracks of constant mass (15 and 50 M$_{\rm Jup}$). Within the limits of the diagrams, the 50-Myr isochrones display a luminosity peak at around 1900 K, which is likely due to deuterium burning in the interior of 12--15-M$_{\rm Jup}$ objects (some nuclear fusion activity maintains constant luminosity and temperature for a while), whereas more massive objects have already burnt their deuterium content and are now cooling off rapidly. 

As seen in Figure~\ref{lumteff}, it is worth mentioning that the observed luminosity--$T_{\rm eff}$ trend (or slope) is well reproduced by the theory, and that none of our sources, except for  J0033$-$1521, appears to be extremely young (i.e., less than 10 Myr).  Based on Figure~\ref{lumteff}, we provided in Table~\ref{residuals} the likely age and mass ranges for each target assuming they are all single and the \citet{baraffe03} models. The quoted age and mass intervals (we adopted 1 M$_\odot \sim$ 1000 M$_{\rm Jup}$) account for the wide variety of feasible ages. The tracks by \citet{burrows93} tend to yield slightly younger ages and smaller masses than \citet{baraffe03}; however, the results agree with those of Table~\ref{residuals} at the 1-$\sigma$ level, and the conclusions of this paper do not change if any of these models is used. 

J0045$+$1634, J0355$+$1133, J0501$-$0010, G\,196$-$3B, J1726$+$1538, and J2208$+$2921 have ages between $\approx$10 and $\approx$500 Myr and masses ranging from 0.011 to 0.045 M$_\odot$, while J0241$-$0326, J1022$+$5825, and J1552$+$2948 are characterized by their older ages ($\ge$500 Myr) and higher masses (typically $\ge$0.050 M$_\odot$). The former group of objects thus consist of low-mass brown dwarfs or sources close to the planet borderline with typical ages of $\approx$50 Myr (J0045$+$1634, G\,196$-$3B, and J2208$+$2921) and $\approx$120 Myr (J0355$+$1133, J0501$-$0010, and J1726$+$1538). For the compatibility between the spectroscopic lithium observations and these results see Section~\ref{litio}.  If any object in the sample were an equal-mass binary, its luminosity would decrease by 0.3 dex in Figure~\ref{lumteff} and its age would increase up significantly. However, various radial velocity and high spatial resolution imaging works in the literature (see also Section~\ref{companions}) have not revealed the presence of a second, massive object in relatively close orbits around J0045$+$1634, J0355$+$1133, J1726$+$1538, and J2208$+$2921 \citep{reid01,reid06,bouy03,bernat10,blake10,stumpf10}. Our age estimate for J0355$+$1133 (120\,$^{+380}_{-70}$ Myr) extracted from the HR diagram of Figure~\ref{lumteff} is in full agreement with the age determination (50--150 Myr) of \citet{faherty12} based on the likely membership of the L5 dwarf in the young AB Doradus moving group (also see \citealt{liu13b}). In addition, the derived age range of G\,196$-$3B (50\,$^{+450}_{-40}$ Myr) overlaps with the results dynamically inferred by \citet{osorio10}, where it was shown that the L3 dwarf may have been related to the clusters $\alpha$~Persei (more likely about 85 Myr ago) or Collinder 65 (about 20--30 Myr ago).  This adds support to our method of analysis.

As explained in \citet{kirk05} and \citet{cruz09}, the $\beta$ and $\gamma$ appended to the L subtypes indicate intermediate- and very low-gravity spectra, respectively. Based on evolutionary models, the lower the atmospheric gravity the younger age and/or smaller mass \citep{burrows93,baraffe03}. However, there is no strong correlation between the $\beta$/$\gamma$ classification and the ages and masses derived for the target sample. With the Stefan-Boltzmann law we inferred the radii of the targets, and using the equation of gravity and the masses previously derived, we obtained their surface gravities. The results are listed in Table~\ref{residuals}. The uncertainties associated with the radii take into account the errors in luminosity and $T_{\rm eff}$. As for the surface gravities, we provide the probable gravity ranges considering the errors in the radii and the mass intervals of the targets. All ten dwarfs have radii between 9\%~and 19\%~the size of the Sun and intermediate-to-high surface gravities (log\,$g$ $\sim$ 4.0--5.5 cm\,s$^{-2}$) as expected for compact low-mass objects. We conclude that about 60--70\%~of the sample has most likely ages in the interval $\sim$10--500 Myr and intermediate surface gravities (log\,$g$ $\approx$ 4.5 cm\,s$^{-2}$). Therefore, their spectra can be used as a reference for intermediate-gravity cool atmospheres of L types. If we adopt a gravity classification as follows: very low gravity accounts for log\,$g$\,$<$\,4.5, intermediate for 4.5\,$\le$\,log\,$g$\,$<$\,5.0, and high (or ``field'') for log\,$g\,\ge5.0$ (cm\,s$^{-2}$), there is an overall consistency with the results of \citet{allers13} summarized in Table~\ref{summary}. Only J0241$-$0326 and J1552$+$2948 have differing ``gravity class'' assignments: While \citet{allers13} argued that these two dwarfs have very low and intermediate gravities, respectively, we found values closer to log\,$g$\,$\approx$\,5.1--5.5 (cm\,s$^{-2}$), which agrees with a mass above the lithium burning boundary and the actually observed lithium depletion in their atmospheres.

\subsection{Lithium and deuterium\label{litio}}
\citet{chabrier97} argued that, for solar metallicity, 50\%~of Li\,{\sc i} is burned in a dwarf of 0.055 M$_\odot$. Since then, this mass value has been used as the minimum burning mass for  Li\,{\sc i}. Inspection of more recent models by the same group \citep{chabrier00a,baraffe02} reveals that a 0.055-M$_\odot$ brown dwarf has depleted 50\%~of its lithium content at the age of 500 Myr, and about 60\%~at 1 Gyr. Brown dwarfs more massive than 0.055 M$_\odot$ burn Li efficiently at ages below a few hundred Myr. At the age of 120 Myr, the Li depletion boundary is located at the substellar borderline (0.072 M$_\odot$, see \citealt{rebolo96,stauffer98,martin98}). \citet{chabrier97} also pointed out that increasing metallicity yields more efficient depletion; the Li minimum burning mass is thus shifted to lower and higher masses for metal-rich and metal-poor dwarfs, respectively. 

Given our estimated masses and ages from Section~\ref{hrdiagram} (after the assumption that all objects are single), three dwarfs (J0241$-$0326, J1022$+$5825, and J1552$+$2948) should have depleted Li severely and should not show the associated absorption feature at 670.8 nm in their optical spectra. This fact agrees with our GTC spectroscopic observations and the results from the literature: \citet{cruz09} were able to set strong constraints ($pEW$ $<$ 2 \AA) on the presence of lithium in the atmospheres of these three dwarfs; such upper limit in $pEW$ lies well below the Li\,{\sc i} median strength ($pEW$ $\sim$ 4--5 \AA) for spectral types L1--L2 as illustrated in Figure~7 of \citet{kirk00}. 

The remaining seven dwarfs in our sample have estimated masses typically below the lithium burning mass limit; therefore, they should have preserved a significant amount of their primordial lithium stock. Indeed,  our GTC spectra and the spectra available in the literature \citep[and references therein]{rebolo98,cruz09} do confirm the presence of strong lithium lines in all seven but J0033$-$1521. \citet{cruz09} provided an upper limit of $pEW$ $<$ 1 \AA~on the Li\,{\sc i} line strength of J0033$-$1521, which was based on the good quality Keck spectrum available for this L4 dwarf. Furthermore, its optical spectrum does not have the strong VO bands seen in other low-gravity L0 and L1 dwarfs, which may indicate an age higher than the one inferred from the HR diagram. The lack of lithium, the shape of the optical spectral features and the position of J0033$-$1521 in the HR diagram of Figure~\ref{lumteff} cannot be reconciled. Either our parallax and/or $T_{\rm eff}$ (spectral type) determinations are mistaken  (warmer temperature and/or closer distance would be needed) and/or the dwarf is a multiple system of similar mass components. The PARSEC project of \citet{andrei11} will provide an independent parallax measurement for J0033$-$1521. Regarding the multiple nature of this dwarf, very little information is found in the literature. \citet{bouy03} and \citet{gizis03} discarded the presence of massive companions at projected separations $>$0\farcs6 (or 2.4 AU at the distance of the source)  using {\sl HST} high-spatial resolution images. Our analysis of the astrometric data did not reveal any periodic variation due to the presence of a massive companion in orbital periods of 76--1095 d. Interestingly, the spectral type of J0033$-$1521 obtained from the near-infrared wavelengths is L1 \citep{allers13}, three subtypes earlier than the optical classification. This has an impact in the $T_{\rm eff}$ derivation, making the object about 400 K warmer and shifting it to older ages (50--1000 Myr) and higher masses (0.025--0.072 M$_\odot$) in the HR diagrams, in better agreement with the spectroscopic observations. Furthermore, \citet{allers13} discussed that  J0033$-$1521 shows no signs of youth in any near-infrared band, and they classified it as a high-gravity dwarf.

\begin{figure*}   
\center
\includegraphics[width=4.3cm]{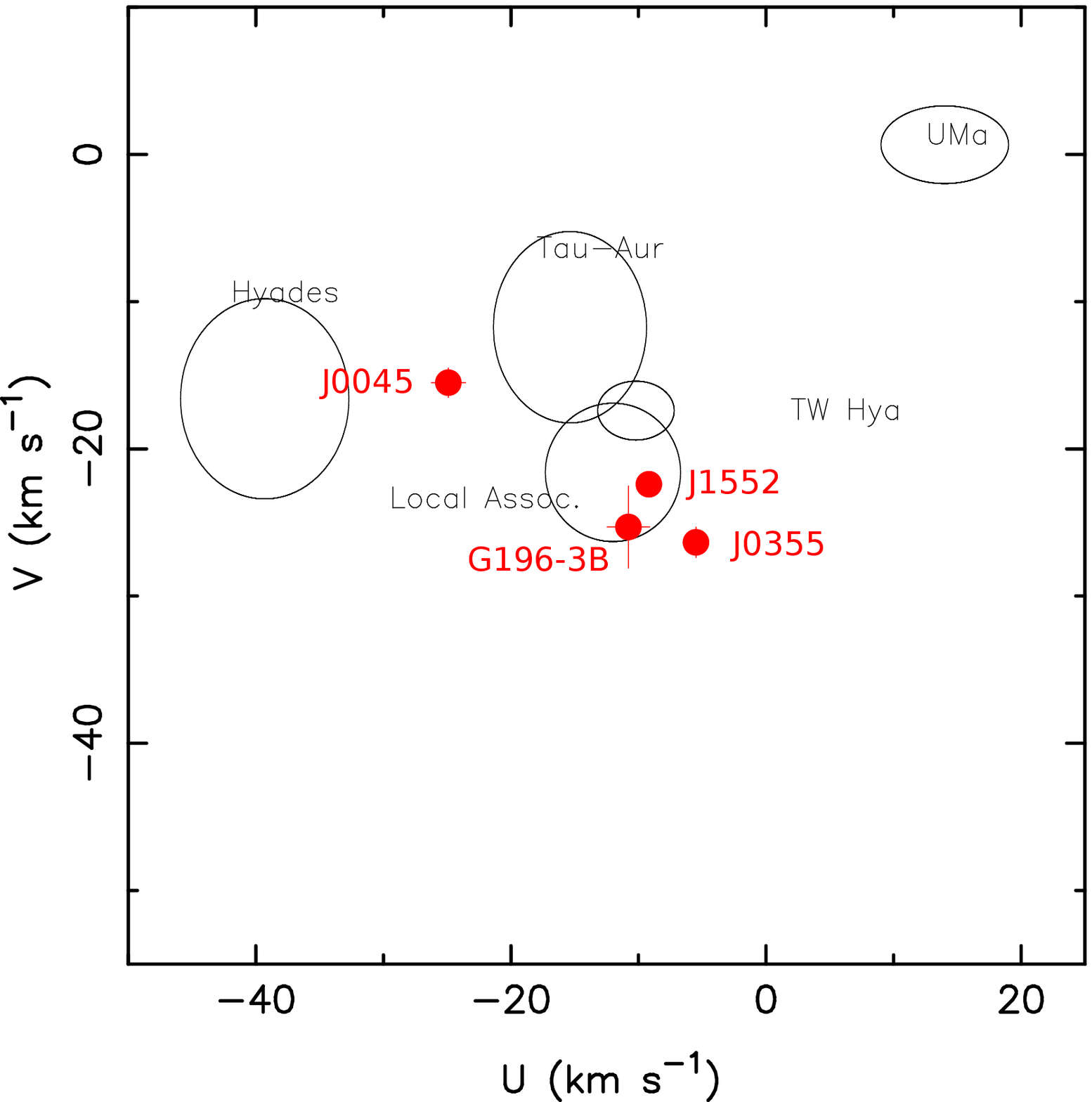} ~~
\includegraphics[width=4.3cm]{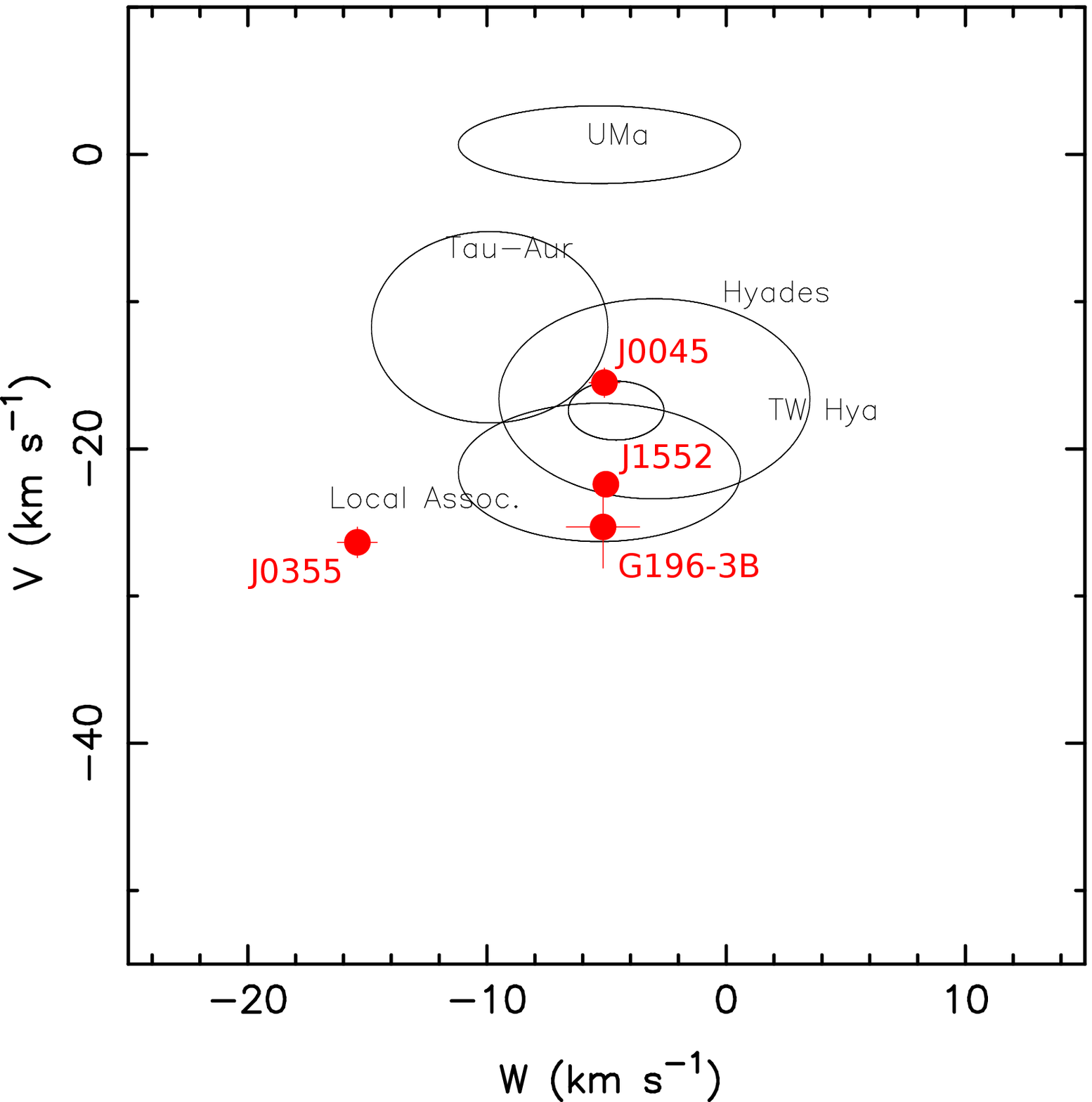} ~~
\includegraphics[width=4.3cm]{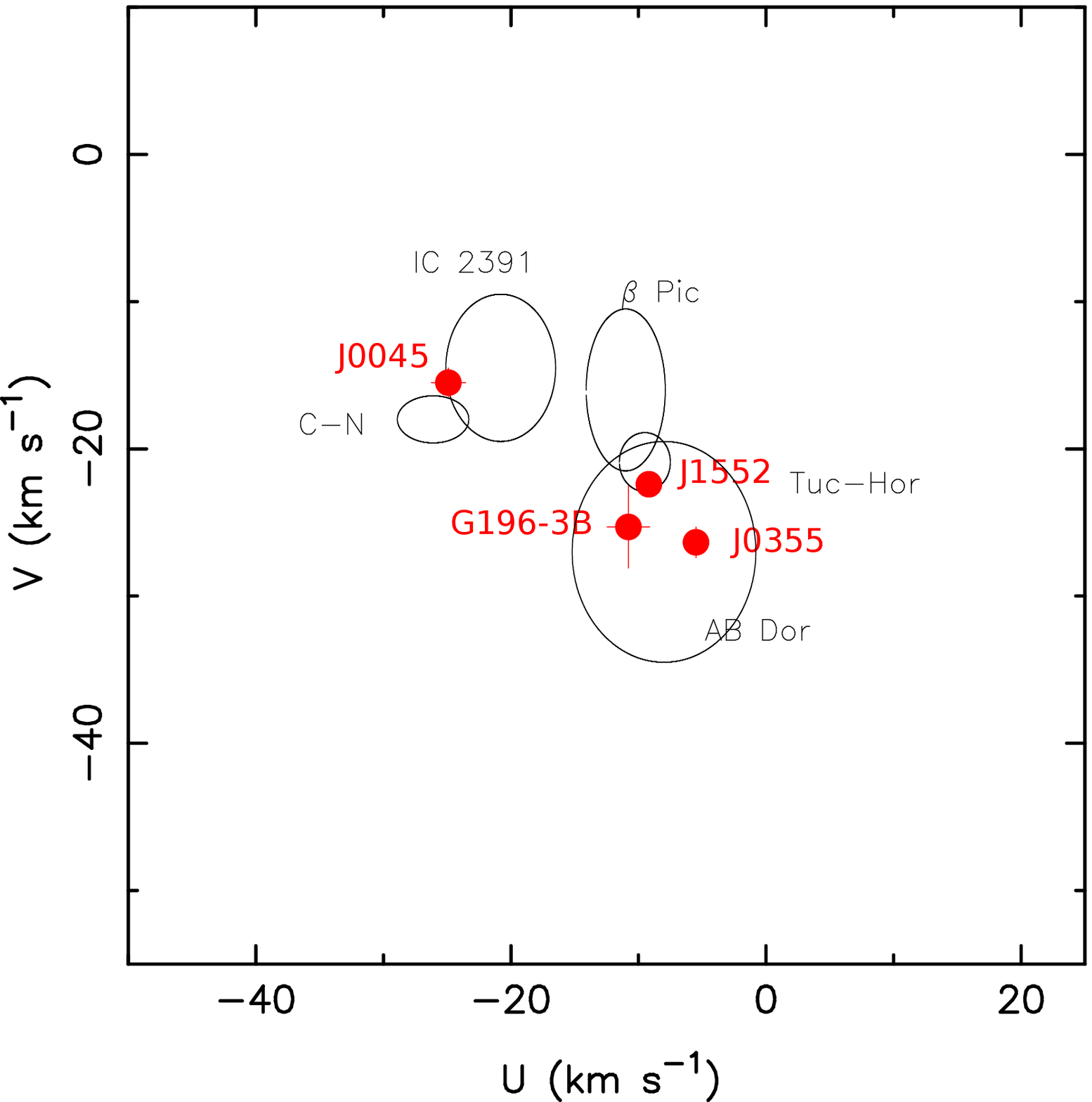} ~~
\includegraphics[width=4.3cm]{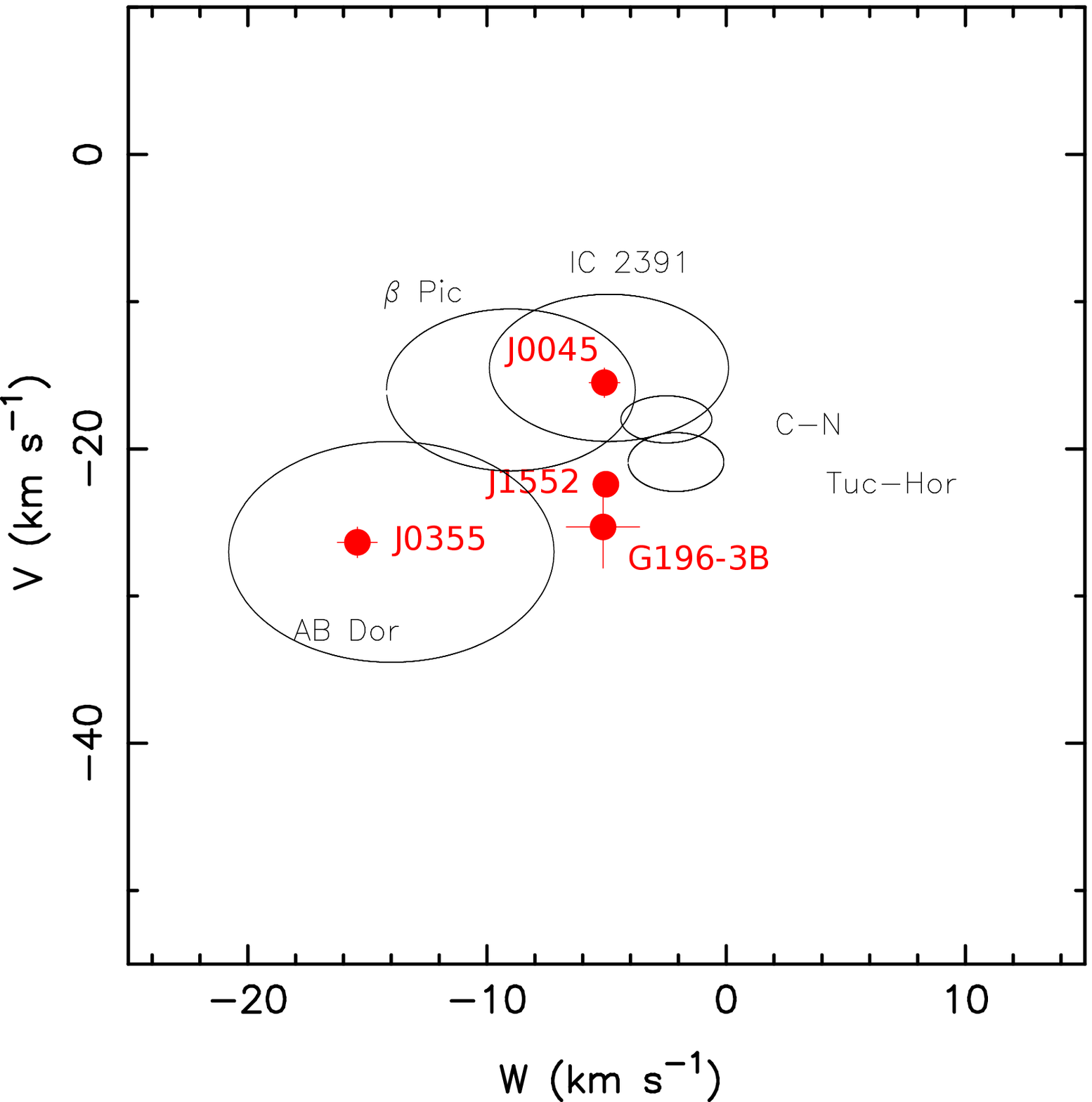} ~~
\caption{Galactic space velocities of some L dwarfs in our sample (filled dots). Note that J1022$+$5825 lies beyond the limits of the plots and is not shown. Error bars are shown (some are of the same size as the symbols). Overplotted are the ellipsoids of known young star associations and moving groups. Galactocentric $U$ velocity is positive toward the Galactic center. 
\label{uvw}}
\end{figure*}

The deuterium burning-mass limit at around 12 M$_{\rm Jup}$ \citep{burrows93,saumon96,chabrier97,chabrier00b} is widely used to separate brown dwarfs and planets. Similarly to the lithium test, the deuterium test was proposed as a way to determine the age and/or mass of very young low-mass objects \citep{bejar99}. Sources with masses above the deuterium burning-mass boundary efficiently burn this element in time scales of less than 50 Myr. A few objects in our sample may have a mass close to this borderline and an age short enough for deuterium to be present in their atmospheres:  J0045$+$1634, G\,196$-$3B, and J2208$+$2921} (see Figure~\ref{lumteff}). If they are confirmed to be single dwarfs with a mass of $\approx$15 M$_{\rm Jup}$, these three sources, particularly the nearest ones (J0045$+$1634 and G\,196$-$3B) would become promising candidates in the nearby field to search for deuterated molecules that independently confirm their low mass and young age. We note, however, that according to the deuterium burning curves computed by \citet{chabrier00b}, they may have depleted this element by a factor of around 100 at the age of $\sim$50 Myr, thus making the ``deuterium'' observations challenging.


\subsection{Space velocities \label{space}}
Membership in stellar moving groups may also help constrain/confirm the ages of our sample targets. In addition to parallax and proper motion, radial velocity is a fundamental ingredient to estimate the three components of the Galactic space velocity, $U$, $V$, and $W$. Spectroscopic radial velocities are available in the literature for J0045$+$1634, J0355$+$1133, J1022$+$5825, and J1552$+$2948 \citep{blake07,blake10,seifahrt10}. As explained in \citet{osorio10}, G\,196$-$3B has no radial velocity measurement, but \citet{basrietal00} obtained a velocity for the primary star, which is the value we used in this paper. We applied the equations by \citet{johnson87} to derive the $UVW$ velocities listed in Table~\ref{tabuvw} and displayed in Figure~\ref{uvw}. The uncertainties associated to all three Galactic velocities come from the proper motion, parallax, and radial velocity error bars. With the only exception of J1022$+$5825, the space velocities of the remaining L dwarfs are consistent with the Galaxy young disk kinematics according to the classification made by \citet{eggen90} and \citet{leggett92}. The $UVW$ velocities of J1022$+$5825 reported here differ from those given by \citet{seifahrt10}, \citet{blake10}, and \citet{allers13} because of our larger proper motion measurement (see Section~\ref{literature}). Nevertheless, the high velocities suggest that this dwarf may kinematically belong to the old population of the Galaxy \citep{leggett92}, which agrees with the results of \citet{blake10}. As indicated in Table~\ref{residuals}, we estimated an age range of 100--1000 Myr for J1022$+$5825, implying that according to its location in the HR diagram, this dwarfs is not very old. It might have been ejected at a high velocity from its original birth place, or it might have gained high velocity through dynamical interactions/encounters with other massive objects, in which case the space velocity analysis is not representative of the dwarf age.

\begin{table}
\caption{Space velocities. \label{tabuvw}}
\begin{tabular}{llrrr}
\hline  \hline
\multicolumn{1}{l}{Object} & 
\multicolumn{1}{l}{SpT} & 
\multicolumn{1}{c}{$U$} & 
\multicolumn{1}{c}{$V$} &
\multicolumn{1}{c}{$W$} \\
\multicolumn{1}{l}{} & 
\multicolumn{1}{r}{} & 
\multicolumn{1}{c}{(km\,s$^{-1}$)} & 
\multicolumn{1}{c}{(km\,s$^{-1}$)} &
\multicolumn{1}{c}{(km\,s$^{-1}$)} \\
\hline
J0045$+$1634 & L2$\beta$  &  $-$24.9$\pm$1.3 &  $-$15.5$\pm$1.0 &  $-$5.1$\pm$0.6  \\
J0355$+$1133 & L5$\gamma$ &   $-$5.5$\pm$0.5 &  $-$26.4$\pm$1.0 & $-$15.4$\pm$0.8  \\
G\,196--3B   & L3$\beta$  &  $-$10.8$\pm$1.7 &  $-$25.3$\pm$2.8 &  $-$5.2$\pm$1.5  \\
J1022$+$5825 & L1$\beta$  &  $-$79.1$\pm$2.0 &  $-$81.3$\pm$2.4 &  $-$1.8$\pm$1.6  \\ 
J1552$+$2948 & L0$\beta$  &   $-$9.2$\pm$0.9 &  $-$22.4$\pm$0.7 &  $-$5.0$\pm$0.4  \\  
\hline
\end{tabular}
\end{table}

Figure~\ref{uvw} also illustrates the ellipsoids corresponding to well-known young stellar moving groups of the solar neighborhood (data from \citealt{zuckerman04,torres08}). Our data confirm the claims by \citet{faherty12}, \citet{osorio10}, and \citet{gagne14} that J0355$+$1133, G\,196$-$3B, and J0045$+$1634 are likely members of the young AB Doradus, Local Association, and Argus moving groups, respectively. AB Doradus ($\sim$70--120 Myr, \citealt{luhman05}) is one of the closest moving groups to the Sun, while the Argus Association, including IC\,2391 (ages ranging from 30 through 250 Myr, \citealt{makarov00}), lies on average more distant from the Sun but some of its members are within 70 pc (see \citealt{zuckerman11}).  \citet{gagne14} also discussed that G\,196$-$3B has a high probability of being an AB Doradus member and that J2208$+$2921 possibly belongs to the $\beta$ Pictoris moving group ($\sim$20 Myr, \citealt{binks14}). The memberships of J0045$+$1634, J0355$+$1133, G\,196$-$3B, and J2208$+$2921 in the aforementioned young moving groups agree with the age estimations given in Table~\ref{residuals}.

As seen in Figure~\ref{uvw}, the space velocities of J1552$+$2948 point to its likely membership in the Local Association group, a coherent kinematic stream of young stars (all below 300 Myr, \citealt{eggen92}) with embedded clusters and associations such as the Pleiades (120 Myr), $\alpha$ Persei (50--80 Myr), and IC\,2602 ($\sim$70 Myr). Only the oldest ages are barely compatible with the position of this dwarf in Figure~\ref{lumteff} and the age range provided in Table~\ref{residuals}. Similarly, \citet{gagne14} argued that J0241$-$0326 is a likely member of the $\sim$30-Myr Tucana-Horologium moving group. The lack of lithium and our age estimation are not consistent with such a young age for the L0 dwarf.

\subsection{Astrometric companions \label{companions}}
The multi-epoch astrometric residuals were examined for any periodic perturbations that might reveal unseen companions. If any candidate is found, the characterization of its orbital parameters would require additional observations. We note, however, that the observing strategy was optimized for the parallax derivations. 

The OMEGA2000 astrometric residuals of each L dwarf were subjected to a time-series analysis using the Lomb-Scargle \citep{scargle82} and the Plavchan \citep{plavchan08} periodograms\footnote{We used our codes and the codes provided by the NASA Exoplanet Archive: http://exoplanetarchive.ipac.caltech.edu/cgi-bin/Periodogram/nph-simpleupload}. The later method is useful to detect periodic time-series patterns that are not well described by the Lomb-Scargle algorithm, e.g., large eccentricity orbits that clearly deviate from a sinusoidal-shaped periodic signal. The periodograms were computed for $\alpha$ and $\delta$ separately. We explored frequencies up to two times the effective Nyquist frequency, i.e., periods between 38--97 d and 1095 d. The periods investigated for each target are shown in Table~\ref{residuals}. The lower period limit is different from dwarf to dwarf, and it mostly depends on the number of monthly data points available per object: the larger the data number, the smaller the lower period limit that can be searched. The longest period considered was three years, nearly the observational baseline. If any companion is present, its signal must appear at the same frequency/period in both d$\alpha$ and d$\delta$, unless the inclination of the orbit is close to 90\degr, in which case a strong peak happens in only one coordinate. Observations were taken typically one per month, and the sinusoidal parallax pattern repeats itself annually; these two effects may produce false peaks in the periodograms at frequencies around 30 d and 1 yr and their corresponding harmonics.  

\begin{figure*}
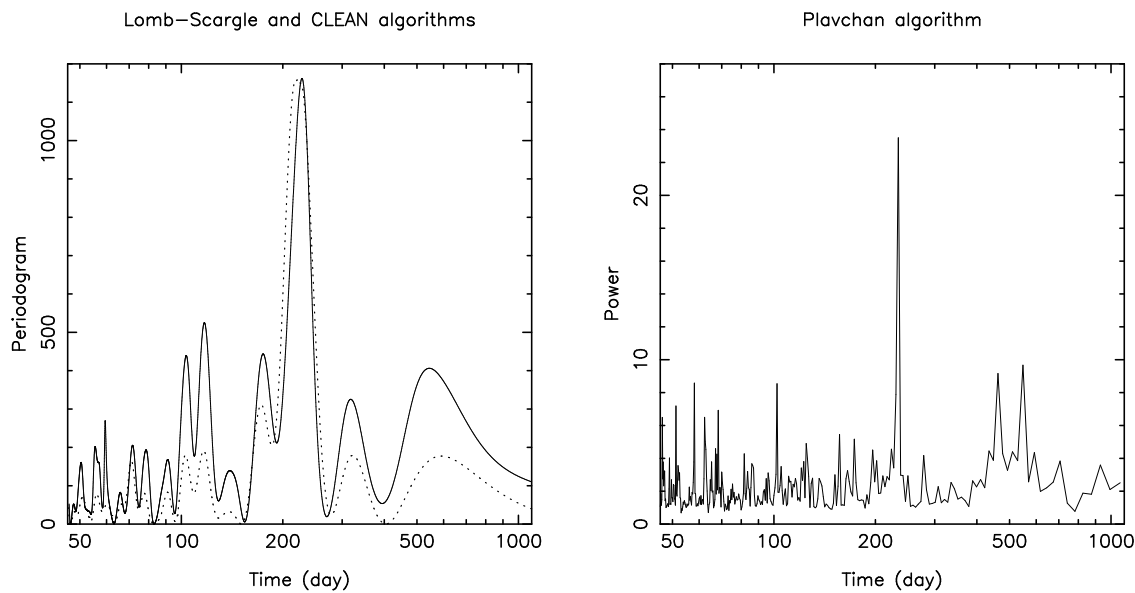
   
\center
\includegraphics[width=7cm]{g196_LS_clean_alpha.ps} ~~~~~~~
\includegraphics[width=7cm]{g196_plavchan_alpha.ps}
\caption{Periodograms computed from the d$\alpha$ astrometric residuals of G\,196$-$3B (only OMEGA2000 data). The highest peaks are found at 228.46 d (left panel, solid line, Lomb-Scargle algorithm, \citealt{scargle82}) and 233.88 d (right panel, Plavchan algorithm, \citealt{plavchan08}). The ``cleaned'' periodogram produced by the algorithm of \citet{roberts87} and normalized to the Lomb-Scargle peak power is plotted as a dotted line in the left panel.
\label{period}}
\end{figure*}

We found no significant peaks in the periodograms calculated for the sample, except for G\,196$-$3B, whose Lomb-Scargle and Plavchan periodograms are shown in Figure~\ref{period}. Both algorithms yielded very similar peaks at 228.46 d (Lomb-Scargle) and 233.88 d (Plavchan), although the confidence that this peak is real is only 87\%. To have a third determination of G\,196$-$3B's periodogram, we also executed the {\sc clean} algorithm of \citet{roberts87} by using a low gain of 0.1 and a moderate number of iterations (5). This algorithm basically deconvolves the spectral window from the discrete Fourier power spectrum. The ``cleaned'' periodogram, which has a peak at around 222 d, is shown in Figure~\ref{period} in comparison with the Lomb-Scargle power spectrum. Coincidentally, G\,196$-$3B shows the largest astrometric residuals scatter in the sample of L dwarfs (Table~\ref{residuals}). The uncertainty of the Lomb-Scargle period is $\pm$14 d, which comes from the FWHM of the peak. The signal of the peaks appears only in $\alpha$ and is likely diluted in $\delta$; as discussed in Section~\ref{method}, our astrometric data have better precision in $\alpha$ than in $\delta$.  Figure~\ref{phase} shows the d$\alpha$ residuals of G\,196$-$3B folded over the 228.46-d period; the OMEGA2000 data show a distinctive pattern that is also followed by the NOTCam astrometry within the quoted uncertainties. The amplitude of the astrometric signal is about 11 mas. 

The detailed interpretation and analysis of the 228.46-d period is beyond the scope of this paper. There are several factors that may contribute to the astrometric signal (activity, pulsations, enhanced aliases, uncontrolled systematics, ...), one is the presence of a companion in which case the observed periodicity is the reflex motion of the primary. Given the measured parallax and the reflex displacement, we estimated that the minimum separation between the putative components of G\,196$-$3B would be 0.27 AU for an equal-mass binary. If confirmed with further observations, G\,196$-$3 would turn out to be a young multiple system formed by a low mass star and a binary substellar companion. Multiple systems where two close brown dwarfs orbit stars are already known in the solar neighborhood (e.g., GJ\,569Bab, \citealt{martin00,osorio04}, and $\epsilon$ Ind B, \citealt{volk03,mccaughrean04}). Current direct imaging techniques do not have the capability to resolve G\,196$-$3B at the distance of the system; radial velocity observations, on the other hand, are feasible and highly required to confirm or discard the 228.46-d period. Because we cannot rule out any uncontrolled systematics (e.g., the 228.64-d period corresponds to a factor of 2.5 the Nyquist wavelength), G\,196$-$3B is not to be considered a binary until follow-up spectroscopic observations are available.  L\'opez-Mart\'\i~\& Zapatero Osorio (2014, submitted) studied the OMEGA2000 photometric light curve of G\,196$-$3B finding that this dwarf is variable in time scales of months, which may impact the astrometry.

\subsubsection{Minimum detectable mass of astrometric companions}
We estimated the minimum mass of companions around the remaining L dwarfs as well as around G\,196$-$3B that could have been detected in our study by assuming the primary masses discussed in Section~\ref{hrdiagram} (Table~\ref{residuals}), the absolute parallaxes derived in Section~\ref{method}, a circular and face-on orbit (for simplicity), and the mass function given by the equation:
\begin{equation}
\frac{M_c}{(M + M_c)^{2/3}} = \frac{1}{P^{2/3}} \frac{\Delta({\rm d}\alpha)}{\pi}
\end{equation}
where the masses are in solar units ($M_c$ and $M$ are the minimum detectable mass of the companion and the mass of the primary, respectively), $P$ is the period in years (1095 d in our study), and $\Delta({\rm d}\alpha)$ is the minimum detectable astrometric perturbation. We did not attempt to derive the minimum detectable mass of an astrometric companion around J0033$-$1521 because of the disagreement between our age determination and the spectroscopic observations (see Section~\ref{litio}). At the 1-$\sigma$ level, we adopted the scatter of the residuals of each target as the minimum detectable perturbation; this is a conservative assumption since it has been indicated that a signal must be about 92\%~of the magnitude of the average residual in order to be detected \citep{bartlett07,bartlett09}. Table~\ref{residuals} provides the minimum masses of the companions for each target. Globally, the astrometric data of eight dwarfs in the sample (all except for J1022$+$5825 and G\,196$-$B, see Table~\ref{residuals}) do not show evidence for companions more massive than $\sim$15 M$_{\rm Jup}$ at the largest investigated periods. At shorter orbital periods, the minimum masses increase with $P^{-2/3}$.

{Noteworthy are the minimum mass constraints of any putative companions around J0355$+$1133 and J0501$-$0010, which we estimated at 2.4 and 4.7 M$_{\rm Jup}$ for orbits of 1095 d. More massive planets up to 12 M$_{\rm Jup}$ can be discarded for shorter orbital periods down to$\sim$135 d (J0355$+$1133) and $\sim$ 380 d (J0501$-$0010). Using aperture masking interferometry and laser guide star adaptive optics of the 200-inch telescope on the Palomar Observatory, \citet{bernat10} reported the detection of a companion candidate around J0355$+$1133 at separation 82.5 mas and 2.1:1 contrast in the $K_s$-band, suggesting that both companion and primary have similar masses. At the distance of J0355$+$1133, the projected physical separation would result in 0.74 AU, and each component would have a mass of about 0.04--0.05 M$_\odot$ (after splitting the object luminosity into two sources in Fig$.$~\ref{lumteff}), implying and orbital period of $\ge$2 yr.  The small dispersion of J0355$+$1133's astrometric residuals in $\alpha$ (Table~\ref{residuals}) during three consecutive years of observations hints at a mass ratio $q < 0.1$, which is not consistent with \citet{bernat10} finding. Yet, our astrometric results would be compatible with \citet{bernat10} candidate if the orbital period of the putative binary were beyond the limits of our study or the inclination of the orbit is such that its astrometric projection mainly lies along the Declination axis. Other searches for companions around J0355$+$1133 based on Hubble Space Telescope (HST) high-spatial resolution imaging and radial velocity measurements \citep{reid06,blake10} yielded no companion candidate detection.

\begin{figure}   
\center
\includegraphics[width=9cm]{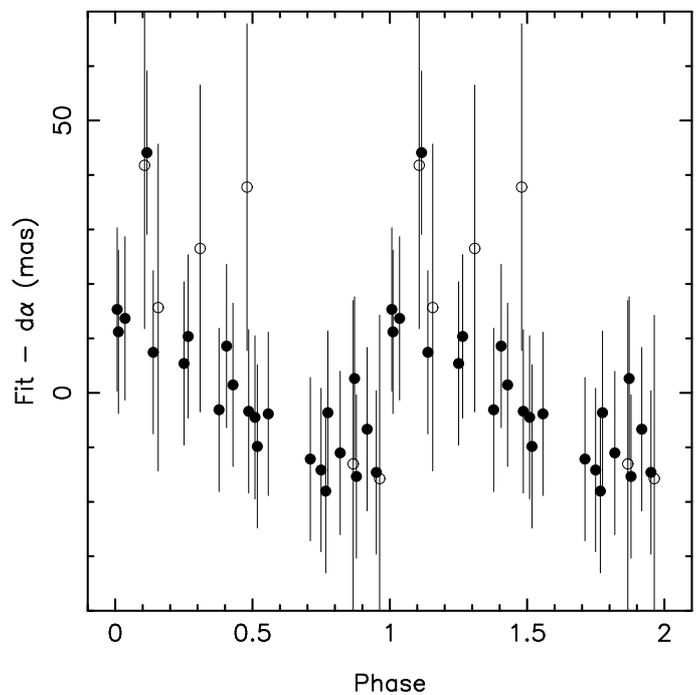}
\caption{Astrometric residuals curve of G\,196$-$3B folded in phase using the period of 228.46 d. OMEGA2000 data are depicted with solid dots, and NOTCam data are plotted as open circles. Two cycles are shown.
\label{phase}}
\end{figure}


\section{Summary and conclusions \label{conclusions}}
Using $J$- and $K_s$-band imaging data obtained with a typical cadence of one month over a time scale of nearly three years between 2010 January and 2012 December, we derived the trigonometric parallaxes and proper motions for a sample of ten field L dwarfs (L0--L5) from the catalog of \citet{cruz09}, whose spectral features and red near- and mid-infrared colors are indicative of low-gravity atmospheres. The sample represents 44\%~of \citet{cruz09} objects. Parallaxes and proper motions were measured with typical accuracies of 1--4 mas and $\sim$10 mas\,yr$^{-1}$. All ten sources are located at nearby distances, between 9 and 47 pc, and show significant proper motion ($\mu \ge$70 mas\,yr$^{-1}$). To put our sources in context, we built various magnitude--spectral type diagrams and the luminosity--temperature plane by obtaining absolute magnitudes, luminosities, and effective temperatures from our trigonometric parallaxes and the observed photometry available in the literature. Our sample of ``young'' L dwarfs was compared with the recent catalog of late-M--L--T objects by \cite{dupuy12} in the $M_J$ and $M_{K_{s}}$ versus spectral type planes, finding that all of our sources lie along the field sequence of ``high-gravity'' dwarfs or between the sequences defined by the 3-Myr $\sigma$~Orionis cluster and the field, as expected for young ages.  This result contrasts with the latest studies by \citet{faherty12,faherty13}, who announced that a significant fraction of ``young'' field L dwarfs are underluminous by 0.2--1.0 mag at $M_J$, $M_H$, and $M_K$ bands. There are two objects (J0501$-$0010 and J0355$+$1133) in common with \citet{faherty12,faherty13}; our distances are larger than those obtained by these authors, thus making the two L dwarfs brighter in $M_J$ and $M_{K_{s}}$ than previously reported. Young, red L dwarfs may appear fainter at short wavelengths and brighter at long ones; this effect arises from their very red colors (indicative of rather dusty atmospheres or, less likely, the presence of disks) and not from their net integrated luminosity.

To estimate the likely ages and masses of the sample, all ten sources were placed in the HR diagram and their positions were compared to solar metallicity stellar and substellar evolutionary models. Assuming that our targets are single, we determined that six out of ten (J0045$+$1634, J0355$+$1133, J0501$-$0010, G\,196$-$3B, J1726$+$1538, and J2208$+$2921) have likely ages and masses in the intervals $\approx$10--500 Myr and $\approx$11--45 M$_{\rm Jup}$, respectively, thus supporting their young-to-intermediate age and substellar nature. According to the lithium depletion curves of the evolutionary models, all six dwarfs must have preserve significant amounts of lithium in their atmospheres. Our spectroscopic observations obtained at optical wavelengths with a spectral resolution of about $R \sim 300$ confirmed the presence of strong Li\,{\sc i} absorption at 670.8 nm in J0045$+$1634 (for which only an upper limit on the strength of the atomic line was reported in the literature), J0355$+$1133, and G\,196$-$3B. \citet{cruz09} presented the lithium detection in the spectra of J0501$-$0010, J1726$+$1538, and J2208$+$2921. Therefore, lithium observations and our mass and age derivations are consistent with each other. There is one exception, J0033$-$1521, which has no lithium absorption despite its location in the HR diagram indicating an age below 10 Myr and a very low mass. The near-infrared spectral type of J0033$-$1521 suggests that this dwarf is significantly warmer than typical L4 dwarfs, thus bringing all independent observations of J0033$-$1521 into better agreement. The remaining three dwarfs in the sample (J0241$-$0326, J1022$+$5825, and J1552$+$2948) have locations in the HR diagram indicative of ages older than 500 Myr and masses above the lithium burning-mass limit. This is consistent with the lack of lithium absorption observed in their optical spectra. The $H$-band peaked-shape seen in the spectra of many of our targets indicates that this particular spectral feature, which is a signpost of youth, may persist up to ages of $\approx$120--500 Myr and intermediate-to-high gravities. 

Finally, our multi-epoch astrometric data (d$\alpha$) were explored for any periodic signal that might reveal unseen companions. For most targets we were able to set the minimum detectable mass of any possible astrometric companion in face-on, circular orbits with periods between $\sim$60--90 d and 3 yr. No candidates more massive than typically $\sim$25 M$_{\rm Jup}$ were found around any of the targets. A tentative signal is, however, observed in G\,196$-$3B, which deserves further data for confirmation. 


\begin{acknowledgements}
We are thankful to Jos\'e A$.$ Caballero for fruitful discussions on proper motions, and for checking the proper motions of a few targets using tools of the Virtual Observatory. We also thank the referee (Sandy Leggett) for useful comments that have improved some sections of the manuscript. Based on observations collected at the Centro Astron\'omico Hispano Alem\'an (CAHA) at Calar Alto, operated jointly by the Max-Planck Institut f\"ur Astronomie and the Instituto de Astrof\'\i sica de Andaluc\'\i a (CSIC); observations made with the Nordic Optical Telescope, operated on the island of La Palma jointly by Denmark, Finland, Iceland, Norway, and Sweden, in the Spanish Observatorio del Roque de los Muchachos of the Instituto de Astrof\'\i sica de Canarias; observations made with the William Herschel Telescope operated on the island of La Palma by the Isaac Newton Group in the Spanish Observatorio del Roque de los Muchachos of the Instituto de Astrof\'\i sica de Canarias; and observations made with the Gran Telescopio de Canarias (GTC), instaled in the Spanish Observatorio del Roque de los Muchachos of the Instituto de Astrof\'\i sica de Canarias, in the island of La Palma. This publication makes use of data products from the Wide-field Infrared Survey Explorer, which is a joint project of the University of California, Los Angeles, and the Jet Propulsion Laboratory/California Institute of Technology, funded by the National Aeronautics and Space Administration. This research has been supported by the Spanish Ministry of Economics and Competitiveness under the projects AYA2011-30147-C03-03 and AYA2010-20535. 
\end{acknowledgements}

\bibliographystyle{aa} 
\bibliography{2.bib}

\begin{thebibliography}{118}
\expandafter\ifx\csname natexlab\endcsname\relax\def\natexlab#1{#1}\fi

\bibitem[{Allers \& Liu(2013)}]{allers13}
Allers, K.~N. \& Liu, M.~C. 2013, ApJ, 771, 79

\bibitem[{Allers {et~al.}(2010)Allers, Liu, Dupuy, \& Cushing}]{allers10}
Allers, K.~N., Liu, M.~C., Dupuy, T.~J., \& Cushing, M.~C. 2010, ApJ, 715, 561

\bibitem[{Andrei {et~al.}(2011)Andrei, Smart, Penna, d'Avila, Bucciarelli,
  Camargo, Crosta, Dapr\`a, Goldman, Jones, Lattanzi, Nicastro, Pinfield,
  da~Silva~Neto, \& Teixeira}]{andrei11}
Andrei, A.~H., Smart, R.~L., Penna, J.~L., {et~al.} 2011, AJ, 141, 54

\bibitem[{Baraffe {et~al.}(2002)Baraffe, Chabrier, Allard, \&
  Hauschildt}]{baraffe02}
Baraffe, I., Chabrier, G., Allard, F., \& Hauschildt, P.~H. 2002, A\&A, 382,
  563

\bibitem[{Baraffe {et~al.}(2003)Baraffe, Chabrier, Barman, Allard, \&
  Hauschildt}]{baraffe03}
Baraffe, I., Chabrier, G., Barman, T.~S., Allard, F., \& Hauschildt, P.~H.
  2003, A\&A, 402, 701

\bibitem[{Bartlett(2007)}]{bartlett07}
Bartlett, J.~L. 2007, Ph.D. thesis, Univ. Virginia

\bibitem[{Bartlett {et~al.}(2009)Bartlett, Ianna, \& Begam}]{bartlett09}
Bartlett, J.~L., Ianna, P.~A., \& Begam, M.~C. 2009, PASP, 121, 365

\bibitem[{Basri(2000)}]{basri00}
Basri, G. 2000, ARA\&A, 38, 485

\bibitem[{Basri {et~al.}(2000)Basri, Mohanty, Allard, Hauschildt, Delfosse,
  Mart{\'i}n, Forveille, \& Goldman}]{basrietal00}
Basri, G., Mohanty, S., Allard, F., {et~al.} 2000, ApJ, 538, 363

\bibitem[{Beichman {et~al.}(2013)Beichman, Gelino, Kirkpatrick, Barman, Marsh,
  Cushing, \& Wright}]{beichman13}
Beichman, C.~A., Gelino, C.~R., Kirkpatrick, J.~D., {et~al.} 2013,
  arXiv1301.1669

\bibitem[{B\'ejar {et~al.}(1999)B\'ejar, Zapatero~Osorio, \& Rebolo}]{bejar99}
B\'ejar, V. J.~S., Zapatero~Osorio, M.~R., \& Rebolo, R. 1999, ApJ, 521, 671

\bibitem[{Bernat {et~al.}(2010)Bernat, Bouchez, Ireland, Tuthill, Martinache,
  Angione, Burruss, Cromer, Dekany, Guiwits, Henning, Hickey, Kibblewhite,
  McKenna, Moore, Petrie, Roberts, Shelton, Thicksten, Trinh, Tripathi, Troy,
  Truong, Velur, \& Lloyd}]{bernat10}
Bernat, D., Bouchez, A.~H., Ireland, M., {et~al.} 2010, ApJ, 715, 724

\bibitem[{Bihain {et~al.}(2006)Bihain, Rebolo, B~\'ejar, Caballero,
  Bailer-Jones, Mundt, Acosta-Pulido, \& Manchado~Torres}]{bihain06}
Bihain, G., Rebolo, R., B~\'ejar, V. J.~S., {et~al.} 2006, A\&A, 458, 805

\bibitem[{Bihain {et~al.}(2010)Bihain, Rebolo, Zapatero~Osorio, B\'ejar, \&
  Caballero}]{bihain10}
Bihain, G., Rebolo, R., Zapatero~Osorio, M.~R., B\'ejar, V. J.~S., \&
  Caballero, J.~A. 2010, A\&A, 519, 93

\bibitem[{Binks \& Jeffries(2014)}]{binks14}
Binks, A., S. \& Jeffries, R.~D. 2014, MNRAS, 438, L11

\bibitem[{Blake {et~al.}(2010)Blake, Charbonneau, \& Whitle}]{blake10}
Blake, C.~H., Charbonneau, D., \& Whitle, R.~J. 2010, ApJ, 723, 684

\bibitem[{Blake {et~al.}(2007)Blake, Charbonneau, Whitle, Marley, \&
  Saumon}]{blake07}
Blake, C.~H., Charbonneau, D., Whitle, R.~J., Marley, M.~S., \& Saumon, D.
  2007, ApJ, 666, 1198

\bibitem[{Bouy {et~al.}(2003)Bouy, Brandner, Mart{\'i}n, Delfosse, Allard, \&
  Basri}]{bouy03}
Bouy, H., Brandner, W., Mart{\'i}n, E.~L., {et~al.} 2003, AJ, 126, 1526

\bibitem[{Bowler {et~al.}(2013)Bowler, Liu, Shkolnik, \& Dupuy}]{bowler13}
Bowler, B.~P., Liu, M.~C., Shkolnik, E., \& Dupuy, T.~J. 2013, ApJ, 774, 55

\bibitem[{Burgasser {et~al.}(2008)Burgasser, Vrba, L\'epine, Munn, Luginbuhl,
  Henden, Guetter, \& Canzian}]{burgasser08}
Burgasser, A.~J., Vrba, F.~J., L\'epine, S., {et~al.} 2008, ApJ, 672, 1159

\bibitem[{Burrows {et~al.}(1993)Burrows, Hubbard, Saumon, \&
  Lunine}]{burrows93}
Burrows, A., Hubbard, W.~B., Saumon, D., \& Lunine, J.~I. 1993, ApJ, 406, 158

\bibitem[{Burrows {et~al.}(2006)Burrows, Sudarsky, \& Hubeny}]{burrows06}
Burrows, A., Sudarsky, D., \& Hubeny, I. 2006, ApJ, 640, 1063

\bibitem[{Casewell {et~al.}(2008)Casewell, Jameson, \& Burleigh}]{casewell08}
Casewell, S.~L., Jameson, R.~F., \& Burleigh, M.~R. 2008, MNRAS, 390, 1517

\bibitem[{Cepa(1998)}]{cepa98}
Cepa, J. 1998, Ap\&SS, 263, 369

\bibitem[{Chabrier \& Baraffe(1997)}]{chabrier97}
Chabrier, G. \& Baraffe, I. 1997, A\&A, 327, 1039

\bibitem[{Chabrier {et~al.}(2000{\natexlab{a}})Chabrier, Baraffe, Allard, \&
  Hauschildt}]{chabrier00a}
Chabrier, G., Baraffe, I., Allard, F., \& Hauschildt, P.~H. 2000{\natexlab{a}},
  ApJ, 542, 464

\bibitem[{Chabrier {et~al.}(2000{\natexlab{b}})Chabrier, Baraffe, Allard, \&
  Hauschildt}]{chabrier00b}
Chabrier, G., Baraffe, I., Allard, F., \& Hauschildt, P.~H. 2000{\natexlab{b}},
  ApJ, 542, L119

\bibitem[{Cruz {et~al.}(2009)Cruz, Kirkpatrick, \& Burgasser}]{cruz09}
Cruz, K.~L., Kirkpatrick, J.~D., \& Burgasser, A.~J. 2009, AJ, 137, 3345

\bibitem[{Cushing {et~al.}(2011)Cushing, Kirkpatrick, Gelino, Griffith,
  Skrutskie, Mainzer, Marsh, Beichman, Burgasser, Prato, Smcoe, ;arley, Saumon,
  Freedman, Eisenhardt, \& Wright}]{cushing11}
Cushing, M.~C., Kirkpatrick, J.~D., Gelino, C.~R., {et~al.} 2011, ApJ, 743, 50

\bibitem[{Cushing {et~al.}(2008)Cushing, Marley, Saumon, Kelly, Vacca, Rayner,
  Freedman, Lodders, \& Roellig}]{cushing08}
Cushing, M.~C., Marley, M.~S., Saumon, D., {et~al.} 2008, ApJ, 678, 1372

\bibitem[{Cushing {et~al.}(2005)Cushing, Rayner, \& Vacca}]{cushing05}
Cushing, M.~C., Rayner, J.~T., \& Vacca, W.~D. 2005, ApJ, 623, 1115

\bibitem[{Dahn {et~al.}(2002)Dahn, Harris, Vrba, Guetter, Canzian, Henden,
  Levine, Luginbuhl, Monet, Monet, Pier, Stone, Walker, Burgasser, Gizis,
  Kirkpatrick, Liebert, \& Reid}]{dahn02}
Dahn, C.~C., Harris, H.~C., Vrba, F.~J., {et~al.} 2002, AJ, 124, 1170

\bibitem[{Dupuy \& Liu(2012)}]{dupuy12}
Dupuy, T.~J. \& Liu, M.~C. 2012, ApJS, 201, 19

\bibitem[{Eggen(1990)}]{eggen90}
Eggen, O.~J. 1990, AJ, 100, 1159

\bibitem[{Eggen(1992)}]{eggen92}
Eggen, O.~J. 1992, AJ, 103, 1302

\bibitem[{Faherty {et~al.}(2012)Faherty, Burgasser, Walter, Van~der Bliek,
  Shara, Cruz, West, Vrba, \& Anglada-Escud\'e}]{faherty12}
Faherty, J.~K., Burgasser, A.~J., Walter, F.~M., {et~al.} 2012, ApJ, 752, 56

\bibitem[{Faherty {et~al.}(2013)Faherty, Rice, Cruz, Mamajek, \&
  N\'u\~nez}]{faherty13}
Faherty, J.~K., Rice, E.~L., Cruz, K.~L., Mamajek, E.~E., \& N\'u\~nez, A.
  2013, AJ, 145, 2

\bibitem[{Faherty {et~al.}(2009)Faherty, Burgasser, Cruz, Shara, Walter, \&
  Gelino}]{faherty09}
Faherty, J.~L., Burgasser, A.~J., Cruz, K.~L., {et~al.} 2009, AJ, 137, 1

\bibitem[{Filippenko(1982)}]{filippenko82}
Filippenko, A.~V. 1982, PASP, 94, 715

\bibitem[{Filippenko \& Greenstein(1984)}]{filippenko84}
Filippenko, A., V. \& Greenstein, J.~L. 1984, PASP, 96, 530

\bibitem[{Fritz {et~al.}(2010)Fritz, Gillessen, Trippe, Ott, Bartko, Pfuhl,
  Dodds-Eden, Davies, Eisenhauer, \& Genzel}]{fritz10}
Fritz, T., Gillessen, S., Trippe, S., {et~al.} 2010, MNRAS, 401, 1177

\bibitem[{Gagn\'e {et~al.}(2014)Gagn\'e, Lafreni\`ere, Doyon, Malo, \&
  Artigau}]{gagne14}
Gagn\'e, J., Lafreni\`ere, D., Doyon, R., Malo, L., \& Artigau, E. 2014, ApJ,
  783, 121

\bibitem[{Gizis {et~al.}(2012)Gizis, Faherty, Liu, Castro, Vrba, Harris, Aller,
  \& Deacon}]{gizis12}
Gizis, J.~E., Faherty, J.~K., Liu, M.~C., {et~al.} 2012, AJ, 144, 94

\bibitem[{Gizis {et~al.}(2003)Gizis, Reid, Knapp, Liebert, Kirkpatrick,
  Koerner, \& Burgasser}]{gizis03}
Gizis, J.~E., Reid, I.~N., Knapp, G.~R., {et~al.} 2003, AJ, 125, 3302

\bibitem[{Golimowski {et~al.}(2004)Golimowski, Leggett, Marley, Fan, Geballe,
  Knapp, Vrba, Henden, Luginbuhl, Guetter, Munn, Canzian, Zheng, Tsvetanov,
  Chiu, Glazebrook, Hoversten, Schneider, \& Brinkmann}]{golimowski04}
Golimowski, D.~A., Leggett, S.~K., Marley, M.~S., {et~al.} 2004, AJ, 127, 3516

\bibitem[{Green(1985)}]{green85}
Green, R.~M. 1985, Spherical Astronomy (University of Cambridge)

\bibitem[{Jameson {et~al.}(2008)Jameson, Casewell, Bannister, Lodieu,
  Keresztes, Dobbie, \& Hodgkin}]{jameson08}
Jameson, R.~F., Casewell, S.~L., Bannister, N.~P., {et~al.} 2008, MNRAS, 384,
  1399

\bibitem[{Johnson \& Soderblom(1987)}]{johnson87}
Johnson, D. R.~H. \& Soderblom, D.~R. 1987, AJ, 93, 864

\bibitem[{Johnson(1966)}]{johnson66}
Johnson, H.~L. 1966, ARA\&A, 4, 193

\bibitem[{Kirkpatrick(2005)}]{kirk05}
Kirkpatrick, J.~D. 2005, ARA\&A, 43, 195

\bibitem[{Kirkpatrick {et~al.}(2008)Kirkpatrick, Cruz, Barman, Burgasser,
  Looper, Tinney, Gelino, Lowrance, Liebert, \& Carpenter}]{kirk08}
Kirkpatrick, J.~D., Cruz, K.~L., Barman, T.~S., {et~al.} 2008, ApJ, 689, 1295

\bibitem[{Kirkpatrick {et~al.}(2011)Kirkpatrick, Cushing, Gelino, Griffith,
  Skrutskie, Marsh, Wright, Mainzer, Eisenhardt, McLean, Thompson, Bauer,
  Benford, Bridge, Lake, Petty, Standford, Tsai, Bailey, Beichman, Bloom,
  Bochanski, Burgasser, Capak, Cruz, Lartaltepe, Knox, Manohar, Masters,
  Morales-Calder\'on, Prato, Rodigas, Salvato, Schurr, Scoville, Simcoe,
  Stapelfeldt, Stern, Stock, \& Vacca}]{kirk11}
Kirkpatrick, J.~D., Cushing, M.~C., Gelino, C.~R., {et~al.} 2011, ApJS, 197, 19

\bibitem[{Kirkpatrick {et~al.}(2012)Kirkpatrick, Gelino, Cushing, Mace,
  Griffith, Skrutskie, Marsh, Wright, Eisenhardt, McLean, Mainzer, Burgasser,
  Tinney, Parker, \& Salter}]{kirk12}
Kirkpatrick, J.~D., Gelino, C.~R., Cushing, M.~C., {et~al.} 2012, ApJ, 753, 156

\bibitem[{Kirkpatrick {et~al.}(1993)Kirkpatrick, Kelly, Rieke, \&
  Liebert}]{kirk93}
Kirkpatrick, J.~D., Kelly, D.~M., Rieke, G.~H., \& Liebert, J. 1993, ApJ, 402,
  643

\bibitem[{Kirkpatrick {et~al.}(1999)Kirkpatrick, Reid, Liebert, Cutri, Nelson,
  Beichman, Dahn, Monet, Gizis, \& Skrutskie}]{kirk99}
Kirkpatrick, J.~D., Reid, I.~N., Liebert, J., {et~al.} 1999, ApJ, 519, 802

\bibitem[{Kirkpatrick {et~al.}(2000)Kirkpatrick, Reid, Liebert, Gizis,
  Burgasser, Monet, Dahn, Nelson, \& Williams}]{kirk00}
Kirkpatrick, J.~D., Reid, I.~N., Liebert, J., {et~al.} 2000, AJ, 120, 447

\bibitem[{Leggett(1992)}]{leggett92}
Leggett, S.~K. 1992, ApJS, 82, 351

\bibitem[{Liu {et~al.}(2011)Liu, Deacon, Magnier, Dupuy, Aller, Bowler,
  Redstone, Goldman, Burgett, Chambers, Hodapp, Kaiser, Kudritzki, Morgan,
  Price, Tonry, \& Wainscoat}]{liu11}
Liu, M.~C., Deacon, N.~R., Magnier, E.~A., {et~al.} 2011, ApJ, 740, L32

\bibitem[{Liu {et~al.}(2013b)Liu, Dupuy, \& Allers}]{liu13b}
Liu, M.~C., Dupuy, T.~J., \& Allers, K.~N. 2013b, Astron. Nachr., 334, 85

\bibitem[{Liu {et~al.}(2013a)Liu, Magnier, Deacon, Allers, Dupuy, Kotson,
  Aller, Burgett, Chambers, Draper, Hodapp, Jedicke, Kaiser, Kudritzki,
  Metcalfe, Morgan, Price, Tonry, \& Wainscoat}]{liu13a}
Liu, M.~C., Magnier, E.~A., Deacon, N.~R., {et~al.} 2013a, ApJ, 777, L20

\bibitem[{Lodieu {et~al.}(2012)Lodieu, Burningham, Day-Jones, Scholz, Marocco,
  Koposov, Barrado~y Navascu\'es, Lucas, Cruz, Lillo, Jones, P\'erez-Garrido,
  Ruiz, Pinfield, Rebolo, B\'ejar, Boudreault, Emerson, Banerji,
  Gonz\'alez-Solares, Hodgkin, McMahon, Canty, \& Contreras}]{lodieu12}
Lodieu, N., Burningham, B., Day-Jones, A., {et~al.} 2012, A\&A, 548, 53

\bibitem[{Looper {et~al.}(2008)Looper, Gelino, Burgasser, \&
  Kirkpatrick}]{looper08}
Looper, D.~L., Gelino, C.~R., Burgasser, A.~J., \& Kirkpatrick, J.~D. 2008,
  ApJL, 685, 1183

\bibitem[{Lucas {et~al.}(2001)Lucas, Roche, Allard, \& Hauschildt}]{lucas01}
Lucas, P.~W., Roche, P.~F., Allard, F., \& Hauschildt, P.~H. 2001, MNRAS, 326,
  695

\bibitem[{Luhman(2012)}]{luhman12rev}
Luhman, K.~L. 2012, ARA\&A, 50, 65

\bibitem[{Luhman {et~al.}(2012)Luhman, Loutrel, McCurdy, Mace, Melso, Star,
  Young, Terrien, McLean, Kirkpatrick, \& Rhode}]{luhman12}
Luhman, K.~L., Loutrel, N.~P., McCurdy, N.~S., {et~al.} 2012, ApJ, 760, 152

\bibitem[{Luhman {et~al.}(2009)Luhman, Mamajek, Allen, \& Cruz}]{luhman09}
Luhman, K.~L., Mamajek, E.~E., Allen, P.~R., \& Cruz, K.~L. 2009, ApJ, 703, 399

\bibitem[{Luhman {et~al.}(2005)Luhman, Stauffer, \& Mamajek}]{luhman05}
Luhman, K.~L., Stauffer, J.~R., \& Mamajek, E.~E. 2005, ApJ, 628, 69

\bibitem[{Makarov \& Urban(2000)}]{makarov00}
Makarov, V.~V. \& Urban, S. 2000, MNRAS, 317, 289

\bibitem[{Marley {et~al.}(2010)Marley, Saumon, \& Goldblatt}]{marley10}
Marley, M.~S., Saumon, D., \& Goldblatt, C. 2010, ApJ, 723, L117

\bibitem[{Marocco {et~al.}(2010)Marocco, Smart, Jones, Burningham, Lattanzi,
  Leggett, Lucas, Tinney, Adamson, Evans, Lodieu, Murray, Pinfield, \&
  Tamura}]{marocco10}
Marocco, F., Smart, R.~L., Jones, H. R.~A., {et~al.} 2010, A\&A, 524, 38

\bibitem[{Marsh {et~al.}(2013)Marsh, Wright, Kirkpatrick, Gelino, Cushing,
  Griffith, Skrutskie, \& Eisenhardt}]{marsh13}
Marsh, K.~A., Wright, E.~L., Kirkpatrick, J.~D., {et~al.} 2013, ApJ, 762, 119

\bibitem[{Mart{\'i}n {et~al.}(1998)Mart{\'i}n, Basri, Gallegos, Rebolo,
  Zapatero~Osorio, \& B\'ejar}]{martin98}
Mart{\'i}n, E.~L., Basri, G., Gallegos, J.~E., {et~al.} 1998, ApJ, 499, L61

\bibitem[{Mart{\'i}n {et~al.}(1999)Mart{\'i}n, Delfosse, Basri, Goldman,
  Forveille, \& Zapatero~Osorio}]{martin99}
Mart{\'i}n, E.~L., Delfosse, X., Basri, G., {et~al.} 1999, AJ, 118, 2466

\bibitem[{Mart{\'in} {et~al.}(2000)Mart{\'in}, Koresko, Kulkarni, Lane, \&
  Wizinowich}]{martin00}
Mart{\'in}, E.~L., Koresko, C.~D., Kulkarni, S.~R., Lane, B.~F., \& Wizinowich,
  P.~L. 2000, ApJ, 529, L37

\bibitem[{Mart{\'i}n {et~al.}(1996)Mart{\'i}n, Rebolo, \&
  Zapatero~Osorio}]{martin96}
Mart{\'i}n, E.~L., Rebolo, R., \& Zapatero~Osorio, M.~R. 1996, ApJ, 469, 706

\bibitem[{McCaughrean {et~al.}(2004)McCaughrean, Close, Scholz, Lenzen, Biller,
  Brandner, Hartung, \& Lodieu}]{mccaughrean04}
McCaughrean, M.~J., Close, L.~M., Scholz, R.~D., {et~al.} 2004, A\&A, 413, 1029

\bibitem[{McLean {et~al.}(2003)McLean, McGovern, Burgasser, Kirkpatrick, Prato,
  \& Kim}]{mclean03}
McLean, I.~S., McGovern, M.~R., Burgasser, A.~J., {et~al.} 2003, ApJ, 596, 561

\bibitem[{Metchev \& Hillenbrand(2006)}]{metchev06}
Metchev, S. \& Hillenbrand, L. 2006, ApJ, 651, 1166

\bibitem[{Mohanty {et~al.}(2007)Mohanty, Jayawardhana, \& Mamajek}]{mohanty07}
Mohanty, S., Jayawardhana, R., H.~N., \& Mamajek, E. 2007, ApJ, 657, 1064

\bibitem[{Monet {et~al.}(1992)Monet, Dahn, Vrba, Harris, Pier, Luginbuhl, \&
  Ables}]{monet92}
Monet, D.~G., Dahn, C.~C., Vrba, F.~J., {et~al.} 1992, AJ, 103, 638

\bibitem[{Pe\~na Ram{\'i}rez {et~al.}(2012)Pe\~na Ram{\'i}rez, B\'ejar,
  Zapatero~Osorio, Petr-Gotzens, \& Mart{\'i}n}]{pena12}
Pe\~na Ram{\'i}rez, K., B\'ejar, V. J.~S., Zapatero~Osorio, M.~R.,
  Petr-Gotzens, M.~G., \& Mart{\'i}n, E.~L. 2012, ApJ, 754, 30

\bibitem[{Perryman {et~al.}(1997)Perryman, Lindegren, Kovalevsky, Hoeg,
  Bastian, Bernacca, Cr\'ez\'e, Donati, Grenon, Grewing, van Leeuwen, van~der
  Marel, Mignard, Murray, Le~Poole, Schrijver, Turon, Arenou, Froeschlé, \&
  Petersen}]{perryman97}
Perryman, M. A.~C., Lindegren, L., Kovalevsky, J., {et~al.} 1997, A\&A, 323,
  L49

\bibitem[{Plavchan {et~al.}(2008)Plavchan, Jura, Kirkpatrick, Cutri, \&
  Gallagher}]{plavchan08}
Plavchan, P., Jura, M., Kirkpatrick, J.~D., Cutri, R.~M., \& Gallagher, S.~C.
  2008, ApJS, 175, 191

\bibitem[{Rayner {et~al.}(2009)Rayner, Cushing, \& Vacca}]{rayner09}
Rayner, J.~T., Cushing, M.~C., \& Vacca, W.~D. 2009, ApJS, 185, 289

\bibitem[{Rebolo {et~al.}(1996)Rebolo, Mart{\'i}n, Basri, Marcy, \&
  Zapatero~Osorio}]{rebolo96}
Rebolo, R., Mart{\'i}n, E.~L., Basri, G., Marcy, G.~W., \& Zapatero~Osorio,
  M.~R. 1996, ApJ, 469, L53

\bibitem[{Rebolo {et~al.}(1992)Rebolo, Mart{\'i}n, \& Magazz\`u}]{rebolo92}
Rebolo, R., Mart{\'i}n, E.~L., \& Magazz\`u, A. 1992, ApJ, 389, L83

\bibitem[{Rebolo {et~al.}(1998)Rebolo, Zapatero~Osorio, Madruga, B\'ejar,
  Arribas, \& Licandro}]{rebolo98}
Rebolo, R., Zapatero~Osorio, M.~R., Madruga, S., {et~al.} 1998, Science, 282,
  1309

\bibitem[{Reid {et~al.}(2001)Reid, Gizis, Kirkpatrick, \& Koerner}]{reid01}
Reid, I.~N., Gizis, J.~E., Kirkpatrick, J.~D., \& Koerner, D.~W. 2001, AJ, 121,
  489

\bibitem[{Reid {et~al.}(2006)Reid, Lewitus, Allen, Cruz, \& Burgasser}]{reid06}
Reid, I.~N., Lewitus, E., Allen, P.~R., Cruz, K.~L., \& Burgasser, A.~J. 2006,
  AJ, 132, 891

\bibitem[{Roberts {et~al.}(1987)Roberts, Lehar, \& Dreher}]{roberts87}
Roberts, D.~H., Lehar, J., \& Dreher, J.~W. 1987, AJ, 93, 968

\bibitem[{Saumon {et~al.}(1996)Saumon, Hubbard, Burrows, Guillot, Lunine, \&
  Chabrier}]{saumon96}
Saumon, D., Hubbard, W.~B., Burrows, A., {et~al.} 1996, ApJ, 460, 993

\bibitem[{Scargle(1982)}]{scargle82}
Scargle, J.~D. 1982, ApJ, 263, 835

\bibitem[{Schilbach {et~al.}(2009)Schilbach, Roser, \& Scholz}]{schilbach09}
Schilbach, E., Roser, S., \& Scholz, R.-D. 2009, A\&A, 493, L27

\bibitem[{Scholz {et~al.}(2012)Scholz, Bihain, Schnurr, \& Stormo}]{scholz12}
Scholz, R.-D., Bihain, G., Schnurr, O., \& Stormo, J. 2012, A\&A, 541, 163

\bibitem[{Schweitzer {et~al.}(2001)Schweitzer, Gizis, Hauschildt, Allard, \&
  Reid}]{schweitzer01}
Schweitzer, A., Gizis, J.~E., Hauschildt, P.~H., Allard, F., \& Reid, I.~N.
  2001, ApJ, 555, 368

\bibitem[{Seifahrt {et~al.}(2010)Seifahrt, Reiners, \& Almaghrbi}]{seifahrt10}
Seifahrt, A., Reiners, A., \& Almaghrbi, K. A.~M., B.~G. 2010, A\&A, 512, A37

\bibitem[{Sherry {et~al.}(2008)Sherry, Walter, Wolk, \& Adams}]{sherry08}
Sherry, W.~H., Walter, F.~M., Wolk, S.~J., \& Adams, N.~R. 2008, AJ, 135, 1616

\bibitem[{Skrutskie {et~al.}(2006)Skrutskie, Cutri, Stiening, Weinberg,
  Schneider, Carpenter, Beichman, Capps, Chester, Huchra, Liebert, Lonsdale,
  Monet, Price, Seitzer, Jarrett, Kirkpatrick, Gizis, Howard, Evans, Fowler,
  Fullmer, Hurt, Light, Kopan, Marsh, McCallon, Tam, Van~Dyk, \&
  Wheelock}]{skrutskie06}
Skrutskie, M.~F., Cutri, R.~M., Stiening, R., {et~al.} 2006, AJ, 131, 1163

\bibitem[{Stauffer {et~al.}(1998)Stauffer, Schultz, \&
  Kirkpatrick}]{stauffer98}
Stauffer, J.~R., Schultz, G., \& Kirkpatrick, J.~D. 1998, ApJ, 499, L199

\bibitem[{Stephens {et~al.}(2009)Stephens, Leggett, Cushing, Marley, Saumon,
  Geballe, Golimowski, Fan, \& Noll}]{stephens09}
Stephens, D.~C., Leggett, S.~K., Cushing, M.~C., {et~al.} 2009, ApJ, 702, 154

\bibitem[{Stone(2002)}]{stone02}
Stone, R.~C. 2002, PASP, 114, 1070

\bibitem[{Stumpf {et~al.}(2010)Stumpf, Brandner, Joergens, Henning, Bouy,
  K{\"o}hler, \& Kasper}]{stumpf10}
Stumpf, M.~B., Brandner, W., Joergens, V., {et~al.} 2010, ApJ, 724, 1

\bibitem[{Testi(2009)}]{testi09}
Testi, L. 2009, A\&A, 503, 639

\bibitem[{Tinney {et~al.}(2003)Tinney, Burgasser, \& Kirkpatrick}]{tinney03}
Tinney, C.~G., Burgasser, A.~J., \& Kirkpatrick, J.~D. 2003, AJ, 126, 975

\bibitem[{Todorov {et~al.}(2010)Todorov, Luhman, \& McLeod}]{todorov10}
Todorov, K., Luhman, K.~L., \& McLeod, K.~K. 2010, ApJ, 714, L84

\bibitem[{Torres {et~al.}(2008)Torres, Quast, Melo, \& Sterzik}]{torres08}
Torres, C. A.~O., Quast, G.~R., Melo, C. H.~F., \& Sterzik, M.~F. 2008,
  Handbook of Star Forming Regions, Vol. II: The Southern Sky (ed. B. Reipurth.
  ASP Monograph Publications 5; San Francisco, CA: ASP), 757

\bibitem[{van Leeuwen(2009a)}]{vanLeeuwen09a}
van Leeuwen, F. 2009a, A\&A, 497, 209

\bibitem[{van Leeuwen(2009b)}]{vanLeeuwen09b}
van Leeuwen, F. 2009b, A\&A, 500, 505

\bibitem[{Volk {et~al.}(2003)Volk, Blum, Walker, \& Puxley}]{volk03}
Volk, K., Blum, R., Walker, G., \& Puxley, P. 2003, IAU Circ., 8188

\bibitem[{Vrba {et~al.}(2004)Vrba, Henden, Luginbulh, Guetter, Munn, Canzian,
  Burgasser, Kirkpatrick, Geballe, Golimowski, Knapp, Leggett, Schneider, \&
  Brinkmann}]{vrba04}
Vrba, F.~J., Henden, A.~A., Luginbulh, C.~B., {et~al.} 2004, AJ, 127, 2948

\bibitem[{Wright {et~al.}(2010)Wright, Eisenhardt, Mainzer, Ressler, Cutri,
  Jarrett, McMillan, Skrutskie, Stanford, Cohen, Walker, Mather, Leisawitz,
  Gautier, McLean, Benford, Lonsdale, Blain, Mendez, Irace, Duval, Liu, Royer,
  Heinrichsen, Howard, Shannon, Kendall, Walsh, Larsen, Cardon, Schick,
  Schwalm, Abid, Fabinsky, Naes, \& Tsai}]{wright10}
Wright, E.~L., Eisenhardt, P. R.~M., Mainzer, A.~K., {et~al.} 2010, AJ, 140,
  1868

\bibitem[{Wright {et~al.}(2011)Wright, Mainzer, Gelino, \&
  Kirkpatrick}]{wright11}
Wright, E.~L., Mainzer, A., Gelino, C., \& Kirkpatrick, J.~D. 2011,
  arXiv:1104.2569

\bibitem[{Zapatero~Osorio {et~al.}(2011)Zapatero~Osorio, B\'ejar, Goldman,
  Caballero, Rebolo, Acosta-Pulido, Manchado, \& Pe\~na Ram\'\i~rez}]{osorio11}
Zapatero~Osorio, M.~R., B\'ejar, V. J.~S., Goldman, B., {et~al.} 2011, ApJ,
  740, 4

\bibitem[{Zapatero~Osorio {et~al.}(2002)Zapatero~Osorio, B\'ejar, Pavlenko,
  Rebolo, Allende~Prieto, Mart{\'i}n, \& Garc{\'i}a~L\'opez}]{osorio02}
Zapatero~Osorio, M.~R., B\'ejar, V. J.~S., Pavlenko, Y., {et~al.} 2002, A\&A,
  384, 937

\bibitem[{Zapatero~Osorio {et~al.}(2004)Zapatero~Osorio, Lane, Pavlenko,
  Mart{\'i}n, Britton, \& Kulkarni}]{osorio04}
Zapatero~Osorio, M.~R., Lane, B.~F., Pavlenko, Y., {et~al.} 2004, ApJ, 615, 958

\bibitem[{Zapatero~Osorio {et~al.}(2010)Zapatero~Osorio, Rebolo, Bihain,
  B\'ejar, Caballero, \& \'Alvarez}]{osorio10}
Zapatero~Osorio, M.~R., Rebolo, R., Bihain, G., {et~al.} 2010, ApJ, 715, 1408

\bibitem[{Zuckerman {et~al.}(2011)Zuckerman, Rhee, Song, \&
  Bessell}]{zuckerman11}
Zuckerman, B., Rhee, J.~H., Song, I., \& Bessell, M.~S. 2011, ApJ, 732, 61

\bibitem[{Zuckerman \& Song(2004)}]{zuckerman04}
Zuckerman, B. \& Song, I. 2004, ARA\&A, 42, 685

\end{thebibliography}


\addtocounter{table}{-8}
\begin{landscape}
\begin{table*}
\caption{List of targets and infrared photometry. \label{targets}}
\tiny
\begin{tabular}{llrrrrrrrrr}
\hline  \hline
Object & 
SpT\tablefootmark{a} & 
\multicolumn{1}{c}{$J$\tablefootmark{b}} & 
\multicolumn{1}{c}{$K_s$\tablefootmark{b}}  &  
\multicolumn{1}{c}{$W1$\tablefootmark{c}}  &  
\multicolumn{1}{c}{$W2$\tablefootmark{c}}  &  
\multicolumn{1}{c}{$W3$\tablefootmark{c}}  &
\multicolumn{1}{c}{$[3.6]$\tablefootmark{d}}  &  
\multicolumn{1}{c}{$[4.5]$\tablefootmark{d}}  &  
\multicolumn{1}{c}{$[5.8]$\tablefootmark{d}}  &  
\multicolumn{1}{c}{$[8.0]$\tablefootmark{d}}  \\
  &   & 
\multicolumn{1}{c}{(mag)} & 
\multicolumn{1}{c}{(mag)} &  
\multicolumn{1}{c}{(mag)} &  
\multicolumn{1}{c}{(mag)} &  
\multicolumn{1}{c}{(mag)} &  
\multicolumn{1}{c}{(mag)} &  
\multicolumn{1}{c}{(mag)} &  
\multicolumn{1}{c}{(mag)} &  
\multicolumn{1}{c}{(mag)} \\
\hline
2MASS\,J00332386$-$1521309     & L4$\beta$  & 15.29$\pm$0.06 & 13.41$\pm$0.04 & 12.80$\pm$0.03 & 12.48$\pm$0.03 & 11.89$\pm$0.25 & 12.54$\pm$0.02 & 12.48$\pm$0.02 & 12.21$\pm$0.03 & 12.04$\pm$0.03 \\
2MASS\,J00452143$+$1634446\tablefootmark{e} & L2$\beta$  & 13.06$\pm$0.02 & 11.37$\pm$0.02 & 10.77$\pm$0.02 & 10.39$\pm$0.02 &  9.74$\pm$0.04 &                &                &                &                \\
2MASS\,J02411151$-$0326587     & L0$\gamma$ & 15.80$\pm$0.06 & 14.04$\pm$0.05 & 13.64$\pm$0.03 & 13.26$\pm$0.03 & 12.77$\pm$0.42 & 13.39$\pm$0.02 & 13.24$\pm$0.02 & 13.04$\pm$0.03 & 12.77$\pm$0.03 \\
2MASS\,J03552337$+$1133437 & L5$\gamma$ & 14.05$\pm$0.02 & 11.53$\pm$0.02 & 10.53$\pm$0.02 &  9.94$\pm$0.02 &  9.29$\pm$0.04 &                &                &                &                \\
2MASS\,J05012406$-$0010452\tablefootmark{f} & L4$\gamma$ & 14.98$\pm$0.04 & 12.96$\pm$0.04 & 12.05$\pm$0.02 & 11.52$\pm$0.02 & 10.95$\pm$0.11 & 11.77$\pm$0.02 & 11.52$\pm$0.02 & 11.22$\pm$0.03 & 11.03$\pm$0.03 \\
G\,196--3B\tablefootmark{g}             & L3$\beta$  & 14.83$\pm$0.05 & 12.78$\pm$0.03 &                &                &                & 11.66$\pm$0.02 & 11.47$\pm$0.04 & 11.10$\pm$0.06 & 10.93$\pm$0.04 \\
2MASS\,J10224821$+$5825453     & L1$\beta$  & 13.50$\pm$0.03 & 12.16$\pm$0.03 & 11.76$\pm$0.02 & 11.50$\pm$0.02 & 11.20$\pm$0.11 &                &                &                &                \\
2MASS\,J15525906$+$2948485     & L0$\beta$  & 13.48$\pm$0.03 & 12.02$\pm$0.03 & 11.54$\pm$0.02 & 11.21$\pm$0.02 & 10.66$\pm$0.05 &                &                &                &                \\
2MASS\,J17260007$+$1538190 & L3$\beta$  & 15.67$\pm$0.07 & 13.66$\pm$0.05 & 13.07$\pm$0.03 & 12.69$\pm$0.03 & 11.56$\pm$0.16 & 12.76$\pm$0.02 & 12.64$\pm$0.02 & 12.41$\pm$0.03 & 12.20$\pm$0.03 \\
2MASS\,J22081363$+$2921215 & L3$\gamma$ & 15.80$\pm$0.09 & 14.15$\pm$0.07 & 13.35$\pm$0.03 & 12.89$\pm$0.03 &                & 13.08$\pm$0.02 & 12.89$\pm$0.02 & 12.62$\pm$0.03 & 12.33$\pm$0.03 \\                           
\hline
\end{tabular}
\tablefoot{
\tablefoottext{a}{Based on optical spectra from \citet{cruz09}. As also explained in \citet{kirk05}, the $\beta$ and $\gamma$ appended to the L subtypes indicate intermediate- and very low-gravity spectra, respectively. } 
\tablefoottext{b}{2MASS photometry \citep{skrutskie06}.}
\tablefoottext{c}{{\sl WISE} photometry.}
\tablefoottext{d}{{\sl Spitzer} photometry.}
\tablefoottext{e}{$W4$\,=\,8.42$\pm$0.26 mag. Other sources are not detected in the $W4$ filter.}
\tablefoottext{f}{{\sl Spitzer} [24]\,=\,10.78$\pm$0.15 mag. }
\tablefoottext{g}{{\sl Spitzer} [24]\,=\,10.55$\pm$0.10 mag. }
}
\end{table*}
\end{landscape}


\addtocounter{table}{1}
\longtab{2}{
\begin{longtable}{llcclll}
\caption{Log of the astrometric observations. \label{obslog}}\\
\hline\hline
Object & Obs$.$ date & Exposure & FWHM & Instrument & Filter & Comment \\
            &                     & (s)           & ($"$)    &                  &           &                \\
\hline
\endfirsthead
\caption{Continued.}\\
\hline \hline
Object & Obs$.$ date & Exposure & FWHM & Instrument & Filter & Comment \\
            &                     & (s)           & ($"$)    &                  &           &                \\
\hline
\endhead
\hline
\endfoot
J0033$-$1521 & 2010 Jan 31 &  6$\times$25$\times$2     &  1.4    &  OMEGA2000 & $K_s$ & \\
             & 2010 Jun 04 &  4$\times$9$\times$60     &  0.8    &  NOTCam    & $J$   & \\
             & 2010 Jun 26 &  6$\times$25$\times$2     &  1.2    &  OMEGA2000 & $K_s$ &  \\
             & 2010 Jul 13 &  4$\times$9$\times$60     &  0.8    &  NOTCam    & $J$   & \\
             & 2010 Sep 17 &  252$\times$2             &  0.6    &  LIRIS     & $K_s$ &  \\
             & 2010 Sep 29 &  3$\times$9$\times$60     &  1.0    &  NOTCam    & $J$   & \\
             & 2010 Oct 30 &  3$\times$9$\times$60     &  0.8    &  NOTCam    & $J$   & \\
             & 2011 Aug 13 &  6$\times$25$\times$2     &  1.2    &  OMEGA2000 & $K_s$ &  \\
\hline
\end{longtable}
}

\addtocounter{table}{1}
\longtab{2}{
\begin{longtable}{lrrccccl}
\caption{Relative astrometric measurements. \label{astrodata}}\\
\hline\hline
\multicolumn{1}{l}{JD$-$240000} & 
\multicolumn{1}{c}{d$\alpha$} &
\multicolumn{1}{c}{d$\delta$} &
\multicolumn{1}{c}{err d$\alpha$} &
\multicolumn{1}{c}{err d$\delta$} &
\multicolumn{1}{c}{Air mass} &
\multicolumn{1}{c}{UT date} &
\multicolumn{1}{c}{Instrument} \\
\multicolumn{1}{l}{} & 
\multicolumn{1}{c}{(mas)} &
\multicolumn{1}{c}{(mas)} &
\multicolumn{1}{c}{(mas)} &
\multicolumn{1}{c}{(mas)} &
\multicolumn{1}{c}{} &
\multicolumn{1}{c}{} &
\multicolumn{1}{c}{} \\
\hline
\endfirsthead
\caption{Continued.}\\
\hline \hline
\multicolumn{1}{l}{JD$-$240000} & 
\multicolumn{1}{c}{d$\alpha$} &
\multicolumn{1}{c}{d$\delta$} &
\multicolumn{1}{c}{err d$\alpha$} &
\multicolumn{1}{c}{err d$\delta$} &
\multicolumn{1}{c}{Air mass} &
\multicolumn{1}{c}{UT date} &
\multicolumn{1}{c}{Instrument} \\
\multicolumn{1}{l}{} & 
\multicolumn{1}{c}{(mas)} &
\multicolumn{1}{c}{(mas)} &
\multicolumn{1}{c}{(mas)} &
\multicolumn{1}{c}{(mas)} &
\multicolumn{1}{c}{} &
\multicolumn{1}{c}{} &
\multicolumn{1}{c}{} \\
\hline
\endhead
\hline
\endfoot
\hline
\multicolumn{8}{c}{J0033$-$1521} \\
 5228.2694    &         0.00 &      0.00 &   25 & 25  &   2.195 &  2010 Jan 31  & OMEGA2000   \\
 5351.7104    &    $+$183.84 &   $+$0.49 &   50 & 50  &   3.116 &  2010 Jun 04  & NOTCam   \\
 5373.6298    &    $+$161.10 &  $+$67.86 &   25 & 25  &   3.004 &  2010 Jun 26  & OMEGA2000   \\
 5390.6961    &    $+$206.58 &  $+$23.66 &   50 & 50  &   1.558 &  2010 Jul 13  & NOTCam   \\
 5456.5701    &    $+$224.33 &   $+$4.30 &   60 & 60  &   1.396 &  2010 Sep 17  & LIRIS \\
 5468.5106    &    $+$215.99 &  $-$20.79 &   50 & 50  &   1.446 &  2010 Sep 29  & NOTCam   \\
 5500.4733    &    $+$214.59 &  $-$29.75 &   50 & 50  &   1.390 &  2010 Oct 30  & NOTCam   \\
\hline
\end{longtable}
\tablefoot{
\tablefoottext{*}{NOTCam and OMEGA2000 normalization dates.}
}
}


\Online

\begin{appendix}\label{appendix}
The apparent astrometric trajectories of all targets and their best fits, except for the two shown in the main text, are provided in Figures~\ref{astro2} and~\ref{astro3}. The astrometric residuals as a function of observing epoch are depicted in Figures~\ref{res2} and~\ref{res3}. Astrometric residuals as a function of observing air mass are illustrated in Figures~\ref{airmass2} and~\ref{airmass3}.

\end{appendix}
\end{document}